\documentclass[a4paper,10pt]{article}

\usepackage{jheppub} 

\usepackage[T1]{fontenc} 
\usepackage{bbold}
\usepackage{amsmath}
\usepackage{empheq}
\usepackage{booktabs}
\usepackage{slashed}
\usepackage{url}

\newcommand{\rd}{{\rm d}}
\newcommand{\phs}{{\phi^{*}_\eta}}
\newcommand{\re}{{\rm e}}
\newcommand{\rC}{{\rm C}}
\newcommand{\rG}{{\rm G}}

\newcommand{\as}{\alpha_s}

\newcommand{\GeV}{\;\mathrm{GeV}}

\newcommand{\eps}{\epsilon}

\newcommand{\pth}{p_t^{H}}

\usepackage{xspace}
\newcommand{\nnnllp}{N$^3$LL$^\prime$\xspace}
\newcommand{\nnllp}{NNLL$^\prime$\xspace}
\newcommand{\nllp}{NLL$^\prime$\xspace}
\newcommand{\nnnnll}{N$^4$LL\xspace}
\newcommand{\nnnll}{N$^3$LL\xspace}
\newcommand{\nnnlo}{N$^3$LO\xspace}
\newcommand{\nnll}{NNLL\xspace}
\newcommand{\radish}{\textsc{RadISH}\xspace}
\newcommand{\nnlojet}{\textsc{NNLOjet}\xspace}

\newcommand{\pt}{p_{t}}

\newcommand{\mll}{\ensuremath{M_{\ell\ell}}\xspace}
\newcommand{\ptll}{\ensuremath{\pt^{\ell\ell}}\xspace}
\newcommand{\ptyy}{\ensuremath{\pt^{\gamma\gamma}}\xspace}

\newcommand{\dZ}{\rd{\cal Z}[\{R', k_i\}]}

\def\lq{\left[} 
\def\rq{\right]} 
\def\rg{\right\}} 
\def\lg{\left\{} 
\def\({\left(} 
\def\){\right)}

\newcommand{\logMtwomurtwo[1]}{\ln^{#1}{\frac{M^2}{\mu_R^2}}}
\newcommand{\logQtwomurtwo[1]}{\ln^{#1}{\frac{Q^2}{\mu_R^2}}}

\newcommand{\MSbar}{$\rm \overline{MS}$}
\newcommand{\mz}{M_Z}
\newcommand{\gammaz}{\Gamma_Z}
\newcommand{\mubar}{\bar{\mu}}

\newcommand{\Amp}[1]{\mathcal{M}^{(#1)}}
\newcommand{\JJq}[2]{[J\hspace{-0.15cm}J^{(#1)}_{#2}(q)]}
\newcommand{\JJl}[2]{[J\hspace{-0.15cm}J^{(#1)}_{#2}(\ell)]}

\usepackage{xcolor}

\title{\boldmath Fiducial Higgs and Drell-Yan distributions at N$^3$LL$^\prime$+NNLO with RadISH}

\author[a]{Emanuele Re,}
\author[b]{Luca Rottoli,}
\author[c]{and Paolo Torrielli.}

\affiliation[a]{LAPTh, Universit\'e Grenoble Alpes, Universit\'e Savoie Mont Blanc, CNRS, F-74940 Annecy, France}
\affiliation[b]{Department of Physics, University of Z\"urich, CH-8057 Z\"urich, Switzerland}
\affiliation[c]{Dipartimento di Fisica and Arnold-Regge Center, Universit\`a di Torino, and INFN, Sezione di Torino,
Via P. Giuria 1, I-10125 Torino, Italy}

\emailAdd{emanuele.re@lapth.cnrs.fr}
\emailAdd{luca.rottoli@physik.uzh.ch}
\emailAdd{torriell@to.infn.it}

\abstract{
We present state-of-the-art predictions for transverse observables relevant to
colour-singlet production at the LHC, in particular the transverse momentum of
the colour singlet in gluon-fusion Higgs production and in neutral Drell-Yan
lepton-pair production, as well as the $\phi^*_\eta$ observable in Drell Yan.
We perform a next-to-next-to-next-to-leading logarithmic (N$^3$LL) resummation
of such observables in momentum space according to the \textsc{RadISH} formalism,
consistently including in our prediction all constant terms of relative order
$\alpha_s^3$ with respect to the Born, thereby achieving N$^3$LL$^\prime$ accuracy.
The calculation is fully exclusive with respect to the Born kinematics, which
allows the application of arbitrary fiducial selection cuts on the decay products
of the colour singlet.
We supplement our results with a transverse-recoil prescription, accounting for dominant
classes of subleading-power corrections in a fiducial setup.
The resummed predictions are matched with fixed-order differential spectra at
next-to-next-to-leading order (NNLO) accuracy.
A phenomenological comparison is carried out with 13 TeV LHC data relevant to the
Higgs to di-photon channel, as well as to neutral Drell-Yan lepton-pair production.
Overall, the inclusion of ${\cal O}(\alpha_s^3)$ constant terms, and to a lesser extent
of transverse-recoil effects, proves beneficial for the comparison of theoretical
predictions to data, leaving a residual theoretical uncertainty in the resummation
region at the 2\,-\,5\% level for Drell-Yan observables, and 5\,-\,7\% in Higgs
production.
}

\preprint{LAPTH-015/21, ZU-TH 17/21}

\allowdisplaybreaks

\begin{document} 
\maketitle
\flushbottom

\section{Introduction}

The experimental data collected in Run I and II at the Large Hadron Collider (LHC) has so
far shown no significant deviation from the predictions of the Standard Model (SM) of
particle physics.
Since signals of new physics could emerge as tiny distortions in the spectra of sensitive
observables with respect to the SM baseline, the availability of very accurate theoretical
calculations, chiefly at the differential level, is of paramount importance.

Processes featuring a colour-singlet system in the final state, such as Drell-Yan (DY)
production or Higgs (H) gluon-fusion production, play a central role in the LHC precision
programme.
In particular, observables which depend only on the total transverse momentum of the associated
QCD radiation represent an especially favourable environment both from the theoretical and the
experimental viewpoint. On the one hand they feature comparatively low complexity, allowing one
to push perturbation theory to its limits; on the other hand, their little sensitivity to multi-parton interactions and non-perturbative modelling allows a particularly clean comparison between
the theoretical predictions and the extremely precise experimental data, thereby challenging the
accuracy of the former.

QCD corrections for Drell-Yan production are available at very high accuracy.
The total cross section is known fully differentially in the Born variables up to
next-to-next-to-leading order (NNLO) accuracy~\cite{Hamberg:1990np,vanNeerven:1991gh,
Anastasiou:2003yy,Melnikov:2006di,Melnikov:2006kv,
Catani:2010en,Catani:2009sm,Gavin:2010az,Anastasiou:2003ds}; the inclusive cross section
has been recently computed at next-to-NNLO (\nnnlo) for neutral DY mediated by a virtual
photon \cite{Duhr:2020seh}, and for charged DY \cite{Duhr:2020sdp}.
Very recently, N$^3$LO predictions within fiducial cuts have been presented in \cite{Camarda:2021ict}.
Differential distributions for the singlet's transverse momentum $\pt$ and for the $\phs$ observable
\cite{Banfi:2010cf} are available up to NNLO QCD both for $Z$ and $W$ production~\cite{Ridder:2015dxa,
Ridder:2016nkl,Gehrmann-DeRidder:2016jns,Gauld:2017tww,Boughezal:2015ded,Boughezal:2016isb,
Boughezal:2015dva,Boughezal:2016dtm,Gehrmann-DeRidder:2017mvr}. 

Fixed-order predictions for Higgs production in gluon fusion are also available at very high precision.
The inclusive cross section is known at \nnnlo accuracy in QCD in the heavy-top-quark limit~\cite{deFlorian:1999zd,Harlander:2002wh,Anastasiou:2002yz,Ravindran:2003um,Ravindran:2002dc,
Anastasiou:2015vya,Anastasiou:2016cez,Mistlberger:2018etf}.
Within this approximation, the Higgs rapidity distribution was computed at N$^3$LO
in~\cite{Cieri:2018oms,Dulat:2018bfe}, and the first fully-differential computation
at N$^3$LO was presented in Ref.~\cite{Chen:2021isd}. Predictions for the fiducial
cross section at N$^3$LO also appeared lately~\cite{Billis:2021ecs}.
The $p_t$ distribution is known at NNLO accuracy~\cite{Boughezal:2015dra,Boughezal:2015aha,
Caola:2015wna,Chen:2016zka} in the heavy-top-quark limit, and the impact of finite quark-mass effects
has been computed at NLO~\cite{Harlander:2012hf,Melnikov:2017pgf,Lindert:2017pky,Lindert:2018iug,Neumann:2018bsx,
Caola:2018zye,Jones:2018hbb}.

It is well know that fixed-order predictions must be supplemented with the all-order resummation of
enhanced logarithmic contributions which arise in the phase-space region dominated by soft and/or
collinear QCD radiation; by denoting with $v$ a generic dimensionless transverse observable, i.e.~one
not depending on the radiation's rapidity (for instance $\pt/M$ or $\phs$, $M$ being the mass of the
colour singlet), such a region corresponds to the $v \to 0$ limit.
The resummation of $v$ spectra in colour-singlet production is customarily performed in impact-parameter
$b$-space, where the phase-space constraints factorise~\cite{Parisi:1979se,Collins:1984kg}.
Using the $b$-space formalism, the $p_t$ distribution in Higgs production has been resummed at
next-to-next-to-leading logarithmic (NNLL) accuracy in Refs.~\cite{Bozzi:2003jy,Bozzi:2005wk,
deFlorian:2012mx}, within the approach of~\cite{Collins:1984kg,Catani:2000vq}, and in
Ref.~\cite{Becher:2012yn} using Soft-Collinear Effective Theory (SCET);
\nnnll resummation was considered in Ref.~\cite{Chen:2018pzu,Becher:2020ugp}.
As for DY, $p_t$ and $\phs$ have been resummed in $b$-space at NNLL in
Refs.~\cite{Bozzi:2010xn,Becher:2010tm,Banfi:2012du,GarciaEchevarria:2011rb,Kang:2017cjk}
and at next-to-\nnll (\nnnll) accuracy in Refs.~\cite{Becher:2019bnm,Becher:2020ugp,
Bertone:2019nxa,Bacchetta:2019sam,Ebert:2020dfc}.

As an alternative to $b$-space resummation, the \radish framework for the resummation of
transverse observables in momentum space
has been introduced in Refs.~\cite{Monni:2016ktx,Bizon:2017rah}, which bases the resummation on
a flexible Monte Carlo (MC) formulation (see also Ref.~\cite{Ebert:2016gcn} for a study of
direct-space $p_t$ resummation in SCET).
Resummed predictions at \nnnll accuracy within the \radish formalism have been presented for
Higgs production at the inclusive level in Ref.~\cite{Bizon:2017rah} and within fiducial cuts in
Ref.~\cite{Bizon:2018foh}. For Drell-Yan production, \nnnll \radish predictions for both $p_t$ and $\phs$
have been achieved in Refs.~\cite{Bizon:2018foh,Bizon:2019zgf}, and also considered in \cite{Alioli:2021qbf}.
\nnnll results for generic colour-singlet production, see for instance \cite{Wiesemann:2020gbm},
are available through the automated MATRIX+\radish interface \cite{Grazzini:2017mhc,Kallweit:2020gva}.
Moreover, the momentum-space formulation is at the core of recent applications in the context of
matching NNLO calculations with parton-shower simulations (NNLO+PS)
\cite{Monni:2019whf,Monni:2020nks,Alioli:2021qbf}.

In this article we consider again the Higgs $p_t$ distribution in gluon fusion, and the di-lepton
$p_t$ and $\phs$ distributions in DY, and present state-of-the-art resummed predictions in
which we consistently supplement known \nnnll results with the inclusion of all constant terms of
relative order $\alpha_s^3$ in the resummation, reaching so-called `primed' accuracy \nnnllp.
While the \nnnlo hard functions for DY and for Higgs production in the $m_{\rm top} \rightarrow \infty$
limit have been known for some time \cite{Chetyrkin:1997un,Schroder:2005hy,Gehrmann:2010ue}, reaching \nnnllp
accuracy for these processes requires, as also done in \cite{Billis:2021ecs,Camarda:2021ict}, to supplement
the ingredients deduced in \cite{Catani:2011kr,Catani:2012qa,Gehrmann:2014yya,Luebbert:2016itl,
Echevarria:2016scs,Li:2016ctv,Vladimirov:2016dll,Moch:2017uml,Moch:2018wjh,Lee:2019zop,Luo:2019bmw,Henn:2019swt,
Bruser:2019auj,Henn:2019rmi,vonManteuffel:2020vjv} with the quark and gluon
transverse-momentum dependent (TMD) beam functions at \nnnlo{}, which were recently
obtained via two independent calculations in Refs.~\cite{Luo:2019szz,Ebert:2020yqt,Luo:2020epw}.
Our predictions are further improved by the inclusion of transverse-recoil effects, which we achieve
by implementing in \radish the prescription of Ref.~\cite{Catani:2015vma}.

We combine our resummed \nnnllp results with fixed-order differential spectra at NNLO accuracy from
\nnlojet \cite{Ridder:2015dxa,Ridder:2016nkl,Gehrmann-DeRidder:2016jns,Chen:2016zka}, and we present
matched \nnnllp{}+NNLO predictions within fiducial cuts in comparison with 13 TeV LHC experimental
data relevant to Drell-Yan di-lepton production~\cite{Aad:2019wmn}, and to Higgs di-photon
production~\cite{ATLAS-CONF-2019-029}.
\\

This manuscript is structured as follows: in Sec.~\ref{sec:rad_review} we review the \radish formalism for
resummation in momentum space, up to \nnnll order; Sec.~\ref{sec:primed} details the consistent inclusion of
constant ${\cal O}(\as^3)$ terms, necessary to reach \nnnllp accuracy, and of transverse-recoil effects;
in Sec.~\ref{sec:validation} we report on the tests we have performed to validate the correct implementation
of the new contributions; phenomenological results at the LHC are presented in Sec.~\ref{sec:pheno_results},
and we give our conclusions in Sec.~\ref{sec:conclusions}. We collect in Appendix \ref{app:luminosities}
some formulae relevant for resummation up to \nnnllp, while Appendix \ref{app:axial} discusses
subtleties related to the axial-vector structure of the three-loop DY form factor.

\section{Momentum-space resummation in \radish}
\label{sec:rad_review}
The \radish approach, developed in Refs.~\cite{Monni:2016ktx,Bizon:2017rah},
is designed to resum recursively infrared and collinear (rIRC) safe observables
\cite{Banfi:2004yd} in momentum space.
This is achieved by exploiting the factorisation properties of QCD squared
matrix elements to devise a Monte Carlo formulation of the all-order calculation,
effectively resumming large logarithms by generating soft and/or collinear
radiation as an event generator of definite logarithmic accuracy.

The starting point is the cumulative probability for observable
$V(\{\tilde p\},k_1,\dots,k_n)$ (which, without loss of generality we assume
as dimensionless) to be smaller than a certain value $v$
\begin{equation}
\label{eq:cumulative}
\Sigma(v)
\, \equiv \,
\int_0^v \rd V\; \frac{\rd \Sigma(V)}{\rd V}
\, ,
\end{equation}
where $\{\tilde p\}= \tilde p_1, \tilde p_2$ are the Born momenta of the incoming partons,
and $k_1,\dots,k_n$ are the momenta of radiated QCD partons.
Even though the formalism is in principle extendible to \emph{generic} rIRC safe observables,
in the present article, as was done in Refs.~\cite{Monni:2016ktx,Bizon:2017rah}, we focus on
inclusive transverse observables: the former condition means $V(\{\tilde{p}\},k_1,\dots, k_n)
= V(\{\tilde{p}\},k_1+\dots +k_n)$, while the latter specifies that for a single soft
emission $k$ collinear to leg $\ell$ the observable can be parametrised as
\begin{equation}
V(\{\tilde p\},k)
\, \equiv \,
V(k)
\, = \,
d_\ell \, g_\ell(\phi)
\left(\frac{k_t}M\right)^a
\, ,
\end{equation}
where $M$ is the mass of the considered colour singlet, $k_t$ is the transverse
momentum of $k$ with respect to the beam axis, $g_\ell(\phi)$ is a generic function
of the angle $\phi$ between $\vec k_t$ and a reference direction $\vec n$,
orthogonal to the beam axis, $d_\ell$ is a normalisation factor, and $a>0$.
For definiteness, the rescaled transverse momentum $p_t/M$ of the colour-singlet
system features $d_\ell = g_\ell(\phi) = a = 1$, while $\phs$ corresponds to
$d_\ell = a = 1$, $g_\ell(\phi) = |\sin(\phi)|$.

In the soft limit, the cumulative cross section in \eqref{eq:cumulative} can be
cast to all orders as
\begin{equation}
\label{eq:Sigma-2}
\Sigma(v)
\, = \,
\int \rd \Phi_B \, {\cal V}(\Phi_B) \,
\sum_{n=0}^{\infty} \int\prod_{i=1}^n [\rd k_i] \, 
|{\cal M}(\{\tilde p\},k_1,\dots,k_n)|^2 \,
\Theta\big(v-V(\{\tilde p\},k_1,\dots,k_n)\big)
\, ,
\end{equation}
where ${\cal M}$ is the renormalised matrix element for $n$ real emissions
(the case with $n = 0$ reduces to the Born contribution), $[\rd k_i]$ denotes
the phase space for the $i$-th emission with momentum $k_i$, and the $\Theta$
function represents the measurement function for the observable under study.
By $\Phi_B$ we denote the Born phase space, while ${\cal V}(\Phi_B)$ is the
all-order virtual form factor relevant to the considered $q\bar q$ or $gg$
reaction.

The rIRC safety of the observable allows one to establish a well defined
logarithmic counting for the squared amplitude \cite{Banfi:2004yd,Banfi:2014sua},
and to systematically identify the terms that contribute at a given logarithmic
order. In particular, $|{\cal M}|^2$ can be conveniently expanded in
$n$-particle-correlated ($n$PC) blocks \cite{Bizon:2017rah}, defined as the
contributions to the emission of $n$ partons that cannot be factorised in
terms of lower-multiplicity squared amplitudes. $n$PC blocks with higher
$n$ and loop order are logarithmically suppressed with respect to blocks
with lower $n$ and number of loops, so that an $n$PC block at $l$ loops just
enters at N$^{n+l-1}$LL accuracy.

The cumulative cross section in \eqref{eq:Sigma-2} contains exponentiated
virtual IRC divergences in $\mathcal V(\Phi_B)$, as well as real singularities
in the multi-radiative squared matrix element.
Such singularities are handled by introducing a resolution scale $q_0$ on the
transverse momentum $k_t$ of radiation: rIRC safety ensures that blocks
with total $k_t<q_0$, dubbed {\it unresolved}, contribute negligibly to
the observable's value, and can be discarded in the evaluation of the
measurement function; unresolved radiation thus exponentiates and regularises
the divergences contained in $\mathcal V(\Phi_B)$ at all orders.
On the other hand, blocks harder than the resolution scale, referred to
as {\it resolved}, must be generated exclusively, as they are constrained
by the measurement function.
The dependence of the prediction upon $q_0$ is guaranteed by rIRC safety
to be power-like, hence the $q_0\to0$ limit can be safely taken.
For the observables considered in this paper, which solely depend on the total
transverse momentum of QCD radiation, it is convenient to set the resolution
scale to $\eps k_{t1}$, where $0<\eps \ll 1$, while $k_{t1}$ is the total
transverse momentum of the hardest resolved block. We point out that the same
resolution scale can be applied for the resummation of different observables,
thereby allowing a flexible Monte Carlo implementation where multiple different
resummations can be performed in a single framework, such as for instance the
recently-introduced double-differential resummation of Higgs and leading-jet
transverse momentum in gluon fusion \cite{Monni:2019yyr}.

After performing the above described set of operations, the all-order result
for the cumulative cross section takes a particularly compact form in Mellin
space, where convolutions with parton densities reduce to algebraic products.
We introduce Mellin moments of generic functions $g(z)$ as $g_{N} = \int_0^1 \rd z \, \,
z^{N-1} \, \, g(z)$, and define ${\bf f}$ as the array containing the $2n_f+1$
parton densities, ($n_f$ being the number of light flavours), whose
DGLAP~\cite{Gribov:1972ri,Altarelli:1977zs,Dokshitzer:1977sg} evolution
between scales $\mu_0$ and $\mu$ reads
\begin{eqnarray}
\label{eq:DGLAP}
{\bf  f}_{N}(\mu)
& = &
{\cal P}
\exp\left[-\int_{\mu}^{\mu_0}\frac{\rd k_t}{k_t} \,
\frac{\as(k_t)}{\pi} \, 
{\bf\Gamma}_{N}(\as(k_t)) \right] \, 
{\bf  f}_{N}(\mu_0)
\, ,
\nonumber\\
\Big[{\bf \Gamma}_{N}(\as)\Big]_{ab}
& = &
\int_0^1 \rd z \, z^{N-1} \, \hat{P}_{f(a) f(b)}(z,\as)
\, = \,
\sum_{n=0}^\infty
\left(\frac{\as}{2\pi}\right)^n
\Big[{\bf \Gamma}^{(n)}_{N}(\as)\Big]_{ab}
\, ,
\nonumber\\
\hat{P}_{ij}(z,\as)
& = &
\sum_{n=0}^\infty
\left(\frac{\as}{2\pi}\right)^n
\hat{P}^{(n)}_{ij}(z)
\, ,
\end{eqnarray}
with $\cal P$ the path-ordering symbol, $\hat{P}_{f(a)f(b)}$ the regularised
collinear splitting functions, and $f(a)$ the flavour of the $a$-th entry
of ${\bf f}$. For notational simplicity, for the time being we consider
only flavour-conserving kernels, so to make the ${\bf \Gamma}$ matrix diagonal
and drop the path ordering; we will relax this assumption by the end of the
section.

The cumulative cross section differential in the Born variables can be written,
with the convention of \cite{Bizon:2017rah}, as
\begin{eqnarray}
\label{eq:hadxs}
\frac{\rd\Sigma(v)}{\rd\Phi_B}
& = &
\int_{{\cal C}_1} \frac{\rd N_1}{2\pi i} \int_{{\cal C}_2}\frac{\rd N_2}{2\pi i} \, \,
x_1^{-N_1} \, x_2^{-N_2}
\sum_{c_1, c_2} \frac{\rd|{\cal M}_B|_{c_1c_2}^2}{\rd\Phi_B} \, \, 
{\bf f}^{T}_{N_1}(\mu_0) \, \hat{\bf \Sigma}^{c_1,c_2}_{N_1,N_2}(v) \, {\bf f}_{N_2}(\mu_0)
\, ,
\nonumber\\
\frac{\rd|{\cal M}_B|_{c_1c_2}^2}{\rd\Phi_B}
& \equiv &
\int {\rd}\Phi'_B \, 
|{\cal M}_B|_{c_1c_2}^2 \,
\delta(x_1-x_1') \, \delta(x_2-x_2') \, \delta(\Omega_B-\Omega_B')
\, ,
\end{eqnarray}
where $|{\cal M}_B|$ is the Born squared matrix element, the sum runs over all allowed
Born flavour combinations, $\Omega_B$ denotes a set of internal phase-space variables
of the colour-singlet system, and the integration contours  ${\cal C}_1$ and ${\cal C}_2$
in the double inverse Mellin transform  lie along the imaginary axis to the right of
all singularities of the integrand. 

The $\hat{\bf \Sigma}$ matrix encodes the effect of parton-density DGLAP evolution
from scale $\mu_0$, as well as that of flavour-conserving radiation evolving the
partonic cross section.
For inclusive observables, its all-order expression under the above assumption on
${\bf\Gamma}_N$ is\footnote{The last two lines of eq.~\eqref{eq:partxs-mellin} reduce
to $\Theta\left(v-V(\{\tilde{p}\},k_1)\right)$ for $n=0$.}
\begin{eqnarray}
\label{eq:partxs-mellin}
\hat{\bf \Sigma}^{c_1,c_2}_{N_1,N_2}(v)
& = &
\bigg[
{\bf C}^{c_1; T}_{N_1}(\as(\mu_0)) \, H(\mu_R) \, {\bf C}^{c_2}_{N_2}(\as(\mu_0))
\bigg] \,
\int_0^{M}\frac{\rd k_{t1}}{k_{t1}} \int_0^{2\pi} \frac{\rd\phi_1}{2\pi} \,
\re^{-{\bf R}(\epsilon k_{t1})}
\nonumber\\
&& 
\times \,
\exp\left[-\sum_{\ell=1}^{2}
\left(\int_{\epsilon k_{t1}}^{\mu_0}\frac{\rd k_t}{k_t} \, \frac{\as(k_t)}{\pi}{\bf\Gamma}_{N_\ell}(\as(k_t))
+
\int_{\epsilon k_{t1}}^{\mu_0}\frac{\rd k_t}{k_t} \, {\bf \Gamma}_{N_\ell}^{(\rC)}(\as(k_{t}))\right)\right]
\nonumber\\
&&
\times \,
\sum_{\ell_1=1}^2
\left({\bf R}_{\ell_1}'\left(k_{t1}\right)
+ \frac{\as(k_{t1})}{\pi}{\bf\Gamma}_{N_{\ell_1}}(\as(k_{t1}))
+ {\bf \Gamma}_{N_{\ell_1}}^{(\rC)}(\as(k_{t1}))\right)
\nonumber\\
&&
\times \,
\sum_{n=0}^{\infty}\frac{1}{n!}
\prod_{i=2}^{n+1}
\int_{\epsilon k_{t1}}^{k_{t1}}\frac{\rd k_{ti}}{k_{ti}}\int_0^{2\pi}
\frac{\rd\phi_i}{2\pi} \, \,
\Theta\left(v-V(\{\tilde{p}\},k_1,\dots, k_{n+1})\right)
\nonumber\\
&&
\times \,
\sum_{\ell_i=1}^2
\left({\bf R}_{\ell_i}'\left(k_{ti}\right)
+ \frac{\as(k_{ti})}{\pi}{\bf\Gamma}_{N_{\ell_i}}(\as(k_{ti}))
+ {\bf \Gamma}_{N_{\ell_i}}^{(\rC)}(\as(k_{ti}))\right)
\, .
\end{eqnarray}
$H(\mu_R)$ represents the finite contribution to the virtual form factor,
evaluated at the renormalisation scale $\mu_R$, and has a perturbative expansion
of the form
\begin{eqnarray}
H(\mu_R)
& = &
1 +
\sum_{n=1}^{\infty}
\left(\frac{\as(\mu_R)}{2\pi}\right)^n
\,
H^{(n)}(\mu_R)
\, .
\end{eqnarray}
${\bf C}^{c_\ell}$ is a
$(2n_f+1)\times(2n_f+1)$ diagonal matrix defined as $[{\bf C}^{c_\ell}]_{a b}
= C_{c_\ell f(a)} \delta_{a b}$ in terms of the collinear coefficient
functions $C_{ij}$. It satisfies a flavour-conserving renormalisation-group
evolution equation stemming from the running of its coupling
\begin{eqnarray}
\label{eq:RGEC}
{\bf C}^{c_\ell}(\as(\mu))
& = &
\exp\left[-\int_\mu^{\mu_0}\frac{\rd k_t}{k_t} \,
{\bf \Gamma}^{(\rm C)}(\as(k_{t}))\right] \,
{\bf C}^{c_\ell}(\alpha_s(\mu_0))
\nonumber\\
& = &
\delta(1-z) \, {\bf 1} +
\sum_{n=1}^\infty
\left(\frac{\as(\mu)}{2\pi}\right)^n
{\bf C}^{(n)}(z)
\, ,
\nonumber\\
{\bf \Gamma}^{(\rC)}(\as(k_{t}))
& = &
2 \, \beta(\as(k_{t})) \,
\frac{\rd \ln{\bf C}^{c_\ell}(\as(k_{t}))}{\rd \as(k_{t})}
\, = \,
\sum_{n=1}^\infty
\left(\frac{\as(k_{t})}{2\pi}\right)^{n+1}
{\bf \Gamma}^{({\rC}, n)}(\as(k_{t}))
\, ,
\end{eqnarray}
where we unambiguously dropped the $c_\ell$ index in ${\bf C}^{(n)}$ and in
${\bf \Gamma}^{(\rC)}$, for the sake of brevity.
The ${\bf R}_{\ell}'$ function encodes the contribution from radiation off leg $\ell$
which conserves the momentum fraction of the incoming partons and the flavour
$c_\ell$ of the emitter, namely $[{\bf R}_{\ell}']_{ab} = R_\ell' \, \delta_{ab}$.
It is related to the Sudakov radiator ${\bf R}$, with entries $[{\bf R}]_{ab}
= R \, \delta_{ab}$, by
\begin{eqnarray}
\label{eq:rad-mat}
R(k_{t1})
\, = \,
\sum_{\ell=1}^2 \,
R_\ell(k_{t1})
& = &
\int_{k_{t1}}^{M}\frac{\rd k_{t}}{k_{t}} \, 
\sum_{\ell=1}^2 \,
{R}_\ell'(k_t)
\, = \,
\int_{k_{t1}}^M
\frac{\rd k_t}{k_t}
\sum_{\ell=1}^2 \,
\left[
A_{\ell}(\as(k_t))\ln\frac{M^2}{k_t^2}
+
B_{\ell}(\as(k_t))
\right]
\, ,
\nonumber\\[5pt]
{R}'(k_{t1})
& = &
\sum_{\ell=1}^2 \,
{R}_\ell'(k_{t1})
\, ,
\qquad \quad
{R}_\ell'(k_{t1})
\, = \,
\frac{\rd R_\ell(k_{t1})}{\rd L}
\, ,
\qquad \quad
L \, = \, \ln\frac{M}{k_{t1}}
\, .
\end{eqnarray}
Finally, the anomalous dimensions $A_\ell$ and $B_\ell$ encode the inclusive probability
$|{\cal M}(k)|_{\rm inc}^2$ \cite{Bizon:2017rah} for a correlated block of arbitrary
multiplicity to have total transverse momentum $k_t$;
they admit a perturbative expansion as
\begin{eqnarray}
&&
A_\ell(\as)
\, = \,
\sum_{n=1}^\infty
\left(\frac{\as}{2\pi}\right)^n \, A_{\ell}^{(n)}
\, ,
\qquad
B_\ell(\as)
\, = \,
\sum_{n=1}^\infty
\left(\frac{\as}{2\pi}\right)^n \, B_{\ell}^{(n)}
\, .
\end{eqnarray}

The structure of \eqref{eq:partxs-mellin} shows the different contributions of resolved
and unresolved radiation. The former, encoded in the third to fifth line, is represented
by an ensemble of emissions (more appropriately: of correlated blocks treated inclusively)
harder than $\epsilon k_{t1}$, with contributions from flavour-diagonal radiation as well
as from exclusive DGLAP-evolution steps.
Conversely, the exponentiated unresolved emissions combine with the all-order virtual
form factor giving rise to the Sudakov exponential $\re^{-{\bf R}(\epsilon k_{t1})}$.
The factor ${\bf C}^{c_1; T}_{N_1}(\as(\mu_0)) \, H(\mu_R) \, {\bf C}^{c_2}_{N_2}(\as(\mu_0))$
encodes the hard-virtual corrections to the form factor, and the collinear coefficient
functions.
The coupling of the latter is evaluated at scale $\mu_0$ and subsequently evolved inclusively
up to $\epsilon k_{t1}$ by the operator containing ${\bf \Gamma}_{N_\ell}^{(\rC)}$ in the
second line of~\eqref{eq:partxs-mellin}. Similarly, the parton densities are DGLAP-evolved
from $\mu_0$ up to $\epsilon k_{t1}$ by ${\bf \Gamma}_{N_\ell}$.

As shown in Ref.~\cite{Catani:2010pd}, for gluon-fusion processes the structure in
eq.~\eqref{eq:partxs-mellin} must be supplemented with the contribution from the
(flavour-diagonal) ${\bf G}$ collinear coefficient functions, describing the azimuthal
correlations with initial-state gluons.
This contribution, starting at ${\cal O}(\as^2)$, i.e.~\nnnll order, is included in the
above formulation by adding to eq.~\eqref{eq:partxs-mellin} an analogous term where
one performs the replacements
\begin{eqnarray}
\left[
{\bf C}^{c_1; T}_{N_1}(\as(\mu_0)) \,\, H(\mu_R) \,\, {\bf C}^{c_2}_{N_2}(\as(\mu_0))
\right]
& \quad \to \quad &
\left[
{\bf G}^{c_1; T}_{N_1}(\as(\mu_0)) \,\, H(\mu_R) \,\, {\bf G}^{c_2}_{N_2}(\as(\mu_0))
\right]
\, ,
\nonumber\\
{\bf \Gamma}_{N_\ell}^{(\rC)}(\as(k_t))
& \quad \to \quad &
{\bf \Gamma}_{N_\ell}^{(\rG)}(\as(k_{t}))
\, ,
\end{eqnarray}
where ${\bf \Gamma}_{N_\ell}^{(\rG)}$ is defined in formal analogy with ${\bf \Gamma}_{N_\ell}^{(\rC)}$
in \eqref{eq:RGEC}. In the following, this contribution is understood whenever not explicitly
reported.\\

The evaluation of eq.~\eqref{eq:partxs-mellin} at this point may be simplified by exploiting again
rIRC safety. The latter ensures that the transverse momenta of all blocks in the resolved
ensemble are parametrically of the same order, as blocks that are significantly softer than
$k_{t1}$ do not contribute to the evaluation of the observable and are accounted for in
the radiator.
All resolved contributions in eq.~\eqref{eq:partxs-mellin} with argument $k_{ti} \, \equiv \,
\zeta_i \, k_{t1}$ can thus be Taylor-expanded about $k_{t1}$, with subsequent terms in the
expansion being more and more logarithmically suppressed, since $\zeta_i$ is of ${\cal O}(1)$.
Analogously, unresolved quantities depending on $\epsilon k_{t1}$ can be expanded about
$k_{t1}$: the ensuing logarithms $\ln(1/\epsilon)$ exactly cancel the logarithmic
$\epsilon$-dependence of the corresponding terms in the resolved radiation, conveniently
achieving an all-order subtraction of IRC divergences.

Aiming for \nnnll accuracy, one needs to retain only the following terms in the Taylor
expansion of the unresolved quantities
\begin{eqnarray}
\label{eq:expanded_unres}
R(\epsilon k_{t1})
& = &
\sum_{j=0}^3
R^{(j)}(k_{t1}) \, \frac1{j!} \, \ln^j\frac1\epsilon \, + \, \dots
\, ,
\nonumber\\
\int_{\epsilon k_{t1}}^{\mu_0} \frac{\rd k_t}{k_t} \,
\frac{\as(k_t)}{\pi} \, 
{\bf\Gamma}_{N_\ell}(\as(k_t))
& = &
\sum_{j=0}^2 \,
\frac{\rd^j}{\rd L^{\,j}} \,
\int_{k_{t1}}^{\mu_0} \frac{\rd k_t}{k_t} \,
\frac{\as(k_t)}{\pi} \,
{\bf\Gamma}_{N_\ell}(\as(k_t)) \,
\frac1{j!} \,
\ln^j\frac1\epsilon
\, + \, \dots
\, ,
\nonumber\\
\int_{\epsilon k_{t1}}^{\mu_0} \frac{\rd k_t}{k_t} \,
{\bf\Gamma}_{N_\ell}^{(\rC)}(\as(k_t))
& = &
\sum_{j=0}^1 \,
\frac{\rd^j}{\rd L^{\,j}} \,
\int_{k_{t1}}^{\mu_0} \frac{\rd k_t}{k_t} \,
{\bf\Gamma}_{N_\ell}^{(\rC)}(\as(k_t)) \,
\frac1{j!} \,
\ln^j\frac1\epsilon
\, + \, \dots
\, ,
\nonumber\\
\int_{\epsilon k_{t1}}^{\mu_0} \frac{\rd k_t}{k_t} \,
{\bf\Gamma}_{N_\ell}^{(\rG)}(\as(k_t))
& = &
\sum_{j=0}^0 \,
\frac{\rd^j}{\rd L^{\,j}} \,
\int_{k_{t1}}^{\mu_0} \frac{\rd k_t}{k_t} \,
{\bf\Gamma}_{N_\ell}^{(\rG)}(\as(k_t)) \,
\frac1{j!} \,
\ln^j\frac1\epsilon
\, + \, \dots
\, ,
\end{eqnarray}
as well as of the resolved contributions, which are suppressed by one logarithmic
order with respect to the corresponding unresolved ones:
\begin{eqnarray}
\label{eq:expanded_res}
R'(k_{ti})
& = &
\sum_{j=0}^2
R^{(j+1)}(k_{t1}) \, \frac1{j!} \, \ln^j\frac1{\zeta_i} \, + \, \dots
\, ,
\nonumber\\
\frac{\as(k_{ti})}{\pi} \,
{\bf\Gamma}_{N_\ell}(\as(k_{ti}))
& = &
\sum_{j=0}^1 \,
\frac{\rd^j}{\rd L^{\,j}} \,
\frac{\as(k_{t1})}{\pi} \,
{\bf\Gamma}_{N_\ell}(\as(k_{t1})) \,
\frac1{j!} \,
\ln^j\frac1{\zeta_i}
\, + \, \dots
\, ,
\nonumber\\
{\bf\Gamma}_{N_\ell}^{(\rC)}(\as(k_{ti}))
& = &
\sum_{j=0}^0 \,
\frac{\rd^j}{\rd L^{\,j}} \,
{\bf\Gamma}_{N_\ell}^{(\rC)}(\as(k_{t1})) \,
\frac1{j!} \,
\ln^j\frac1{\zeta_i}
\, + \, \dots
\, ,
\end{eqnarray}
where $R^{(j)}(k_{t1}) = \rd^j R(k_{t1})/\rd L^{\,j}$, $L=\ln(M/k_{t1})$,
and the ellipses denote neglected \nnnnll terms.
The loop expansion of the involved anomalous
dimensions obeys an analogous perturbative counting.
A further significant simplification stems from the fact that,
at a given logarithmic accuracy, one needs to retain subleading
terms in the above expansions only for a limited number of
resolved blocks. For instance, at N$^k$LL, only up to $k-1$
resolved blocks need to feature a $\ln1/{\zeta_i}$
correction in $R'$, as the simultaneous correction of $k$
factors of $R'$ affects N$^{k+1}$LL.
Unresolved contributions are expanded correspondingly, in
order to cancel the $\epsilon$ divergences of the modified
resolved blocks to the given logarithmic order.

By means of the above expansions, the master formula
\eqref{eq:partxs-mellin}, which is valid to all logarithmic
orders, at \nnnll (and, as we will show in the next section,
at \nnnllp as well) reduces to
\begin{eqnarray}
\label{eq:mellin_expanded_final}
\hat{\bf \Sigma}^{c_1,c_2}_{N_1,N_2}(v)
& = &
\bigg[
{\bf C}^{c_1; T}_{N_1}(\as(\mu_0)) \, H(\mu_R) \, {\bf C}^{c_2}_{N_2}(\as(\mu_0))
\bigg] \,
\int_0^{M}\frac{\rd k_{t1}}{k_{t1}} \int_0^{2\pi} \frac{\rd\phi_1}{2\pi} \,
\nonumber\\
&&
\times \,
\re^{-{\bf R}(k_{t1}) - {\bf R}'(k_{t1})\ln\frac1\epsilon
- \frac1{2!}{\bf R}''(k_{t1})\ln^2\frac1\epsilon
- \frac1{3!}{\bf R}'''(k_{t1})\ln^3\frac1\epsilon}
\nonumber\\
&&
\times \,
\exp \Bigg[
-\sum_{\ell=1}^{2}
\bigg(
\int_{k_{t1}}^{\mu_0} \frac{\rd k_t}{k_t} \, \frac{\as(k_t)}{\pi} \, {\bf \Gamma}_{N_\ell}(\as(k_t))
+
\frac{\rd}{\rd L}
\int_{k_{t1}}^{\mu_0} \frac{\rd k_t}{k_t} \, \frac{\as(k_t)}{\pi} \, {\bf \Gamma}_{N_\ell}(\as(k_t)) \, \ln\frac1\epsilon
\nonumber\\
&&
\hspace{10mm}
+ \,
\frac1{2!} \frac{\rd^2}{\rd L^2}
\int_{k_{t1}}^{\mu_0} \frac{\rd k_t}{k_t} \, \frac{\as(k_t)}{\pi} \, {\bf \Gamma}_{N_\ell}(\as(k_t)) \, \ln^2\frac1\epsilon
\nonumber\\
&&
\hspace{10mm}
+
\int_{k_{t1}}^{\mu_0} \frac{\rd k_t}{k_t} \, {\bf \Gamma}_{N_\ell}^{(\rC)}(\as(k_{t}))
+
\frac{\rd}{\rd L}
\int_{k_{t1}}^{\mu_0} \frac{\rd k_t}{k_t} \, {\bf \Gamma}_{N_\ell}^{(\rC)}(\as(k_{t})) \, \ln\frac1\epsilon
\bigg)
\Bigg]
\nonumber\\
&&
\times \,
\sum_{\ell_1=1}^2
\left(
{\bf R}_{\ell_1}'(k_{t1}) +
\frac{\as(k_{t1})}{\pi} \, {\bf\Gamma}_{N_{\ell_1}}(\as(k_{t1})) +
{\bf \Gamma}_{N_{\ell_1}}^{(\rC)}(\as(k_{t1}))
\right)
\nonumber\\
&&
\times \,
\sum_{n=0}^{\infty}\frac1{n!} \prod_{i=2}^{n+1}
\int_{\epsilon}^1 \frac{\rd \zeta_i}{\zeta_i} \int_0^{2\pi} \frac{\rd\phi_i}{2\pi}
\sum_{\ell_i=1}^2
\Bigg[
{\bf R}_{\ell_i}'(k_{t1}) +
{\bf R}_{\ell_i}''(k_{t1}) \, \ln\frac1{\zeta_i} +
\frac1{2!}{\bf R}_{\ell_i}'''(k_{t1}) \, \ln^2\frac1{\zeta_i}
\nonumber\\
&&
\hspace{10mm}
+ \,
\frac{\as(k_{t1})}{\pi} {\bf \Gamma}_{N_{\ell_i}}(\as(k_{t1})) +
\frac{\rd}{\rd L} \left(\frac{\as(k_{t1})}{\pi} {\bf \Gamma}_{N_{\ell_i}}(\as(k_{t1}))\right) \ln\frac1{\zeta_i}
+
{\bf \Gamma}_{N_{\ell_i}}^{(\rC)}(\as(k_{t1}))
\Bigg]
\nonumber\\
&&
\times \,
\Theta\left(v-V(\{\tilde{p}\},k_1,\dots, k_{n+1})\right)
~~ + ~~
\Big\{
{\bf C} \to  {\bf G}, ~~ {\bf \Gamma}^{(\rC)} \to {\bf \Gamma}^{(\rG)}
\Big \}
\, .
\end{eqnarray}

The final operation is to rewrite eq.~\eqref{eq:mellin_expanded_final} in direct
(as opposed to Mellin) space, which requires little effort at this point, as a very
limited number of exclusive evolutions steps have been retained  in the above expression.
In particular, at \nnnll, only up to two hard-collinear resolved emissions are needed,
and one can relax the above assumption of flavour-conserving real radiation
by including flavour-changing kernels in the DGLAP-evolution contributions
in momentum space.
This amounts to the following identifications, valid at \nnnll:
\begin{eqnarray}
\label{eq:conversions_to_direct_space}
&&
\frac{\rd|{\cal M}_B|_{c_1c_2}^2}{\rd\Phi_B} \, \,
{\bf f}^{T}_{N_1}(k_{t1})
\Bigg[
\sum_{\ell=1}^2
\frac{\as(k_{t1})}{\pi}
{\bf\Gamma}_{N_{\ell}}(\as(k_{t1}))
\Bigg]
{\bf f}_{N_2}(k_{t1})
\nonumber\\
&&
\qquad
\longrightarrow ~~
\frac{\as(k_{t1})}{\pi} \,
\hat{P}(z,\as(k_{t1}))
\otimes
{\cal L}_{\rm NLL}(k_{t1})
~ = ~
-
\partial_L \,
{\cal L}_{\rm NLL}(k_{t1})
\, ,
\nonumber\\[5pt]
&&
\frac{\rd|{\cal M}_B|_{c_1c_2}^2}{\rd\Phi_B} \, \,
{\bf f}^{T}_{N_1}(k_{t1}) \,
{\bf C}^{c_1;T}_{N_1}(\as(k_{t1})) \,
H(\mu_R)
\nonumber\\
&&
\qquad \qquad
\times \,
\Bigg[
\sum_{\ell=1}^2
\left(
\frac{\as(k_{t1})}{\pi}
{\bf \Gamma}_{N_{\ell}}(\as(k_{t1}))
+ \,
{\bf \Gamma}_{N_{\ell}}^{(\rC)}(\as(k_{t1}))
\right)
\Bigg]
{\bf C}^{c_2}_{N_2}(\as(k_{t1})) \,
{\bf f}_{N_2}(k_{t1})
\nonumber\\
&&
\qquad
\longrightarrow ~~
- \, \partial_L \, {\cal L}(k_{t1})
\, ,
\nonumber\\[5pt]
&&
\frac{\rd|{\cal M}_B|_{c_1c_2}^2}{d\Phi_B} \, \,
{\bf f}^{T}_{N_1}(k_{t1})
\Bigg[
\sum_{\ell=1}^2
\frac{\rd}{\rd L}
\left(
\frac{\as(k_{t1})}{\pi}
{\bf\Gamma}_{N_{\ell}}(\as(k_{t1}))
\right)
\Bigg]
{\bf f}_{N_2}(k_{t1})
\nonumber\\
&&
\qquad
\longrightarrow ~~
2 \, \frac{\beta_0}{\pi} \, \as^2(k_{t1}) \, 
\hat{P}^{(0)} \otimes {\cal L}_{\rm NLL}(k_{t1})
\, ,
\nonumber\\[5pt]
&&
\frac{\rd|{\cal M}_B|_{c_1c_2}^2}{\rd\Phi_B} \, \, 
{\bf f}^{T}_{N_1}(k_{t1})
\Bigg[
\sum_{\ell_i=1}^2
\frac{\as(k_{t1})}{\pi}
{\bf\Gamma}_{N_{\ell_i}}(\as(k_{t1}))
\Bigg]
\Bigg[
\sum_{\ell_j=1}^2
\frac{\as(k_{t1})}{\pi}
{\bf\Gamma}_{N_{\ell_j}}(\as(k_{t1}))
\Bigg]
{\bf f}_{N_2}(k_{t1})
\nonumber\\
&&
\qquad
\longrightarrow ~~
\frac{\as^2(k_{t1})}{\pi^2} \,
\hat{P}(z,\as(k_{t1})) \otimes \hat{P}(z,\as(k_{t1})) \otimes {\cal L}_{\rm NLL}(k_{t1})
\nonumber\\
&&
\qquad
\simeq
\frac{\as^2(k_{t1})}{\pi^2} \,
\hat{P}^{(0)} \otimes \hat{P}^{(0)} \otimes {\cal L}_{\rm NLL}(k_{t1})
\, ,
\end{eqnarray}
where we defined $\partial_L \, \equiv \, \rd/\rd L$, $L=\ln(M/k_{t1})$,
$\beta_0$ is the lowest-order contribution to the QCD beta function,
${\cal L}$ is the parton luminosity (see Appendix \ref{app:luminosities}
for its explicit expression at the various logarithmic orders, and
Sec.~\ref{sec:as3const} for a discussion about standard and improved
luminosities in the context of {N$^3$LL$^\prime$\xspace}-accurate predictions), and
\begin{eqnarray} 
\label{eq:Pluminosity-NLL}
\hat{P}^{(0)} \otimes {\cal L}_{\rm NLL}(k_{t1})
& \equiv &
\sum_{c,c'}
\frac{\rd|{\cal M}_{B}|_{cc'}^2}{d\Phi_B} \,
\bigg[
\big(\hat{P}^{(0)} \otimes f \big)_c(k_{t1},x_1) \, f_{c'}(k_{t1},x_2)
\\
&&
\hspace{20mm}
+ \,
f_c(k_{t1},x_1) \, \big(\hat{P}^{(0)} \otimes f\big)_{c'}(k_{t1},x_2)
\bigg]
\, ,
\nonumber\\[5pt]
\hat{P}^{(0)} \otimes \hat{P}^{(0)} \otimes {\cal L}_{\rm NLL}(k_{t1})
& \equiv &
\sum_{c,c'}
\frac{\rd|{\cal M}_{B}|_{cc'}^2}{d\Phi_B} \,
\bigg[
\big(\hat{P}^{(0)} \otimes \hat{P}^{(0)} \otimes f \big)_c (k_{t1},x_1) \, f_{c'}(k_{t1},x_2)
\nonumber\\
&&
\hspace{20mm}
+ \,
f_c(k_{t1},x_1) \,
\big(\hat{P}^{(0)} \otimes \hat{P}^{(0)} \otimes f \big)_{c'}(k_{t1},x_2)
\nonumber\\
&&
\hspace{20mm}
+ \,
2 \,
\big(\hat{P}^{(0)} \otimes f \big)_c (k_{t1},x_1) \,
\big(\hat{P}^{(0)} \otimes f \big)_{c'} (k_{t1},x_2)
\bigg]
\, .
\nonumber
\end{eqnarray}
After expressing the logarithms of $1/\epsilon$ as dummy radiative integrals
\cite{Banfi:2014sua} according to
\begin{eqnarray*}
\ln^k\frac{1}{\epsilon}
& = &
k\int_\epsilon^1
\frac{\rd\zeta}{\zeta} \,
\ln^{k-1}\frac1\zeta
\, ,
\qquad \quad
k\geq 1
\, ,
\end{eqnarray*}
and introducing the average of a function $G(\{\tilde p\},\{k_i\})$ over the
measure $\rd {\cal Z}$
\begin{eqnarray}
\label{eq:dZ}
\int
\dZ \,
G(\{\tilde p\},\{k_i\})
& = &
\epsilon^{R'(k_{t1})} \, 
\sum_{n=0}^{\infty}\frac1{n!} \,
\prod_{i=2}^{n+1}
\int_\epsilon^1 \frac{\rd\zeta_i}{\zeta_i}
\int_0^{2\pi} \frac{\rd\phi_i}{2\pi} \,
R'(k_{t1}) \, G(\{\tilde p\},k_1,\dots,k_{n+1})
\, ,
\nonumber\\
\end{eqnarray}
in which the dependence upon the $\epsilon$ regulator exactly cancels
to all orders, one finally gets at \nnnll
\begin{eqnarray}
\label{eq:master-kt-space}
\frac{\rd\Sigma^{\rm N^3LL}(v)}{\rd\Phi_B}
& = &
\int\frac{\rd k_{t1}}{k_{t1}}
\frac{\rd\phi_1}{2\pi} \, 
\partial_{L}
\left(- \, \re^{-R(k_{t1})} {\cal L}_{\rm N^3LL}(k_{t1}) \right)
\int \dZ \, 
\Theta \left(v-V(\{\tilde{p}\},k_1,\dots, k_{n+1})\right)
\nonumber\\[10pt]
&&
+ \,
\int \frac{\rd k_{t1}}{k_{t1}}
\frac{\rd\phi_1}{2\pi} \,
\re^{-R(k_{t1})}
\int \dZ 
\int_0^1 \frac{\rd\zeta_s}{\zeta_s}
\frac{\rd\phi_s}{2\pi}
\nonumber\\
&&
\times \,
\Bigg\{
\bigg(
R' (k_{t1}) {\cal L}_{\rm NNLL}(k_{t1}) - \partial_L {\cal L}_{\rm NNLL}(k_{t1})
\bigg)
\bigg(
R''(k_{t1}) \ln\frac1{\zeta_s} + \frac{1}{2} R'''(k_{t1}) \ln^2\frac1{\zeta_s}
\bigg)
\nonumber\\
&&
\qquad
- \, R'(k_{t1})
\left(
\partial_L {\cal L}_{\rm NNLL}(k_{t1})
- 2 \, \frac{\beta_0}{\pi} \, \as^2(k_{t1}) \hat{P}^{(0)} \otimes {\cal L}_{\rm NLL}(k_{t1}) \ln\frac1{\zeta_s}
\right)
\nonumber\\
&&
\qquad
+ \,
\frac{\as^2(k_{t1})}{\pi^2} \hat{P}^{(0)} \otimes \hat{P}^{(0)} \otimes {\cal L}_{\rm NLL}(k_{t1})
\Bigg\}
\nonumber\\
&&
\times \,
\bigg[
\Theta\left(v-V(\{\tilde{p}\},k_1,\dots, k_{n+1},k_s)\right) -
\Theta\left(v-V(\{\tilde{p}\},k_1,\dots, k_{n+1})\right)
\bigg]
\nonumber\\[10pt]
&&
+ \,
\frac12
\int\frac{\rd k_{t1}}{k_{t1}}
\frac{\rd\phi_1}{2\pi}
\re^{-R(k_{t1})}
\int \dZ
\int_0^1 \frac{\rd \zeta_{s1}}{\zeta_{s1}}
\frac{\rd\phi_{s1}}{2\pi}
\int_0^1 \frac{\rd \zeta_{s2}}{\zeta_{s2}}
\frac{\rd\phi_{s2}}{2\pi} \,
R'(k_{t1})
\nonumber\\
&&
\times \,
\Bigg\{
{\cal L}_{\rm NLL}(k_{t1})
\big(R''(k_{t1})\big)^2 \ln\frac{1}{\zeta_{s1}} \ln\frac{1}{\zeta_{s2}}
- \partial_L {\cal L}_{\rm NLL}(k_{t1}) R''(k_{t1})
\bigg(
\ln\frac{1}{\zeta_{s1}} + \ln\frac{1}{\zeta_{s2}}
\bigg)
\nonumber\\
&&
\qquad
+ \,
\frac{\as^2(k_{t1})}{\pi^2} \hat{P}^{(0)} \otimes \hat{P}^{(0)} \otimes {\cal L}_{\rm NLL}(k_{t1})
\Bigg\}
\nonumber\\
&&
\times \,
\bigg[
\Theta\left(v-V(\{\tilde{p}\},k_1,\dots,k_{n+1},k_{s1},k_{s2})\right) -
\Theta\left(v-V(\{\tilde{p}\},k_1,\dots,k_{n+1},k_{s1})\right)
\nonumber\\
&&
\qquad
- \,
\Theta\left(v-V(\{\tilde{p}\},k_1,\dots,k_{n+1},k_{s2})\right) +
\Theta\left(v-V(\{\tilde{p}\},k_1,\dots,k_{n+1})\right)
\bigg]
\, ,
\end{eqnarray}
where the explicit factors of $\as(k_{t1})$ are defined as
$\as(k_{t1}) \, = \, \as/(1-2\as\beta_0 L)$, and $\as = \as(\mu_R)$
unless stated otherwise.\\

We conclude this review of the \radish approach with two remarks
on the master formula \eqref{eq:master-kt-space}.
First, we note that the logarithms resummed there are of the form
$L = \ln(M/k_{t1})$. It is convenient to introduce the resummation
scale $Q$, of order $M$, as an auxiliary scale to be varied in
order to probe the size of neglected logarithmic corrections,
and resum logarithms $\ln(Q/k_{t1})$. This is formally achieved
by splitting $L = \ln(Q/k_{t1}) + \ln(M/Q)$, by assuming the
hierarchy $\ln(Q/k_{t1}) \gg \ln(M/Q)$, valid in the IRC limit,
and by expanding $L$ around $\ln(Q/k_{t1})$ at the relevant
logarithmic accuracy.
Second, when the resummed results are matched to a fixed-order
prediction, it is desirable to enforce the former to vanish
in the hard region $k_{t1}\gg Q$ of the $v$ spectrum, reliably
described by the latter. This can be achieved by modifying
the resummed logarithms $\ln(Q/k_{t1})$ by means of power-suppressed
terms, negligible at small $k_{t1}$. A possible choice for
modified logarithms $\tilde L$ is
\begin{eqnarray}
\label{eq:modified-log}
\ln\frac{Q}{k_{t1}}
& ~\longrightarrow~ &
\tilde{L}
\, = \,
\frac{1}{p}
\ln\left[\left(\frac{Q}{k_{t1}}\right)^{p} + 1\right]
\, ,
\end{eqnarray}
where $p$ is a positive real parameter chosen so that
the resummed differential spectrum vanishes faster than
the fixed-order one at large $v$. The above prescription
induces a jacobian $\mathcal J(k_{t1})$,
\begin{eqnarray}
\label{eq:jakobian}
\int \frac{\rd k_{t1}}{k_{t1}}
& ~\longrightarrow~ &
\int_0^\infty \frac{\rd k_{t1}}{k_{t1}} \,
{\cal J}(k_{t1})
\, ,
\nonumber\\
{\cal J}(k_{t1})
& = &
\frac{\left(Q/k_{t1}\right)^p}{1+\left(Q/k_{t1}\right)^p}
\, = \,
1 - \left(\frac{k_{t1}}{Q}\right)^p + \dots
\, ,
\end{eqnarray}
which ensures the absence of subleading-power corrections with
fractional $\as$ powers in the final distribution, still keeping
the $k_{t1}\to 0$ region unmodified. We stress that the procedure
of logarithmic modification is not just a change of variables,
as it does not affect the observable's measurement function.
As a consequence, the final resummed result shows an explicit
$p$ dependence through power-suppressed terms, which however,
after matching, will cancel up to the accuracy of the fixed-order
component.

In the following developments of the article we understand the
procedure of logarithmic modification, which formally corresponds
to considering the logarithmic region $k_{t1} < Q$ and working in
the $p\to\infty$ limit of \eqref{eq:modified-log} and \eqref{eq:jakobian};
moreover we redefine $L \, \equiv \, \ln(Q/k_{t1})$, in order not
to unnecessarily clutter our formulae.

\section{Consistent inclusion of \nnnllp and recoil effects}
\label{sec:primed}
In this section we discuss how the formalism detailed above can be upgraded
to \nnnllp accuracy, which amounts to supplementing the \nnnll result with
the complete set of constant contributions of relative order ${\cal O}(\as^3)$
with respect to the Born.
Such contributions formally pertain to the logarithmic tower $\as^n
L^{n-3}$, namely they are a subset of the \nnnnll correction, however
they are of particular relevance since their inclusion suffices for
the perturbative expansion of the resummed cumulative cross section
$\Sigma(v)$ to correctly encode all terms of order
$\as^n\ln^{2n-6}(1/v)$.

The definition of `primed' accuracy requires to specify more
precisely how these constant terms are actually included in $\Sigma(v)$.
In particular, at N$^k$LL$'$ order, all choices leading to differences
beyond $\as^k$ and N$^k$LL accuracy are legitimate, such as, for
instance, the argument of the coupling constant multiplying the
highermost-order coefficient functions present in the `primed'
luminosity factors.
In this work we adopt as our default `primed' predictions those
obtained by evaluating such a coupling at the scale $k_{t1}$ of
the hardest emission, which is the correct scale one would have
to use for N$^{k+1}$LL accuracy.
In Sec.~\ref{sec:as3const} we will discuss
this choice in more detail, and assess its impact in Sec.~\ref{sec:pheno_results}.
We stress that, although in other formalisms `primed' accuracy may
correspond to encoding different subleading contributions with respect
to ours, we decided not to introduce a new nomenclature for
our improved predictions, given that, regardless of the formalism,
N$^k$LL$^\prime$ results are anyway designed to upgrade N$^k$LL
ones by the inclusion of the whole set of constant $\as^k$ terms.

The resummation formula presented in eq.~\eqref{eq:partxs-mellin} is formally
valid to all logarithmic orders. However, the accuracy of its practical implementation
is limited by the fact that the quantities it features, such as anomalous dimensions
and coefficient functions, are known to finite perturbative order, and by the fact
that, for computational convenience, the expansions detailed in
eq.~\eqref{eq:expanded_unres} and \eqref{eq:expanded_res} have been performed
to arrive at expression \eqref{eq:mellin_expanded_final} in Mellin space and
\eqref{eq:master-kt-space} in momentum space. Achieving full \nnnllp amounts to
lifting the subset of such approximations that affect third-order constant
contributions.

Focusing on the structure of eq.~\eqref{eq:mellin_expanded_final}, and
recalling that the weight of the hardest resolved radiation $k_{t1}$ provides
at least one power of $\as$, it is immediate to verify that the inclusion
of further logarithmic derivatives in the exponent of the second line,
as well as in the resolved ensemble, only affects ${\cal O}(\as^4)$ terms.
We conclude that the structure of eq.~\eqref{eq:mellin_expanded_final} is
sufficient \emph{as is} to achieve \nnnllp accuracy: one just needs to evaluate
its contributions to appropriate perturbative order, and to upgrade the
conversions \eqref{eq:conversions_to_direct_space} to momentum space, which
we address in turn in the next subsections.

\subsection{${\cal O}(\as^3)$ constants from radiator, hard, and coefficient functions}
\label{sec:as3const}
The first source of constant ${\cal O}(\as^3$) terms we consider
is the radiator defined in eq.~\eqref{eq:rad-mat},
which can be rewritten as:
\begin{eqnarray}
\label{eq:rad-mat2}
R(k_{t1})
& = &
\int_{\lambda_M}^{\lambda}
\frac{\rd\tau}{\as\beta_0} \,
\sum_{\ell=1}^2
\bigg[
\frac{2 A_\ell \, \tau}{\as\beta_0}
+
B_\ell - 2 A_\ell \, L_M
\bigg]
\,\, = \,\,
- L \, g_1\big(\lambda\big)
- \sum_{n=0}^\infty
\left(\frac{\as}{\pi}\right)^n
g_{n+2}\big(\lambda\big)
\, ,
\end{eqnarray}
where $L \, = \, \ln(Q/k_{t1})$, $L_M \, = \, \ln(Q/M)$, $\lambda = \as \, \beta_0 \, L$,
$\lambda_M = \as \, \beta_0 \, L_M$, and $\as = \as(\mu_R)$.
The $g_k$ functions encode the resummation of
N$^{k-1}$LL logarithmic towers $\as^nL^{n+2-k}$; they are explicitly reported in Appendix
B of~\cite{Bizon:2017rah} for $k\leq4$.
From the above integral expression one notes that all constant ${\cal O}(\lambda^0)$ terms
in the radiator vanish as $\lambda_M=0$, i.e.~they are proportional to powers of $L_M$.

By introducing the expansion of the $g_k$ functions in powers of $\lambda$
as $g_k(\lambda) = \sum_{n=0}^\infty g_{k,n} \, \lambda^n$, with $g_{1,0}=g_{2,0}=0$,
the constant $g_{k,0}$ can be inferred by solely analysing $g_j$ functions with $j<k$:
this stems from the fact that $g_{k,0}$ is responsible for the cancellation of a
well-defined part of the $Q$-dependence in the N$^{k-2}$LL radiator.
At LL one has
\begin{eqnarray}
L \, g_1(\lambda)
\, = \,
\sum_{n=1}^{\infty}
g_{1,n} \, \as^n \, \beta_0^n \, L^{n+1}
\, = \,
\sum_{n=1}^{\infty}
g_{1,n} \, \as^n \, \beta_0^n \,
\Big[
\ln(M/k_{t1})+L_M
\Big]^{n+1}
\, ,
\end{eqnarray}
where the $Q$-dependence starts at NLL order, in the coefficient of the
$\as^n\ln^{n}(M/k_{t1})$ term.
The latter dependence is compensated by including $g_2$ in the radiator,
so that the $L \, g_1 + g_2$ sum features a $Q$-dependence starting at
NNLL, in the coefficient of $\as^n\ln^{n-1}(M/k_{t1})$.
In turn, the $Q$-dependence in the $\as\ln^{0}(M/k_{t1})$ term of the
$L \, g_1 + g_2$ sum is exactly compensated by the $g_{3,0}$ constant:
\begin{eqnarray}
\frac{\as}\pi \,  g_{3,0}
\, = \, 
-\Big[ L \, g_1 + g_2\Big]_{\mathcal O(\as\ln^{0}(M/k_{t1}))}
\, = \,
- \, \as \beta_0
\Big(
g_{1,1} \, L_M^2 +
g_{2,1} \, L_M
\Big)
\, ,
\end{eqnarray}
from which $g_{3,0}$ can be read off. We note that there is no other constant term contributing
to $g_{3,0}$ since, as observed above, all constants $g_{k,0}$ are proportional to non-zero
powers of $L_M$.

By generalising this argument, the analysis of the $\as^{k-2}\ln^{0}(M/k_{t1})$ term in the radiator
including up to $g_{k-1}$ allows to deduce $g_{k,0}$, yielding the all-order
expression\footnote{The $g_n(\lambda)$ functions to be used for the extraction of $g_{k,0}$
are the ones defined \emph{before} introducing the $\zeta_3$ contributions that will be detailed
shortly in eq.~\eqref{eq:J0Theta} of the main text.}
\begin{eqnarray}
\label{eq:gk0}
g_{k,0}
\, = \,
- \,
g_{1,k-2} \, (\pi \beta_0 L_M)^{k-2} \, L_M
- \,
\sum_{n=2}^{k-1}
g_{n,k-n} \, (\pi \beta_0 L_M)^{k-n}
\, .
\end{eqnarray}
In our approach the constant part of the radiator, containing up to $g_{5,0}$ in the
\nnnllp case, is then expanded in powers of $\as$, which avoids the presence of any
exponentiated constants. Such terms are included in the hard-virtual function $H$
contributing to the luminosity at the various perturbative orders, which thereby
acquires an explicit $Q$-dependence, as detailed in Sec.~\ref{sec:HCGextract}.
\\

Further ${\cal O}(\as^3)$ constant terms in the one-emission contribution to
eq.~\eqref{eq:mellin_expanded_final} originate from the use in \radish of anomalous
dimensions $A_{{\rm CSS},\ell}$, $B_{{\rm CSS},\ell}$ calculated for a $b$-space
resummation, where the Sudakov radiator is usually defined as
\begin{eqnarray}
\label{eq:R_b_space}
R_b
\, = \,
\int_0^M
\frac{\rd k_t}{k_t}
\,
R_{{\rm CSS}}'(k_t)
\,
\Theta(k_t - b_0/b)
\, ,
\end{eqnarray}
with $b_0 = 2 \re^{- \gamma_E}$, whereas the momentum-space radiator, re-expressed in $b$-space, reads
(see Sec.~2.4 of \cite{Bizon:2017rah})
\begin{eqnarray}
R_b
\, = \,
\int_0^M
\frac{\rd k_t}{k_t}
\,
R'(k_t)
\,
\Big(
1-J_0(b k_t)
\Big)
\, .
\end{eqnarray}
The conversion between the Heaviside and the Bessel function is absorbed
into a redefinition of $A_{{\rm CSS},\ell}$, $B_{{\rm CSS},\ell}$,
$H_{{\rm CSS}}$, and $C_{{\rm CSS}}$ by means of the relation \cite{Banfi:2012jm}
\begin{eqnarray}
\label{eq:J0Theta}
1-J_0(b k_{t})
& = &
\bigg[
1 + 
\frac{\zeta_3}{12}\frac{\partial^{\,3}}{\partial L_b^3} +
{\cal O}
\left(\frac{\partial^{\,5}}{\partial L_b^5}\right)
\bigg]
\,
\Theta(k_t-b_0/b)
\, ,
\qquad\quad
L_b
\, = \,
\ln(bM/b_0)
\, ,
\end{eqnarray}
which starts being non-trivial at \nnnll
\cite{Bizon:2017rah}, involving the third logarithmic derivative $R'''$
of the LL radiator function $g_1$. In order to incorporate
${\cal O}(\as^3)$ constant effects, it is necessary to extend this
construction to the third derivative of function $g_2$, as well as to
include the interference of the one-loop hard function $H^{(1)}$
with the third derivative of $g_1$. This results in a constant term
$(\frac{\as}{2\pi})^3\delta H^{(3)}$ with
\begin{eqnarray}
\label{eq:deltaH3}
\delta H^{(3)}
& = &
\frac83 A^{(1)} \pi \beta_0 \zeta_3
\Big[
H^{(1)} + 2 B^{(1)} L_M - 2 A^{(1)} L_M^2 - 
4 \pi d_B \beta_0 \ln\frac{M}{\mu_R}
\Big]
\nonumber\\
&&
+ \,
\frac{16}3 \pi \zeta_3
\Big[
\beta_0 \Big(A^{(2)} + B^{(1)} \pi \beta_0\Big)
+ A^{(1)} \pi \beta_1
- 2 A^{(1)} \pi \beta_0^2 \,
\Big(3 \, L_M + 2 \ln\frac{M}{\mu_R} \Big)
\Big]
\, ,
\end{eqnarray}
where $A^{(1)} = \sum_{\ell=1,2} A^{(1)}_\ell$,
and $B^{(1)} = \sum_{\ell=1,2} B^{(1)}_\ell$, while $d_B$ is the
strong-coupling order of the Born squared amplitude (e.g.~$d_B = 2$
for Higgs production, and $d_B = 0$ for Drell-Yan production). We include
the $\delta H^{(3)}$ constant in the three-loop hard-virtual
function $H^{(3)}$, see Sec.~\ref{sec:HCGextract}.

The conversion described above for the Sudakov radiator applies
analogously to the parton-density and to the coefficient-function
evolution exponent in eq.~\eqref{eq:partxs-mellin}. While the
third derivatve of the latter starts contributing at ${\cal O}
(\as^4)$, the former generates an ${\cal O}(\as^3)$ constant term
$(\frac{\as}{2\pi})^3 \delta C_{ij}^{(3)}$, with
\begin{eqnarray}
\label{eq:deltaC3}
\delta C_{ij}^{(3)}(z)
\, = \,
\frac{16}3 \pi^2 \beta_0^2 \, \zeta_3 \, \hat P_{ij}^{(0)}(z)
\, ,
\end{eqnarray}
that we include into the third-order coefficient function
$C_{ij}^{(3)}$, see Sec.~\ref{sec:HCGextract}.
Finally, we note that subleading terms in eq.~\eqref{eq:J0Theta}
are proportional to the fifth logarithmic derivative of the Sudakov
radiator, hence they start contributing at ${\cal O}(\as^4)$, and
are neglected in this article.
\\

Blocks
${\bf C}^{c_1; T}_{N_1}(\as(\mu_0)) \, H(\mu_R) \,
 {\bf C}^{c_2}_{N_2}(\as(\mu_0))$ and
${\bf G}^{c_1; T}_{N_1}(\as(\mu_0)) \, H(\mu_R) \,
 {\bf G}^{c_2}_{N_2}(\as(\mu_0))$
of eq.~\eqref{eq:partxs-mellin} are another source of constant terms,
included in the luminosity factors of eq.~\eqref{eq:master-kt-space}.
The latter admit a perturbative expansion that, in turn, originates
from the ones of the hard-virtual function $H$, and of the collinear
coefficient functions $C$ and $G$. Such expansions, already introduced
in Sec.~\ref{sec:rad_review}, are reported here for convenience, using
explicit flavour indices:
\begin{eqnarray}
\label{eq:coeff-fun}
H(\mu_R)
& = &
1 \, + \,
\sum_{n=1}^{3}\left(\frac{\alpha_s(\mu_R)}{2\pi}\right)^n
\, H^{(n)}(\mu_R)
\, ,
\nonumber\\
C_{ij}(\alpha_s(\mu))
& = &
\delta(1-z)\, \delta_{ij} \, + \,
\sum_{n=1}^{3}\left(\frac{\alpha_s(\mu)}{2\pi}\right)^n
\, C_{ij}^{(n)}(z)
\, ,
\nonumber\\
G_{gj}(\alpha_s(\mu))
& = &
\sum_{n=1}^{2} \left(\frac{\alpha_s(\mu)}{2\pi}\right)^n
\, G_{gj}^{(n)}(z)
\, ,
\end{eqnarray}
where $\mu$ is the same scale at which parton densities are
evaluated, and $\mu_R$ is the renormalisation scale.
In eq.~\eqref{eq:coeff-fun} we only retained the perturbative
orders needed to assemble an \nnnllp-accurate luminosity, where
for the first time one needs the third-order coefficient and
hard functions $C^{(3)}$ and $H^{(3)}$, respectively, and
the second-order azimuthal coefficient function $G^{(2)}$,
as discussed in the following.

At N$^k$LL order, the luminosity in the first line of the
\radish master formula (e.g.~${\cal L}_{\rm N^3LL}(k_{t1})$
in eq.~\eqref{eq:master-kt-space}) contains all constant
${\cal O}(\as^{k-1})$ terms, which properly pertain to the
N$^k$LL logarithmic tower, whence the subscript labelling
${\cal L}$. The coupling constants of the involved $C$ and
$G$ coefficient functions have to be evaluated at the same
scale at which the parton densities are evaluated,
i.e.~$\mu=k_{t1}$.
On the other hand, when working at ${{\rm N}^k{\rm LL}^\prime}$ accuracy,
one has the freedom to choose whether the ${\cal O}(\as^{k})$ constant
terms included in ${\cal L}_{{\rm N}^k{\rm LL}^\prime}(k_{t1})$ are
evaluated with a fixed scale, e.g.~$\mu=\mu_R$, or a running scale,
for instance $\mu=k_{t1}$: this ambiguity only affects terms starting
from ${\cal O}(\as^{k+1}L)$, namely non-constant N$^{k+1}$LL
contributions beyond accuracy.

The above discussion can be easily illustrated focusing on the lowest
order at which it applies, namely \nllp: the NNLL luminosity reads
\begin{eqnarray}
{\cal L}_{\rm NNLL}(k_{t1})
& = &
\sum_{c, c'}
\frac{\rd|{\cal M}_{B}|_{cc'}^2}{\rd\Phi_B} \,
\sum_{i,j}
\int_{x_1}^{1}\frac{\rd z_1}{z_1}
\int_{x_2}^{1}\frac{\rd z_2}{z_2} \, 
f_i\Big(k_{t1},\frac{x_1}{z_1}\Big) \,
f_j\Big(k_{t1},\frac{x_2}{z_2}\Big)
\\
&&
\times \,
\Bigg\{
\delta_{ci} \, \delta_{c'j} \, \delta(1-z_1) \, \delta(1-z_2)
\left(1+\frac{\alpha_s(\mu_R)}{2\pi} H^{(1)}(\mu_R) \right)
\nonumber\\
&&
\quad
+ \,
\frac{\as(\mu_R)/(2\pi)}{1-2\alpha_s(\mu_R)\beta_0 \ln(\mu_R/k_{t1})}
\Big(
C_{c i}^{(1)}(z_1)\delta(1-z_2)\delta_{c'j}
+ \, \{z_1,c,i \, \leftrightarrow \, z_2,c',j\}
\Big)
\Bigg\}
\, ,
\nonumber
\end{eqnarray}
whereas, at \nllp, one is allowed to define either
${\cal L}_{\rm NLL'}={\cal L}_{\rm NNLL}$ or
\begin{eqnarray}
{\cal L}_{\rm NLL'}(k_{t1})
& = &
\sum_{c, c'}
\frac{\rd|{\cal M}_{B}|_{cc'}^2}{\rd\Phi_B} \,
\sum_{i,j}
\int_{x_1}^{1}\frac{\rd z_1}{z_1}
\int_{x_2}^{1}\frac{\rd z_2}{z_2} \, 
f_i\Big(k_{t1},\frac{x_1}{z_1}\Big) \,
f_j\Big(k_{t1},\frac{x_2}{z_2}\Big)
\nonumber\\
&&
\times \,
\Bigg\{
\delta_{ci} \, \delta_{c'j} \, \delta(1-z_1) \, \delta(1-z_2)
\left(1+\frac{\alpha_s(\mu_R)}{2\pi} H^{(1)}(\mu_R) \right)
\nonumber\\
&&
\quad
+ \,
\frac{\as(\mu_R)}{2\pi}
\Big(
C_{c i}^{(1)}(z_1)\delta(1-z_2)\delta_{c'j}
+ \, \{z_1,c,i \, \leftrightarrow \, z_2,c',j\}
\Big)
\Bigg\}
\, .
\end{eqnarray}
Other choices for the running coupling of the ${\cal O}(\as^k)$ terms
are of course equally allowed, and we consider these two as representative
of the genuine perturbative ambiguity underlying `primed' predictions.
We refer to results obtained with these two different choices as
\emph{with} or \emph{without running coupling}, respectively. 
At any order above \nllp, ${\cal L}_{{\rm N}^k{\rm LL}^\prime}(k_{t1})$
features a similar ambiguity in constant terms of order ${\cal O}(\as^j)$,
with $j<k$, where one is allowed to run the coupling at arbitrary
loop order, provided the latter is $\geq k-j$.
For consistency, in the ${\cal L}_{{\rm N}^k{\rm LL}^\prime}(k_{t1})$
luminosity without running coupling, constant ${\cal O}(\as^j)$ terms
are evolved at $k-j$ loops, while with running coupling they are
evolved at $k-j+1$ loops. Similar considerations
apply to the luminosity factors appearing in the contributions
with one and two special emissions, i.e.~the lines beyond the first
in eq.~\eqref{eq:master-kt-space}.

The full expressions for the upgraded luminosities up to \nnnllp,
with and without running coupling, are given in Appendix \ref{app:luminosities}.
For the phenomenological \nnnllp presented in this paper, we have
considered both options, choosing as our default the one with
running coupling, as it includes a whole tower of correct N$^{k+1}$LL
effects. We will show the effect of this choice quantitatively in
Sec.~\ref{sec:pheno_results}.

\subsection{Extraction of hard and collinear coefficient functions at ${\cal O}(\as^3)$}
\label{sec:HCGextract}

In this subsection we discuss the extraction of the hard-virtual
function $H$, and of the collinear coefficient functions $C$ and $G$
needed for \nnnllp accuracy.

The $H^{(n)}$ coefficient of the hard-virtual function in
eq.~\eqref{eq:coeff-fun} is obtained from the quark and gluon form
factors at $n$ loops. Except for $\zeta_3$-contributions analogous to
\eqref{eq:deltaH3}, it coincides with the $n$-th term of the
perturbative expansion of $C^2_{\rm \overline{MS}}(\as(M))\times
|C(\as(M),M^2,M^2)|^2$, where $C_{\rm \overline{MS}}(\as(M))$ is the
Wilson coefficient of the $ggH$ effective vertex in the \MSbar{}
scheme~\cite{Chetyrkin:1997un,Schroder:2005hy} and $C(\as(M),M^2,M^2)$
is the hard matching coefficient of Ref.~\cite{Gehrmann:2010ue},
evaluated with time-like virtuality.  Explicit expressions for $n=1,2$
are reported in eqs.~(3.30) and (3.31) of Ref.~\cite{Bizon:2017rah}.
The third-order hard coefficients for Higgs and Drell-Yan production
read
\begin{eqnarray}
\label{eq:H3}
H_g^{(3)}(M)
& = &
34369.2 - 3285.9\ n_f  + 19.9\ n_f^2 
+ L_t\ ( 476.6 + 123.0\ n_f +  0.7\  n_f^2 )
\nonumber\\
&&
+ \, L_t^2\ ( 52.2\ + 11.5\ n_f - 0.8\ n_f^2 )
\, ,
\nonumber\\
H_q^{(3)}(M)
& = &
2507.51 - 309.17\ n_f + 6.52\ n_f^2
\, ,
\end{eqnarray}
where $M$ is the mass of the colour singlet, and
$L_t =\ln(M_H^2/m_{\rm top}^2)$, being $M_H$ the Higgs mass.
The above expression for $H_q^{(3)}(M)$ matches eq.~(7.8) of
Ref.~\cite{Gehrmann:2010ue}, \emph{without} the term
proportional to $N_{F,V}$: this term originates from the structure of
the vector and axial-vector couplings of the neutral Drell-Yan
process, and its presence implies that the Born matrix element
cannot be exactly factored out of the hard-virtual coefficient
at two and three loops.
We discuss the physical reasons for this subtlety, and how we handle
it, in Appendix~\ref{app:axial}.

As discussed in Sec.~\ref{sec:as3const}, in the \radish formalism the
shift $\delta H^{(3)}$ defined in eq.~\eqref{eq:deltaH3}, as well as
the constant parts of the radiator in eq.~\eqref{eq:gk0}, are absorbed
in the hard coefficient $H^{(3)}$. After taking into account the explicit
dependence of the hard function upon the renormalisation scale, the
final expression for $H^{(3)}(\mu_R,L_M)$ reads
\begin{eqnarray}\label{eq:H3scaledepg0k}
H^{(3)}(\mu_R,L_M)
& = &
\lq H^{(3)}(M)\rq_{\rm{eq.}\,\eqref{eq:H3}} + \delta H^{(3)}
\nonumber\\
&&
+ \,
\frac{4}{3}
\left({g^3_{3,0}} + 6 {g_{3,0}} {g_{4,0}} + 6 {g_{5,0}}\right)
+ 2 {H^{(1)}(M)}
\left({g^2_{3,0}} + 2 {g_{4,0}}\right)
+ 2 \, {g_{3,0}} \, {H^{(2)}(M)}
\nonumber\\
&&
+ \,
{\logMtwomurtwo[\!]}
\bigg[
-4 \pi {d_B}
\Big(
\beta_0 \left( {g^2_{3,0}} + 2 {g_{4,0}}\right)
+ 2 \pi ({\beta_1} {g_{3,0}} + \pi  {\beta_2})
\Big)
\nonumber\\
&&
\qquad
- \, 4 \pi ({d_B}+1) {H^{(1)}(M)} ({\beta_0} {g_{3,0}} + \pi {\beta_1})
- 2 \pi \beta_0 ({d_B}+2) {H^{(2)}(M)}
\bigg]
\nonumber\\
&&
+ \,
{\logMtwomurtwo[2]}
\bigg[
2 \pi ^2 \beta_0^2 ({d_B}+1) ({d_B}+2) {H^{(1)}(M)}
\nonumber\\
&&
\qquad
+ \, 4 \pi ^2 \beta_0\, {d_B} \,
\Big({\beta_0} ({d_B}+1) {g_{3,0}} + \pi {\beta_1} (2 {d_B}+3)\Big)
\bigg]
\nonumber\\
&&
- \,
{\logMtwomurtwo[3]}
\bigg[
\frac{4}{3} \pi ^3 \beta_0^3 {d_B} ({d_B}+1) ({d_B}+2)
\bigg]
\, ,
\end{eqnarray}
where $H^{(1)}(M)$ and $H^{(2)}(M)$ are the hard coefficients as given in
eqs.~(3.30) and (3.31) of Ref.~\cite{Bizon:2017rah}, deprived of $\zeta_3$
contributions, and the $g_{k,0}$ constants defined in eq.~\eqref{eq:gk0}
depend on $L_M=\ln(Q/M)$ and read:
\begin{eqnarray}
	g_{3,0} 
	&=&
	B^{(1)} L_M-A^{(1)} L_M^2, \\
	g_{4,0}
	&=&
	- \,
	\frac{2}{3} \pi  A^{(1)} \beta_0\nonumber L_M^3
	+L_M^2 \left(\pi  A^{(1)} \beta_0 \logQtwomurtwo[\!]+
	\pi B^{(1)} \beta_0- \frac{A^{(2)}}{2}\right)\\
    &&
    + \,
    L_M \left(\frac{B^{(2)}}{2}-\pi  B^{(1)} \beta_0
    \logQtwomurtwo[\!]\right),\\
    g_{5,0}
    &=&
    -\frac{2}{3} \pi ^2 A^{(1)}\beta_0^2 L_M^4
    +L_M^3
    \left[-\frac{2}{3} \left(\pi ^2 A^{(1)} \beta_1+\pi  A^{(2)}
    \beta_0-2 \pi ^2 B^{(1)} \beta_0^2\right)+\frac{4}{3} \pi ^2 A^{(1)}
    \beta_0^2 \logQtwomurtwo[\!]\right] \nonumber \\
    &&
    + \,
    L_M^2 \left[ \logQtwomurtwo[\!] \left(\pi ^2 A^{(1)}
   \beta_1 + \pi  A^{(2)} \beta_0 - 2\pi ^2 B^{(1)}
   \beta_0^2\right)-\pi ^2 A^{(1)} \beta_0^2
   \logQtwomurtwo[2] \right. \nonumber\\
    &&
    \left.\hspace{1.1cm} +\left(-\frac14 A^{(3)}+\pi ^2 B^{(1)}
    \beta_1+\pi  B^{(2)} \beta_0\right)
    \right] \nonumber\\
    &&
    + \,
    L_M \left[
    -\logQtwomurtwo[\!]
    \left( \pi ^2 B^{(1)} \beta_1+ \pi  B^{(2)} \beta_0\right)+\pi ^2
    B^{(1)} \beta_0^2 \logQtwomurtwo[2]+\frac{B^{(3)}}{4}
    \right].
\end{eqnarray}
Obviously, in eq.~\eqref{eq:H3scaledepg0k}, the dependence of
the hard coefficient $H^{(n)}(M)$ on the process at hand is understood,
as is the case for the anomalous dimensions contained in
$g_{k,0}$. The above expression matches exactly what we implemented in
the \radish code.
\\

The collinear coefficient functions $C^{(3)}$ are extracted from
the transverse-momentum-dependent (TMD) parton-density functions (PDFs).
These, in turn, are obtained combining the TMD beam functions and the
soft function, computed at third order
in~\cite{Luo:2019szz,Ebert:2020yqt,Luo:2020epw}
and~\cite{Li:2016ctv,Vladimirov:2016dll},
respectively. The $G^{(2)}$ function can be instead extracted from
the computation of the linearly-polarised gluon TMD PDFs at two
loops~\cite{Gutierrez-Reyes:2019rug,Luo:2019bmw}.
Since our starting points for the extraction of $C^{(3)}$ and $G^{(2)}$
are Refs.~\cite{Li:2016ctv,Luo:2019bmw,Luo:2019szz,Luo:2020epw}, in order
to make contact with the notation used therein, we recall that, when
expressed in terms of TMD beam and soft functions, the factorisation
formula for transverse-momentum resummation in $b$-space has the schematic
structure\footnote{For the sake of simplicity, we do not specify here the
renormalisation and the rapidity scales upon which $S_\bot$ and $B$ depend,
denoted respectively as $\mu$ and $\nu$ in Refs.~\cite{Li:2016ctv,Luo:2019bmw,
Luo:2019szz,Luo:2020epw}. As will be explained in the main text, we only need
the scale-independent parts of the TMD beam and soft functions.}
\begin{equation}
\label{eq:sigmascet}
\rd\Sigma^{\rm res}(p_t/M)
\, \sim \,
\sum_{c_1, c_2} \rd|{\cal M}_B|^2_{c_1, c_2} \, H
\int \frac{\rd^2\vec{b}_\bot}{(2\pi)^2}
\, \re^{i \vec{b}_\bot\cdot \vec{\pt}}
\,[B\otimes B]_{c_1, c_2}(\vec{b}_\bot)
\, S_\bot(\vec{b}_\bot)
\, ,
\end{equation}
where $H$, $B$, and $S_\bot$ are the hard, beam, and soft
functions, respectively.
Although the exact correspondence between the \radish formalism
and resummation in impact-parameter space has been discussed
elsewhere \cite{Bizon:2017rah}, by comparing eq.~\eqref{eq:partxs-mellin}
with eq.~\eqref{eq:sigmascet} it is easy to see that the $C$
and $G$ functions are to be extracted from the combination
$B(\vec{b}_\bot)S^{1/2}_\bot(\vec{b}_\bot)$.
As explained in Refs.~\cite{Luo:2019hmp,Luo:2019bmw,Luo:2019szz,
Luo:2020epw} (for instance, eq.~(3.8) of Ref.~\cite{Luo:2020epw})
this is also the combination that allows one to define a TMD
parton density that does not depend on either the rapidity
regulator used for the computation of $B$ and $S_\bot$,
or the rapidity scale.

The beam function $B$ admits an Operator Product Expansion (OPE)
onto the collinear PDFs.
After renormalisation, and after the remaining collinear divergences
are reabsorbed into the collinear PDFs, the unpolarised quark and gluon
beam function are defined by the coefficients of the OPE,
i.e.~by the so called \emph{perturbative matching coefficients}
${\cal I}_{ij}(\xi,b_\bot)$, that are reported, up to ${\cal O}(\as^3)$,
in eq.~(A9) of \cite{Luo:2019szz} for the quark case, and in
eq.~(3.4) of Ref.~\cite{Luo:2020epw} for both the quark
and the gluon cases.
In addition to the ${\cal I}_{gi}(\xi,b_\bot)$ coefficient, the tensor
structure of the gluon beam function contains a second term, associated
to a linearly polarised gluon, whose coefficient is denoted by
${\cal I}'_{gi}(\xi,b_\bot)$ (see eq.~(2.4) of \cite{Luo:2020epw}):
such coefficient gives rise to the $G_{gi}$ collinear
function.\footnote{In Ref.~\cite{Bizon:2017rah}, and for all subsequent
\radish results at \nnnll, the $C^{(n)}$, and $G^{(n)}$ functions
were extracted from the results of Refs.~\cite{Catani:2011kr,
Catani:2012qa,Catani:2013tia}. The match between such results
and those we use here (obtained in SCET), can be easily achieved
by comparing for instance eq.~(2.4) of Ref.~\cite{Luo:2020epw}
with eqs.~(14) to (17) of Ref.~\cite{Catani:2013tia}.}

The results of Refs.~\cite{Luo:2019bmw,Luo:2019szz,Luo:2020epw} and of
Ref.~\cite{Li:2016ctv} contain the boundary conditions for the TMD
soft and beam functions, as well as the complete scale-dependent
expressions for $B$ and $S_\bot$, obtained solving their
evolution equations. We are not interested in the latter, as the
evolution of the coefficient functions is obtained in \radish by means
of the anomalous dimensions ${\bf \Gamma}^{(\rC)}$ and ${\bf
  \Gamma}^{(\rG)}$. According to the above discussion, the
$C_{ij}^{(k)}(z)$ coefficient function is extracted by means of the
identity
\begin{equation}
\label{eq:extr_C}
\lq
\sum_{n=0}^3
\(\frac{\as}{4\pi}\)^n I_{ij}^{(n)}(z)
\rq
\sqrt{\exp\left({\sum_{m=1}^3 \(\frac{\as}{4\pi}\)^{m} c_{m}^\bot}\right)}
\, = \,
\sum_{k=0}^3
\(\frac{\as}{2\pi}\)^k C_{ij}^{(k)}(z) \, + \, {\cal O}(\as^4)
\, ,
\end{equation}
where the $I_{ij}^{(n)}(z)$ functions
(eq.~(3.4) and supplemental material of~\cite{Luo:2020epw})
are the boundary conditions for the TMD beam functions,
and the coefficients $c_{m}^\bot$ (eqs.~(10), (10S) and (11S)
of~\cite{Li:2016ctv}), whose overall colour factor is $C_F$
($C_A$) for Drell-Yan (Higgs) production, are the boundary conditions
for $S_\bot$.
Similarly, the $G_{ij}^{(k)}(z)$ collinear coefficient function is
extracted through
\begin{equation}
\label{eq:extr_G}
\lq
\sum_{n=1}^2
\(\frac{\as}{4\pi}\)^n I_{gj}^{\prime(n)}(z)
\rq
\sqrt{\exp\left({\sum_{m=1}^2 \(\frac{\as}{4\pi}\)^{m} c_{m}^\bot}\right)}
\, = \,
\sum_{k=1}^2
\(\frac{\as}{2\pi}\)^k G_{gj}^{(k)}(z) \, + \, {\cal O}(\as^3)
\, ,
\end{equation}
where the $I_{gj}^{\prime(n)}(z)$ functions are given in
eqs.~(2.21) and (2.22) of~\cite{Luo:2019bmw}.

We have extracted $C_{ij}^{(k)}(z)$ and $G_{gj}^{(k)}(z)$ from the
auxiliary \texttt{Mathematica} notebooks provided in
Refs.~\cite{Luo:2019bmw,Luo:2020epw}, and we have inserted them
in the \radish code, following the conventions of
Refs.~\cite{Moch:2004pa,Vogt:2004mw} for the flavour-decomposition
of coefficient functions.
As a cross check, we have verified that we obtain for $C_{ij}^{(2)}(z)$
the same result we extracted in Ref.~\cite{Bizon:2017rah}.
The $C_{ij}^{(3)}(z)$ expressions contain harmonic polylogarithms
(HPLs) of weight up to $5$, which we efficiently evaluate via the
\texttt{fortran} routine \textsc{hplog5}
\cite{Gehrmann:2001pz}.\footnote{We thank Thomas Gehrmann for providing
us with a version of the routine which contains the evaluation of HPLs
of weight 5.}
As far as the numerical implementation is concerned, we perfectly
reproduced Fig.~(2) of \cite{Luo:2019szz}, and we verified that our
\texttt{fortran} implementation matches the numerical results obtained
using \texttt{Mathematica} and the package \textsc{HPL}
\cite{Maitre:2005uu}.
Moreover,
we also checked our implementation by comparing against the N$^3$LO
TMD PDFs results obtained in Refs.~\cite{Ebert:2020yqt}, finding
perfect agreement.

As discussed in Sec.~\ref{sec:as3const}, in the \radish formalism we
absorb in the third-order coefficient function the shift $\delta
C_{ij}^{(3)}(z)$ defined in eq.~\eqref{eq:deltaC3}.
Furthermore, in order to match our resummed results to fixed-order
calculations that feature $\as(\mu_R)$ and $f(\mu_F)$, we write
the factors of $\as(k_{t1})$ and $f(k_{t1})$ appearing in the luminosities
in terms of $\as(\mu_R \, \re^{-L})$ and $f(\mu_F \, \re^{-L})$, respectively,
with $L=\ln(Q/k_{t1})$, absorbing the ensuing constant difference in
the coefficient functions.
This gives rise to an explicit $\mu_{F,R}$ dependence in the latter,
which is reported in eq.~(4.6) of Ref.~\cite{Bizon:2017rah}
for ${C}_{ij}^{(1)}$ and ${C}_{ij}^{(2)}$.
As for $C_{ij}^{(3)}$, we document such a dependence in the following
equation:
\begin{eqnarray}
{C}_{ij}^{(3)}(z,\mu_F,\mu_R,Q)
& = &
\lq
C_{ij}^{(3)}(z)
\rq_{\rm{eq.}\,\eqref{eq:extr_C}}
\, + \,
\delta C_{ij}^{(3)}(z)
\nonumber\\
&&
- \, 4 \pi
\Big(
\beta_0 C^{(2)}_{ij}(z)
+\pi \beta_1 C^{(1)}_{ij}(z)
\Big)
\ln\frac{Q^2}{\mu_R^2}
+
4 \pi^2 \beta_0^2 C^{(1)}_{ij}(z)
\ln^2\frac{Q^2}{\mu_R^2}
\nonumber\\
&&
+ \,
\ln\frac{Q^2}{\mu_F^2}
\bigg[
(C^{(2)} \otimes \hat P^{(0)})_{ij}(z)
+
(C^{(1)} \otimes \hat P^{(1)})_{ij}(z)
+
\hat P^{(2)}_{ij}(z)
\nonumber\\
&&
\qquad
- \,
4 \pi
\Big(
\beta_0(C^{(1)} \otimes \hat P^{(0)})_{ij}(z)
+
\beta_0 \hat P^{(0)}_{ij}(z)
+
\pi \beta_1 \hat P^{(0)}_{ij}(z)
\Big)
\ln\frac{Q^2}{\mu_R^2}
\nonumber\\
&&
\qquad
+ \,
4 \pi^2 \beta_0^2 \hat P^{(0)}_{ij}(z)
\ln^2\frac{Q^2}{\mu_R^2}
\bigg]
\nonumber\\
&&
+ \,
\ln^2\frac{Q^2}{\mu_F^2}
\bigg[
\frac12
(C^{(1)} \otimes \hat P^{(0)} \otimes \hat P^{(0)}
+\hat P^{(1)} \otimes \hat P^{(0)}+\hat P^{(0)} \otimes \hat P^{(1)})_{ij}(z)
\nonumber\\
&&
\qquad
+ \,
\pi \beta_0
(C^{(1)} \otimes \hat P^{(0)}
+ 2 \hat P^{(1)})_{ij}(z)
+
2 \pi^2 \beta_1 \hat P^{(0)}_{ij}(z)
\nonumber\\
&&
\qquad
- \,
2 \pi \beta_0
\Big(
(\hat P^{(0)} \otimes \hat P^{(0)})_{ij}(z)
+
2 \pi \beta_0 \hat P^{(0)}_{ij}(z)
\Big)
\ln\frac{Q^2}{\mu_R^2}
\bigg]
\nonumber\\
&&
+ \,
\ln^3\frac{Q^2}{\mu_F^2}
\bigg[
\frac16
(\hat P^{(0)} \otimes \hat P^{(0)} \otimes \hat P^{(0)})_{ij}(z)
+ \pi \beta_0
(\hat P^{(0)} \otimes \hat P^{(0)})_{ij}(z)
\nonumber\\
&&
\qquad
+ \,
\frac43 \pi^2 \beta_0^2
\hat P^{(0)}_{ij}(z)
\bigg]
\, .
\end{eqnarray}
Analogously, for $G_{gj}^{(2)}(z)$ one has
\begin{eqnarray}
{G}_{gj}^{(2)}(z,\mu_F,\mu_R,Q)
\, = \,
\lq
G_{gj}^{(2)}(z)
\rq_{\rm{eq.}\,\eqref{eq:extr_G}}
+
(G^{(1)} \otimes \hat P^{(0)})_{gj}(z)
\ln\frac{Q^2}{\mu_F^2}
-
2\pi\beta_0 \,
G^{(1)}_{gj}(z)
\ln\frac{Q^2}{\mu_R^2}
\, .
\end{eqnarray}
The above equations exactly match the expressions implemented in
the \radish code.
In order for the next section to be notationally consistent with
the previous ones, we will still denote parton densities and coupling
constant as $f(k_{t1})$ and $\as(k_{t1})$ in the following formulae,
understanding the above discussion.

\subsection{${\cal O}(\as^3)$ constants from multiple resolved emissions}
We now turn to the description of ${\cal O}(\as^3)$ terms stemming from the
two-emission contribution to \eqref{eq:mellin_expanded_final}.
The first immediate correction comes from the last identification in the list
\eqref{eq:conversions_to_direct_space}, where we now need to retain
${\bf\Gamma}^{(1)}_{N_{\ell}}$, and not just ${\bf\Gamma}^{(0)}_{N_{\ell}}$
as done for \nnnll. This yields
\begin{eqnarray}
&&
\frac{\rd|{\cal M}_B|_{c_1c_2}^2}{\rd\Phi_B} \, \, 
{\bf f}^{T}_{N_1}(k_{t1}) \,
\frac{\as^3(k_{t1})}{2\pi^3}
\sum_{\ell_i,\ell_j=1}^2
\bigg[
{\bf\Gamma}^{(0)}_{N_{\ell_i}} \,
{\bf\Gamma}^{(1)}_{N_{\ell_j}}
+
{\bf\Gamma}^{(1)}_{N_{\ell_i}} \,
{\bf\Gamma}^{(0)}_{N_{\ell_j}}
\bigg]
{\bf f}_{N_2}(k_{t1})
\nonumber\\
&&
\qquad
\longrightarrow ~~
\frac{\as^3(k_{t1})}{2\pi^3}
\bigg[
\hat{P}^{(0)} \otimes \hat{P}^{(1)}
+  
\hat{P}^{(1)} \otimes \hat{P}^{(0)}
\bigg]
\otimes {\cal L}_{\rm NLL}(k_{t1})
\, ,
\end{eqnarray}
where we defined
\begin{eqnarray}
\hat{P}^{(0)} \otimes \hat{P}^{(1)} \otimes {\cal L}_{\rm NLL}(k_{t1})
& \equiv &
\sum_{c,c'}
\frac{\rd|{\cal M}_{B}|_{cc'}^2}{\rd\Phi_B} \,
\bigg[
\big(\hat{P}^{(0)} \otimes \hat{P}^{(1)} \otimes f \big)_c (k_{t1},x_1) \, f_{c'}(k_{t1},x_2)
\nonumber\\
&&
\hspace{20mm}
+ \,
f_c(k_{t1},x_1) \,
\big(\hat{P}^{(0)} \otimes \hat{P}^{(1)} \otimes f \big)_{c'}(k_{t1},x_2)
\nonumber\\[7pt]
&&
\hspace{20mm}
+ \,
\big(\hat{P}^{(0)} \otimes f \big)_c (k_{t1},x_1) \,
\big(\hat{P}^{(1)} \otimes f \big)_{c'} (k_{t1},x_2)
\nonumber\\
&&
\hspace{20mm}
+ \,
\big(\hat{P}^{(1)} \otimes f \big)_c (k_{t1},x_1) \,
\big(\hat{P}^{(0)} \otimes f \big)_{c'} (k_{t1},x_2)
\bigg]
\, .
\end{eqnarray}
The following correction features for the first time the contribution of ${\bf\Gamma}^{({\rC})}$:
\begin{eqnarray}
&&
\frac{\rd|{\cal M}_B|_{c_1c_2}^2}{d\Phi_B} \, \, 
{\bf f}^{T}_{N_1}(k_{t1}) \,
\frac{\as^3(k_{t1})}{4\pi^3}
\sum_{\ell_i,\ell_j=1}^2
\bigg[
{\bf\Gamma}^{(0)}_{N_{\ell_i}} \,
{\bf\Gamma}^{({\rC},1)}_{N_{\ell_j}}
+
{\bf\Gamma}^{({\rC},1)}_{N_{\ell_i}} \,
{\bf\Gamma}^{(0)}_{N_{\ell_j}}
\bigg]
{\bf f}_{N_2}(k_{t1})
\nonumber\\
&&
\qquad
\longrightarrow ~~
- \beta_0
\frac{\as^3(k_{t1})}{\pi^2}
\bigg[
\hat{P}^{(0)} \otimes C^{(1)}
+
C^{(1)} \otimes \hat{P}^{(0)}
\bigg]
\otimes {\cal L}_{\rm NLL}(k_{t1})
\, ,
\end{eqnarray}
where we have used the evolution equation \eqref{eq:RGEC} to evaluate
${\bf\Gamma}^{({\rC},1)}  = - 4 \pi \beta_0 {\bf C}^{(1)}$, and
\begin{eqnarray}
\hat{P}^{(0)} \otimes {C}^{(1)} \otimes {\cal L}_{\rm NLL}(k_{t1})
& \equiv &
\sum_{c,c'}
\frac{\rd|{\cal M}_{B}|_{cc'}^2}{\rd\Phi_B} \,
\bigg[
\big(\hat{P}^{(0)} \otimes {C}^{(1)} \otimes f \big)_c (k_{t1},x_1) \, f_{c'}(k_{t1},x_2)
\nonumber\\
&&
\hspace{20mm}
+ \,
f_c(k_{t1},x_1) \,
\big(\hat{P}^{(0)} \otimes {C}^{(1)} \otimes f \big)_{c'}(k_{t1},x_2)
\nonumber\\[7pt]
&&
\hspace{20mm}
+ \,
\big(\hat{P}^{(0)} \otimes f \big)_c (k_{t1},x_1) \,
\big({C}^{(1)} \otimes f \big)_{c'} (k_{t1},x_2)
\nonumber\\
&&
\hspace{20mm}
+ \,
\big({C}^{(1)} \otimes f \big)_c (k_{t1},x_1) \,
\big(\hat{P}^{(0)} \otimes f \big)_{c'} (k_{t1},x_2)
\bigg]
\, .
\end{eqnarray}
Next, an ${\cal O}(\as^3)$ contribution coming from the derivative
of the DGLAP anomalous dimension in the last line of
\eqref{eq:mellin_expanded_final} reads
\begin{eqnarray}
&&
\frac{\rd|M_B|_{c_1c_2}^2}{\rd\Phi_B} \,
{\bf f}^{T}_{N_1}(k_{t1}) \,
\frac{\as^3(k_{t1})}{\pi^2} \,
2 \, \beta_0 \ln\frac1{\zeta_i}
\sum_{\ell_i,\ell_j=1}^2
{\bf\Gamma}^{(0)}_{N_{\ell_i}}\,
{\bf\Gamma}^{(0)}_{N_{\ell_j}}\,
{\bf f}_{N_2}(k_{t1})
\nonumber\\
&&
\qquad
\longrightarrow ~~
\frac{\as^3(k_{t1})}{\pi^2} \,
2 \, \beta_0
\ln\frac1{\zeta_i} \,
\hat{P}^{(0)} \otimes \hat{P}^{(0)} \otimes {\cal L}_{\rm NLL}(k_{t1})
\, .
\end{eqnarray}
Analogously, a constant ${\cal O}(\as^3)$ term is induced by a luminosity
upgrade ${\cal L}_{\rm NLL}(k_{t1}) \to {\cal L}_{\rm NLL'}(k_{t1})$ in
the fifth line of eq.~\eqref{eq:master-kt-space}, where
${\cal L}_{\rm NLL'}(k_{t1})$ was introduced in Sec.~\ref{sec:as3const}.
Conversely, the three contributions
\begin{eqnarray}
&&
\frac{\rd|M_B|_{c_1c_2}^2}{\rd\Phi_B} \,
{\bf f}^{T}_{N_1}(k_{t1}) \,
\frac{\as^2(k_{t1})}{2\pi^2}
\ln\frac1{\zeta_i} \,
\sum_{\ell_i,\ell_j=1}^2
{\bf\Gamma}_{N_{\ell_j}}^{(1)} {\bf R}_{\ell_i}''(k_{t1}) \,
{\bf f}_{N_2}(k_{t1})
\, ,
\nonumber\\
&&
\frac{\rd|M_B|_{c_1c_2}^2}{\rd\Phi_B} \,
{\bf f}^{T}_{N_1}(k_{t1}) \,
\frac{\as^2(k_{t1})}{4\pi^2}
\ln \frac{1}{\zeta_i} \,
\sum_{\ell_i,\ell_j=1}^2
{\bf\Gamma}^{(\rC,1)}_{N_{\ell_j}} \, {\bf R}_{\ell_i}''(k_{t1}) \,
{\bf f}_{N_2}(k_{t1})
\, ,
\nonumber\\
&&\frac{\rd|M_B|_{c_1c_2}^2}{\rd\Phi_B} \,
{\bf f}^{T}_{N_1}(k_{t1}) \,
\frac{\as(k_{t1})}{2\pi}
\ln^2\frac1{\zeta_i}
\sum_{\ell_i,\ell_j=1}^2
{\bf\Gamma}_{N_{\ell_j}}^{(0)} \, {\bf R}_{\ell_i}'''(k_{t1}) \,
{\bf f}_{N_2}(k_{t1})
\, ,
\end{eqnarray}
are already accounted for by the third line of eq.~\eqref{eq:master-kt-space},
and need not be added.
\\

The final terms to be considered are corrections to the three-emission contributions.
They feature a term with three lowest-order DGLAP-evolution matrices
\begin{eqnarray}
&&
\frac{\rd|M_B|_{c_1c_2}^2}{\rd\Phi_B} \,
{\bf f}^{T}_{N_1}(k_{t1}) \,
\frac{\as^3(k_{t1})}{2\pi^3}
\sum_{\ell_i, \ell_j, \ell_k=1}^2
{\bf\Gamma}_{N_{\ell_i}}^{(0)} \,
{\bf\Gamma}_{N_{\ell_j}}^{(0)} \,
{\bf\Gamma}_{N_{\ell_k}}^{(0)} \,
{\bf f}_{N_2}(k_{t1})
\nonumber\\
&&
\qquad
\longrightarrow ~~
\frac{\as^3(k_{t1})}{2\pi^3}
\hat{P}^{(0)} \otimes \hat{P}^{(0)} \otimes \hat{P}^{(0)} \otimes {\cal L}_{\rm NLL}(k_{t1})
\, ,
\end{eqnarray}
where
\begin{eqnarray}
&&\hspace{-20mm}
\hat{P}^{(0)} \otimes \hat{P}^{(0)} \otimes \hat{P}^{(0)} \otimes {\cal L}_{\rm NLL}(k_{t1})
\, \, =
\nonumber\\
& = &
\sum_{c,c'}
\frac{\rd|{\cal M}_{B}|_{cc'}^2}{\rd\Phi_B} \,
\bigg[
\big(\hat{P}^{(0)} \otimes \hat{P}^{(0)} \otimes \hat{P}^{(0)} \otimes f \big)_c (k_{t1},x_1) \, f_{c'}(k_{t1},x_2)
\nonumber\\
&&
\hspace{20mm}
+ \,
f_c(k_{t1},x_1) \,
\big(\hat{P}^{(0)} \otimes \hat{P}^{(0)} \otimes \hat{P}^{(0)} \otimes f \big)_{c'}(k_{t1},x_2)
\nonumber\\[7pt]
&&
\hspace{20mm}
+ \,
3
\big(\hat{P}^{(0)} \otimes \hat{P}^{(0)} \otimes f \big)_c (k_{t1},x_1) \,
\big(\hat{P}^{(0)} \otimes f \big)_{c'} (k_{t1},x_2)
\nonumber\\
&&
\hspace{20mm}
+ \,
3
\big(\hat{P}^{(0)} \otimes f \big)_c (k_{t1},x_1) \,
\big(\hat{P}^{(0)} \otimes \hat{P}^{(0)} \otimes f \big)_{c'} (k_{t1},x_2)
\bigg]
\, ,
\end{eqnarray}
and two terms with the second derivative of the radiator
\begin{eqnarray}
&&
\frac{\rd|M_B|_{c_1c_2}^2}{\rd\Phi_B} \,
{\bf f}^{T}_{N_1}(k_{t1}) \,
\frac{\as^2(k_{t1})}{2\pi^2}
\left(\ln\frac1{\zeta_i} + \ln\frac1{\zeta_j}\right)
\sum_{\ell_i, \ell_j, \ell_k=1}^2
{\bf R}''_{\ell_i} \,
{\bf \Gamma}_{N_{\ell_j}}^{(0)} \,
{\bf \Gamma}_{N_{\ell_k}}^{(0)} \,
{\bf f}_{N_2}(k_{t1})
\nonumber\\
&&
\qquad
\longrightarrow ~~
\frac{\as^2(k_{t1})}{2\pi^2}
\left(\ln \frac1{\zeta_i} + \ln\frac1{\zeta_j}\right)
R''(k_{t1}) \,
\hat{P}^{(0)} \otimes \hat{P}^{(0)} \otimes {\cal L}_{\rm NLL}(k_{t1})
\, ,
\end{eqnarray}
and
\begin{eqnarray}
&&
\frac{\rd|M_B|_{c_1c_2}^2}{\rd\Phi_B} \,
{\bf f}^{T}_{N_1}(k_{t1}) \,
\frac{\as(k_{t1})}{2\pi} \,
\ln\frac1{\zeta_i}\ln\frac1{\zeta_j}
\sum_{\ell_i, \ell_j, \ell_k=1}^2
{\bf R}''_{\ell_i} \,
{\bf R}''_{\ell_j} \,
{\bf\Gamma}_{N_{\ell_k}}^{(0)} \,
{\bf f}_{N_2}(k_{t1})
\nonumber\\
&&
\qquad
\longrightarrow ~~
-
\frac12
\ln\frac1{\zeta_i}\ln\frac1{\zeta_j}
(R''(k_{t1}) )^2 \, \,
\partial_L \, {\cal L}_{\rm NLL}(k_{t1})
\, .
\end{eqnarray}
Collecting all contributions, our final formula for direct-space resummation
at \nnnllp reads
\begin{eqnarray}
\label{eq:master-kt-space-prime}
\frac{\rd\Sigma^{\rm N^3LL'}(v)}{\rd\Phi_B}
& = &
\int\frac{\rd k_{t1}}{k_{t1}}
\frac{\rd\phi_1}{2\pi} \, 
\partial_{L}
\left(- \, \re^{-R(k_{t1})} {\cal L}_{\rm N^3LL'}(k_{t1}) \right)
\int \dZ \, 
\Theta \left(v-V(\{\tilde{p}\},k_1,\dots, k_{n+1})\right)
\nonumber\\[10pt]
&&
+ \,
\int \frac{\rd k_{t1}}{k_{t1}}
\frac{\rd\phi_1}{2\pi} \,
\re^{-R(k_{t1})}
\int \dZ 
\int_0^1 \frac{\rd\zeta_s}{\zeta_s}
\frac{\rd\phi_s}{2\pi}
\nonumber\\
&&
\times \,
\Bigg\{
\bigg(
R' (k_{t1}) {\cal L}_{\rm NNLL}(k_{t1}) - \partial_L {\cal L}_{\rm NNLL}(k_{t1})
\bigg)
\bigg(
R''(k_{t1}) \ln\frac1{\zeta_s} + \frac{1}{2} R'''(k_{t1}) \ln^2\frac1{\zeta_s}
\bigg)
\nonumber\\
&&
\qquad
- \, R'(k_{t1})
\left(
\partial_L {\cal L}_{\rm NNLL}(k_{t1})
- 2 \, \frac{\beta_0}{\pi} \, \as^2(k_{t1}) \hat{P}^{(0)} \otimes {\cal L}_{\rm NLL}(k_{t1}) \ln\frac1{\zeta_s}
\right)
\nonumber\\
&&
\qquad
+ \,
\frac{\as^2(k_{t1})}{\pi^2} \hat{P}^{(0)} \otimes \hat{P}^{(0)} \otimes {\cal L}_{\rm NLL'}(k_{t1})
\nonumber\\
&&
\qquad
+ \,
\frac{\as^3(k_{t1})}{2\pi^3} \,
\bigg[
\hat{P}^{(0)} \otimes \hat{P}^{(1)}
+  
\hat{P}^{(1)} \otimes \hat{P}^{(0)}
\bigg]
\otimes {\cal L}_{\rm NLL}(k_{t1})
\nonumber\\
&&
\qquad
- \,
\beta_0
\frac{\as^3(k_{t1})}{\pi^2} \,
\bigg[
\hat{P}^{(0)} \otimes C^{(1)}
+
C^{(1)} \otimes \hat{P}^{(0)}
\bigg]
\otimes {\cal L}_{\rm NLL}(k_{t1})
\nonumber\\
&&
\qquad
+ \,
\frac{\as^3(k_{t1})}{\pi^2} \,
2 \, \beta_0 \ln\frac1{\zeta_s} \,
\hat{P}^{(0)} \otimes \hat{P}^{(0)} \otimes {\cal L}_{\rm NLL}(k_{t1})
\Bigg\}
\nonumber\\
&&
\times \,
\bigg[
\Theta\left(v-V(\{\tilde{p}\},k_1,\dots, k_{n+1},k_s)\right) -
\Theta\left(v-V(\{\tilde{p}\},k_1,\dots, k_{n+1})\right)
\bigg]
\nonumber\\[10pt]
&&
+ \,
\frac12
\int\frac{\rd k_{t1}}{k_{t1}}
\frac{\rd\phi_1}{2\pi}
\re^{-R(k_{t1})}
\int \dZ
\int_0^1 \frac{\rd \zeta_{s1}}{\zeta_{s1}}
\frac{\rd\phi_{s1}}{2\pi}
\int_0^1 \frac{\rd \zeta_{s2}}{\zeta_{s2}}
\frac{\rd\phi_{s2}}{2\pi} \,
\nonumber\\
&&
\times \,
\Bigg\{
R'(k_{t1})
\Bigg[
{\cal L}_{\rm NLL}(k_{t1})
\big(R''(k_{t1})\big)^2 \ln\frac{1}{\zeta_{s1}} \ln\frac{1}{\zeta_{s2}}
\nonumber\\
&&
\qquad
- \,
\partial_L {\cal L}_{\rm NLL}(k_{t1}) R''(k_{t1})
\bigg(
\ln\frac{1}{\zeta_{s1}} + \ln\frac{1}{\zeta_{s2}}
\bigg)
\nonumber\\
&&
\qquad
+ \,
\frac{\as^2(k_{t1})}{\pi^2} \hat{P}^{(0)} \otimes \hat{P}^{(0)} \otimes {\cal L}_{\rm NLL}(k_{t1})
\Bigg]
\nonumber\\
&&
\qquad
+ \,
\frac{\as^2(k_{t1})}{\pi^2} \,
\left(\ln \frac1{\zeta_{s1}} + \ln\frac1{\zeta_{s2}}\right)
R''(k_{t1}) \,
\hat{P}^{(0)} \otimes \hat{P}^{(0)} \otimes {\cal L}_{\rm NLL}(k_{t1})
\nonumber\\
&&
\qquad
- \,
\ln\frac1{\zeta_{s1}}\ln\frac1{\zeta_{s2}}
(R''(k_{t1}) )^2 \,
\partial_L {\cal L}_{\rm NLL}(k_{t1})
\nonumber\\
&&
\qquad
+ \,
\frac{\as^3(k_{t1})}{\pi^3} \,
\hat{P}^{(0)} \otimes \hat{P}^{(0)} \otimes \hat{P}^{(0)} \otimes {\cal L_{\rm NLL}}(k_{t1})
\Bigg\}
\nonumber\\
&&
\times \,
\bigg[
\Theta\left(v-V(\{\tilde{p}\},k_1,\dots,k_{n+1},k_{s1},k_{s2})\right) -
\Theta\left(v-V(\{\tilde{p}\},k_1,\dots,k_{n+1},k_{s1})\right)
\nonumber\\
&&
\qquad
- \,
\Theta\left(v-V(\{\tilde{p}\},k_1,\dots,k_{n+1},k_{s2})\right) +
\Theta\left(v-V(\{\tilde{p}\},k_1,\dots,k_{n+1})\right)
\bigg]
\, .
\end{eqnarray}
We stress that the comments on the modified logarithms and jacobian factor reported below
eq.~\eqref{eq:master-kt-space} apply unchanged to eq.~\eqref{eq:master-kt-space-prime}
as well.

\subsection{Transverse-recoil effects}
\label{sec:recoil}

In order to realistically simulate the kinematics of the singlet's decay products, we have implemented
in our framework the default transverse-recoil prescription of \cite{Catani:2015vma} to account for the singlet recoiling against initial-state QCD radiation.
The procedure amounts to considering the differential spectrum with respect to observable $v$,
and to boosting its underlying Born kinematics from a rest frame of the singlet (specifically, the Collins-Soper one \cite{Collins:1977iv} in the default
prescription) to the laboratory frame: there the singlet has transverse momentum
equal to $q_t(v)$, where $q_t(v) = M v$ (or $q_t(v) = M v/|\sin\phi|$, with $\phi$ the singlet's azimuthal
angle) if $v = \pt/M$ (or $v=\phs$).
Fiducial selection cuts are then applied on the boosted Born kinematics.

As argued in \cite{Ebert:2020dfc}, see also \cite{Becher:2020ugp}, the inclusion of recoil effects via the prescriptions of \cite{Catani:2015vma} is sufficient to account for all linear power corrections in presence of fiducial cuts, together with their resummation with the same accuracy as the leading-power terms, for observables which are azimuthally symmetric at leading power, such as $\pt/M$.

Let us briefly discuss the technical implementation of recoil effects in the \radish code.
For each $m$-parton contribution to eq.~\eqref{eq:master-kt-space-prime}, as defined by the
$\Theta (v - V(\{\tilde p\}, k_1, \dots, k_m ))$ measurement functions, we evaluate the transverse
momentum $q_t(v)$ of the colour singlet and its azimuthal angle $\phi$, and we apply the
above mentioned boost.
In order to enforce fiducial cuts on the boosted Born system, we modify each measurement function
in eq.~\eqref{eq:master-kt-space-prime} as
\begin{equation}
\Theta (v - V(\{\tilde p\}, k_1, \dots, k_m ) )
\,\, \longrightarrow \,\,
\Theta (v - V(\{\tilde p\}, k_1, \dots, k_m ) ) \,
\Theta_{\rm cuts}(\Phi_{B} , \{ k_1, \dots, k_m \})
\, ,
\end{equation}
where the dependence on $k_1, \dots, k_m$ in $\Theta_{\rm cuts}$ encodes the effect of the boost
(i.e.~$\Theta_{\rm cuts}$ equals 1 or 0 if the boosted Born configuration passes or not the cuts).
On the contrary, in absence of recoil effects, the action of the cuts does not depend on
momenta $k_1, \dots, k_m$: the constraint $\Theta_{\rm cuts} (\Phi_{B}, \{ k_1, \dots, k_m \})$
reduces to $\Theta_{\rm cuts} (\Phi_{B})$ and factorises out of the resummation formula,
therefore eq.~\eqref{eq:master-kt-space-prime} is calculated only for the points which pass
the fiducial cuts.

Finally, in order to match the resummed result with fixed-order predictions, when transverse-recoil
effects are included we also need to modify the perturbative expansion of the resummation.
As detailed in Ref.~\cite{Bizon:2017rah} (see in particular Sec.~4.2), in the default code
the latter expansion is computed at the cumulative level, and expressed as a combination of
classes of `master' integrals.
Since the recoil procedure entails boosts on the differential spectrum, we now first compute the
derivative of the expansion at a given value $v$, and then apply fiducial cuts on the boosted
kinematics, consistently with what is done in the resummation component.

\section{Validation}
\label{sec:validation}

In this section we discuss the tests we performed to validate our implementation of N$^3$LL$^\prime$
effects in the \radish code.

A first robust check is achieved by comparing the $\as^3$ expansion of the momentum-space
resummation formula for $\pt$ with the analogous expression derived starting from the cumulative
$\pt$ cross section in $b$-space:
\begin{equation}
\label{eq:Sigma_cum_b_space}
\frac{\rd \Sigma^{\rm N^3LL^\prime}(\pt)}{\rd \Phi_B}
\, = \, 
\int_0^\infty \rd b \, \pt \,
J_1 (b\pt) \, \re^{-R_b} \, {\cal L}_{\rm N^3LL^\prime}(b_0/b)
\, ,
\end{equation}
where $J_1$ is the second Bessel function, and $R_b$ is the radiator
as written in \eqref{eq:R_b_space} in terms of anomalous dimensions
$A_{{\rm CSS},\ell}$, $B_{{\rm CSS},\ell}$.
We stress that this test has the virtue of allowing to assess at the
analytic level the correctness of the $\delta H^{(3)}$ and
$\delta C_{ij}^{(3)}$ terms derived in Sec.~\ref{sec:as3const}.

The inverse Fourier transform \eqref{eq:Sigma_cum_b_space} can be
calculated by Taylor-expanding the radiator and the luminosity factor
around $b = b_0/\pt$ at the appropriate order.
This allows to write the cumulative cross section as
\begin{equation}
\label{eq:b_space_test}
\frac{\rd \Sigma^{\rm N^3LL^\prime}(\pt)}{\rd \Phi_B}
\, = \,
\sum_n \, c_n(\pt)
\int_0^\infty \rd b \, \pt \,
J_1 (b\pt) \, \ln^n(b\pt/b_0) \,
\left(\frac {b\pt}{b_0}\right)^{-R'_{\rm CSS}(\pt)}
\, ,
\end{equation}
where $R_{\rm CSS}'$ was introduced in eq.~\eqref{eq:R_b_space},
and $c_n(\pt)$ are coefficients encoding luminosity and
radiator information.
The integrals in eq.~\eqref{eq:b_space_test} are then readily obtained
as derivatives with respect to $R'_{\rm CSS}(\pt)$ of the generating functional
\begin{equation}
{\cal F} \big[ R_{\rm CSS}'(\pt)\big]
\, = \,
\int_0^\infty \rd b \, \pt \,
J_1 (b\pt) \, \left(\frac {b\pt}{b_0}\right)^{-R_{\rm CSS}'(\pt)}
\, = \,
\re^{-\gamma_E R_{\rm CSS}'(\pt)} \,
\frac{\Gamma \Big[1-R_{\rm CSS}'(\pt)/2\Big]}{\Gamma \Big[1+R_{\rm CSS}'(\pt)/2\Big]}
\, .
\end{equation}

This procedure provides an analytic expression to be directly compared
with the momentum-space expansion, which is written (see the discussion in
Sec.~4.2 of Ref.~\cite{Bizon:2017rah}) as a linear combination of classes
of `master' integrals.
Such master integrals are evaluated analytically up to
${\cal O}(\as^2)$\footnote{For this test we have considered the momentum-space expansion
in terms of un-modified logarithms, differing only by power corrections. This yields
much simpler expressions for the master integrals, significantly enhancing the stability
of the test.}, while we resorted to high-accuracy numerical integration for
those entering at ${\cal O}(\as^3)$.
By comparing the two expressions for each relevant combination of $A_{\ell}^{(n)}$
and $B_{\ell}^{(n)}$ anomalous dimensions, and retaining full renormalisation,
factorisation, and resummation scale dependence, we achieved complete analytic agreement
at order $\as^2$, and numerical agreement at or below the permyriad level for all
terms entering at $\as^3$, which is the numerical accuracy level of the master
integrals.
An analogous check was performed in the case of the $\phs$ expansion, finding similar
agreement.

As a further stringent test, we have numerically checked that the $\mu_R$, $\mu_F$,
and $Q$ dependence of our N$^k$LL$^\prime$ cumulative results cancels exactly at order
${\cal O}(\as^k)$, and is of relative order ${\cal O}(\as^{k+1} L)$ with respect to
the Born, i.e.~a pure N$^{k+1}$LL effect.
In order to perform this test, we evaluate our expressions in the small-coupling regime
$\alpha_{s} \ll 1$ with a set of analytic toy PDFs~\cite{Vogt:2004ns}.
If the dependence on e.g.~the renormalisation scale $\mu_R$ is implemented
correctly, one must obtain
\begin{equation}
\Delta_{{\rm N}^k{\rm LL}'}(v; \as)
\, \equiv \,
\Sigma_{{\rm N}^k{\rm LL}'}(v; \as; \mu_R = \lambda M)
-
\Sigma_{{\rm N}^k{\rm LL}'}(v; \as; \mu_R = M)
\, = \,
{\cal O}(\as^{d_B+k+1} L),
\end{equation}
where $d_B=2$ ($d_B=0$) for Higgs (DY) production, and $\lambda$ is
an ${\cal O}(1)$ rescaling factor.
For sufficiently small $\as$ values, we have tested the exact cancellation
of the scale dependence by confronting the ratio $\Delta_{{\rm N}^k{\rm LL}'}
(v; \kappa\as)/\Delta_{{\rm N}^k{\rm LL}'} (v; \as)$ against its expected
scaling $\kappa^{d_B+k+1}$. A similar test has been successfully performed
on the expansion of the resummation formula in powers of $\as$.
We have also explicitly checked that the artificial introduction of small bugs
in the coefficients of the scale-dependent terms results in clearly visible
violations of the test, whose successful outcome then strongly corroborates
the robustness of our implementation.

Finally, as an internal self-consistency test, we compare the resummed result for
$\Sigma_{{\rm N}^k{\rm LL}'}(v)$ to its ${\cal O}(\as^k)$ expansion in the asymptotic
$v \gg 1$ limit. Owing to the presence of modified logarithms, the two expressions
are expected and numerically checked to coincide in such a limit, which also
ensures the absence of residual exponentiated constants in the resummed expressions.

\section{Phenomenological results at the LHC}
\label{sec:pheno_results}
In this section we present predictions up to \nnnllp\!\!+NNLO\footnote{We stress
that the fixed-order nomenclature refers to the perturbative accuracy of the differential
$\pt$ spectrum. For instance, NNLO includes terms of relative order $\as^3$ with respect to the
singlet production Born cross section.} relevant for neutral Drell-Yan lepton-pair
production, and for gluon-fusion Higgs production and decay to a photon pair, at the 13 TeV LHC.
For both processes we consider inclusive and fiducial setups, the latter allowing a direct comparison
with experimental data, without relying on Monte Carlo modelling for acceptances.
We stress that the availability of theoretical results at the fiducial level is
guaranteed by the fact that our resummmation formalism is fully differential with respect to the Born
phase-space variables.

In principle, the availability of an \nnnllp resummation would allow us to obtain results for the \nnnlo
fiducial Drell-Yan and Higgs cross sections by means of a slicing technique such as $q_T$-subtraction
\cite{Catani:2007vq}. It is however well-known that, especially in presence of symmetric cuts on the
$\pt$ of the singlet's decay products, such a technique requires to push the slicing parameter down to
very small values, requiring an extreme control on the stability of the numerical calculation in
the far IRC regime.
This in turn translates into the necessity of dedicated high-statistics fixed-order predictions, to
minimise possible numerical fluctuations. We thus refrain from quoting fiducial cross sections at \nnnlo
in this article, and leave this development for future studies.

Aiming at reliable predictions across the entire $v$ phase space, we match our resummed results
with fixed-order differential spectra computed with the \nnlojet code, and used in previous works
\cite{Bizon:2018foh,Bizon:2019zgf}.
The matching is designed to reproduce the resummed prediction in the $v\to0$ region, dominated by
soft/collinear QCD radiation, while reducing to the fixed-order calculation in the hard tails $v\gg1$.

In \cite{Bizon:2018foh} we adopted a multiplicative matching at the \emph{cumulative} level.
Besides an improved numerical stability in the $v \rightarrow 0$ limit, where the cancellation between
the fixed-order result and the perturbative expansion of the resummation can be delicate,
a cumulative multiplicative scheme had the advantage, for processes with known total \nnnlo cross section,
of extracting the constant ${\cal O}(\as^3)$ terms from the fixed-order result trough matching.
Since such terms are now included directly in the resummation at \nnnllp accuracy, a multiplicative
scheme is no longer advantageous in this particular respect.
Moreover, as discussed in Sec.~\ref{sec:recoil}, we now include in our framework a transverse-recoil
prescription to improve the kinematical description of the singlet's decay products, which is implemented
at the level of the differential $v$ spectrum.

For these reasons, in this phenomenological study we adopt as our default a \emph{differential} matching belonging
to the additive family, defined as
\begin{equation}
\label{eq:add_matching}
\frac{\rd \Sigma^{{\rm N}^k{\rm LL}^{(\prime)}}_{\rm add}(v)}{\rd v}
\, = \,
\left(
\frac{\rd \Sigma^{{\rm N}^k{\rm LL}^{(\prime)}}(v)}{\rd v}
-
\frac{\rd \Sigma^{{\rm N}^k{\rm LL}^{(\prime)}}_{\rm exp}(v)}{\rd v}
\right)
Z(v)
+
\frac{\rd \Sigma^{{\rm N}^{k-1}{\rm LO}}(v)}{\rd v}
\, ,
\end{equation}
where $v$ is $p_t/M$ or $\phs$, $\rd \Sigma^{{\rm N}^{k-1}{\rm LO}}(v)/\rd v$ is the fixed-order
differential spectrum with respect to $v$ at ${\cal O}(\as^k)$, while
$\rd \Sigma^{{\rm N}^k{\rm LL}^ {(\prime)}}_{\rm exp}/\rd v$ represents the perturbative expansion of the
resummed spectrum at the same order.
The $Z(v)$ factor, that we choose as \cite{Bizon:2017rah}
\begin{equation}
\label{eq:Zmatching}
Z(v)
\, = \,
\Big[1-(v/v_0)^u \Big]^h
\,
\Theta(v_0-v)
\, ,
\end{equation}
is designed to enforce a dampening of the resummation component in the hard region of the
spectrum, while leaving the $v\to0$ limit unaffected.
We set $u=2$ (we stress that $u$ must be $>1$ not to induce linear power corrections), and
$h=3$ as our defaults; we take a central $v_0=1$ ($v_0=1/2$) for $v=p_t/M$ ($v=\phs$),
and consider a variation of $v_0$ in the range $[2/3,3/2]$ around its central value in
order to reliably estimate matching systematics.
Our reference value for the parameter $p$ appearing in the definition \eqref{eq:modified-log}
of modified logarithms is $p = 4$; we have checked that a variation of $p$ by one unit does
not induce significant differences.
We also present results obtained through a multiplicative matching at the differential level,
defined as
\begin{equation}
\label{eq:mult_matching}
\frac{\rd \Sigma^{{\rm N}^k{\rm LL}^{(\prime)}}_{\rm mult}(v)}{\rd v}
\, = \,
\left(
\frac{\rd \Sigma^{{\rm N}^k{\rm LL}^{(\prime)}}(v)/\rd v}
{\rd \Sigma^{{\rm N}^k{\rm LL}^{(\prime)}}_{\rm exp}(v)/\rd v}
\right)^{Z(v)}
\frac{\rd \Sigma^{{\rm N}^{k-1}{\rm LO}}(v)}{\rd v}
\, ,
\end{equation}
where $Z(v)$ is the same function introduced for the additive matching.
Analogously to the additive case, the matching in eq.~\eqref{eq:mult_matching}
only acts at the level of quadratic power corrections for $u=2$.

We stress that all of our resummed calculations feature a Landau singularity arising
from configurations where QCD radiation takes place at transverse-momentum scales
$k_t \sim M \re^{-1/(2\beta_0\as(M))} \sim 0.1$ GeV.
In the predictions we present in the following, we set our results to zero when
the hardest radiation's transverse momentum is below the singularity.
This prescription has a negligible impact on differential spectra for
typical values of $M$ as, due to the vectorial nature of the considered
observables~\cite{Parisi:1979se,Bizon:2017rah}, the $v\to0$ limit is dominated
by radiation at the few-GeV scale, significantly harder than the Landau scale.
We however stress that for a precise description of this kinematic
regime, a thorough study of the impact of non-perturbative corrections, not
included in the present article, would be necessary.

We finally recall that in all predictions shown in the following we adopt the
NNLO DGLAP evolution for parton densities. Although the NNLO corrections to
the evolution are formally of N$^3$LL order, we include them also in the NLL
and NNLL results to ensure an identical treatment of heavy-quark
thresholds. Parton densities are evolved from a scale $\mu_0\sim 1$ GeV upwards
by means of the {\tt Hoppet} package~\cite{Salam:2008qg}, which is used as well
to handle all parton-density and coefficient-function convolutions.

\subsection{Drell-Yan results}
\label{sec:DYresults}
For Drell-Yan phenomenology, we consider $pp$ collisions at $13$ TeV
centre-of-mass energy, and we use the NNLO {\tt NNPDF3.1} PDF
set~\cite{Ball:2017nwa} with $\as(M_Z) = 0.118$ through the LHAPDF
interface~\cite{Buckley:2014ana}.
We adopt the $G_\mu$ scheme with electro-weak parameters taken
from the PDG~\cite{Tanabashi:2018oca}, namely
\begin{eqnarray}
M_Z \, = \, 91.1876\,\GeV \, ,
\qquad
{\rm \Gamma}_Z \, = \, 2.4952\,\GeV \, ,
\qquad
G_F \, = \, 1.1663787  \times  10^{-5}\,\GeV^{-2} \, .
\end{eqnarray}
The fiducial volume is defined by applying the following set of
selection cuts on the lepton pair
\cite{Aad:2019wmn}\footnote{We stress that, due to the 27 GeV cut on
the transverse momenta of the leptons, the fixed-order predictions used
below are slightly different from those employed in \cite{Bizon:2019zgf},
which employed a 25 GeV cut.}:
\begin{eqnarray}
\label{eq:Z_fiducial}
\pt^{\ell^\pm} \, > \, 27\,\GeV \, ,
\qquad
|\eta^{\ell^\pm}| \, < \, 2.5 \, ,
\qquad
66\,\GeV \, < \, \mll \, < \, 116\,\GeV \, ,
\end{eqnarray}
where $\pt^{\ell^\pm}$ are the transverse momenta of the
leptons, $\eta^{\ell^\pm}$ are their pseudo-rapidities in the
hadronic centre-of-mass frame, and $\mll$ is the invariant mass
of the di-lepton system.
We also define an `inclusive' setup by dropping in
eq.~\eqref{eq:Z_fiducial} the cuts on $\pt^{\ell^\pm}$
and $\eta^{\ell^\pm}$.

Factorisation and renormalisation scales are chosen as $\mu_R
= \kappa_R \, M_t$, $\mu_F = \kappa_F \, M_t$, with $M_t =
\sqrt{\mll^2 + {\ptll}^2}$, and $\ptll$ the
di-lepton-system transverse momentum, while the resummation
scale is set to $Q = \kappa_Q \, \mll$.
For the resummed results, the definition of $M_t$ is actually
approximated by $\mll$, which is appropriate up to
quadratic power corrections.
We assess the impact of missing higher-order contributions by
performing a variation of $\mu_R$ and $\mu_F$ by a factor
of 2 around their respective central values whilst keeping $1/2
\leq \mu_R/\mu_F \leq 2$.  In addition, for central $\mu_R$ and
$\mu_F$ we vary the resummation scale $Q$ by a factor of 2 in either
direction. The final uncertainty for resummed results is built
as the envelope of the resulting 9-scale variation, while in the
case of matched results, as anticipated above, the envelope also
includes variations of the $v_0$ parameter in eq.~\eqref{eq:Zmatching}.

\begin{figure*}[t]
\includegraphics[width=0.49\textwidth]{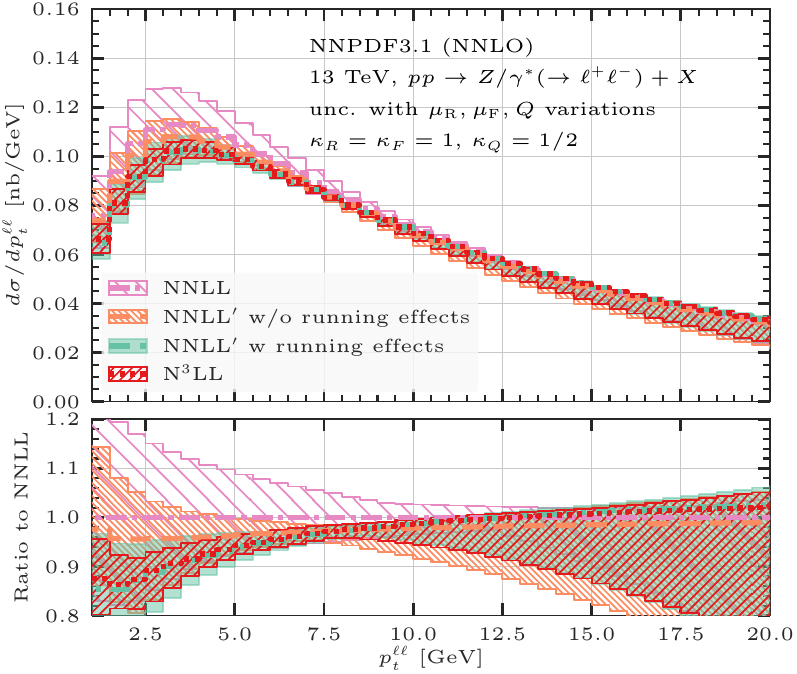}%
\includegraphics[width=0.49\textwidth]{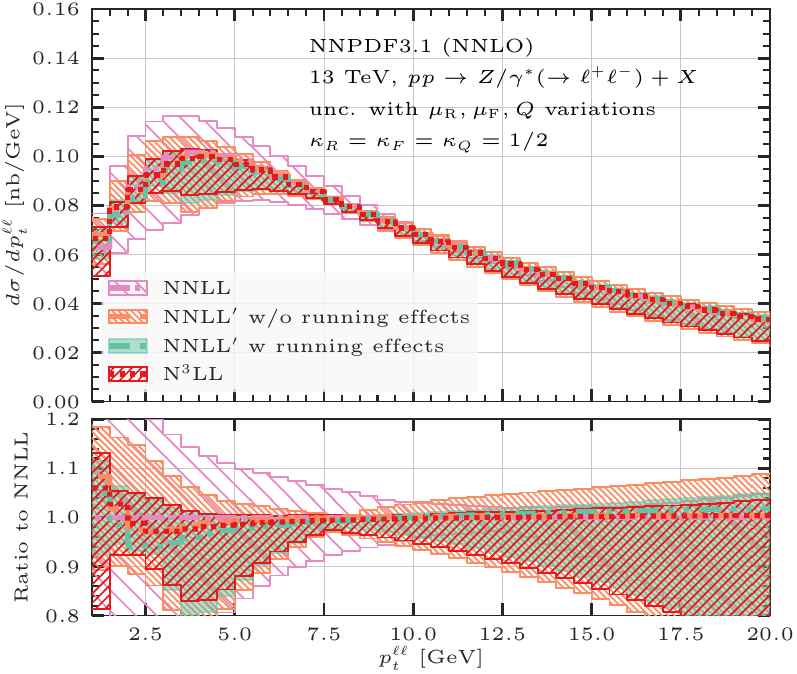}%
\caption{Resummed $\ptll$ spectrum in the inclusive setup at \nnll, \nnllp, \nnnll.
Left panel: central scales $\kappa_R = \kappa_F = 1$, $\kappa_Q = 1/2$.
Right panel: central scales $\kappa_R = \kappa_F = \kappa_Q = 1/2$.}
\label{fig:Zpt_inclusive}
\end{figure*}

In Fig.~\ref{fig:Zpt_inclusive} we show a comparison of pure resummed results
for the di-lepton transverse-momentum $\ptll$ distribution in the inclusive
setup at \nnll (pink), \nnllp without running-coupling effects (orange), \nnllp with
running-coupling effects (green), and \nnnll (red). The plot on the left panel displays
variations around central scales $\kappa_R = \kappa_F = 1$, $\kappa_Q = 1/2$,
while the right panel features central scales $\kappa_R = \kappa_F = \kappa_Q
= 1/2$.
Both plots clearly show the benefits of the inclusion of `primed' effects
on the \nnll predictions at the level of central value and theoretical-uncertainty
bands, especially in terms of shapes.
We note that \nnllp predictions, both with and without running-coupling effects,
are significantly closer to the full \nnnll result than the \nnll one is, although
the pattern of comparison somewhat depends on the chosen central-scale setup, with
the running-coupling option closer to full \nnnll on the left, and the opposite
on the right panel.
The uncertainty band of the \nnllp predictions is also significantly reduced below
10 GeV with respect to the \nnll one. The band relevant to the running-coupling
option is smaller than the the non-running one, which is generally expected
since the former encodes correct higher-order running-coupling information,
absent in the latter. We note that across the entire $\ptll$ range the former
band is also very similar to the \nnnll one, and moreover, in all cases does
it contain the central \nnnll prediction, yielding a reliable estimate of the
impact of missing higher-order terms.
The difference between the two \nnllp results may become non
negligible at very small $\ptll$ for certain scale setups (especially so when
the central $\kappa_Q$ is different from the central $\kappa_R$, $\kappa_F$ values),
which is also qualitatively expected as due to the approaching to a strong-coupling
regime; in all cases the discrepancy is covered by the uncertainty band of the
\nnllp with running-coupling option, which faithfully assesses the ambiguity
related to the inclusion of beyond-accuracy running-coupling effects.
In the following we choose the running-coupling option of `primed' results as
the default for our phenomenological study.

\begin{figure*}[t]
\includegraphics[width=0.49\textwidth]{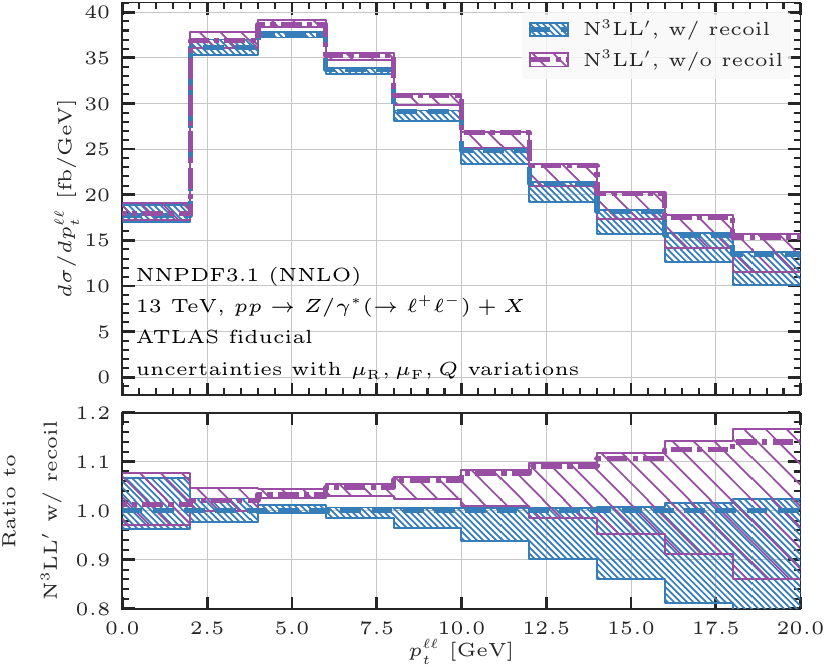}%
\includegraphics[width=0.49\textwidth]{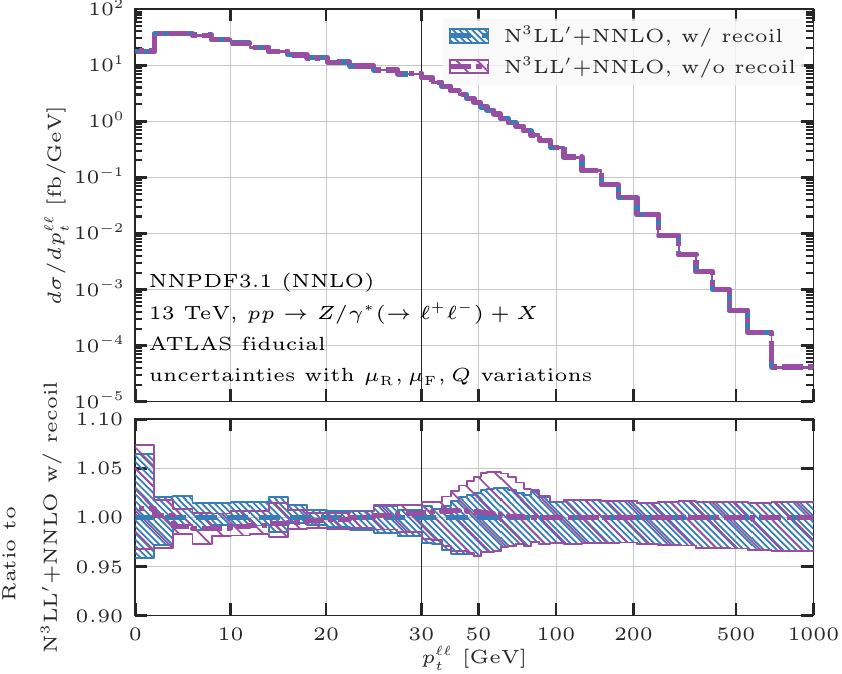}%
\caption{\nnnllp (left) and \nnnllp\!\!+NNLO (right) $\ptll$ spectra in the fiducial ATLAS
setup with (blue) and without (purple) recoil effects.
In the right plot, the $x$ axis is linear up to 30~GeV and logarithmic above.}
\label{fig:Zpt_rec_vs_norec}
\end{figure*}

In Fig.~\ref{fig:Zpt_rec_vs_norec} we assess the effect of the recoil prescription
detailed in Sec.~\ref{sec:recoil} on fiducial $\ptll$ predictions at \nnnllp accuracy
(where not explicitly stated, we employ the running-coupling option), both without
(left panel) and with (right panel) additive matching \eqref{eq:add_matching} to the
fixed NNLO differential result. The uncertainty band stems from variations around
central scales $\kappa_R = \kappa_F = 1$, $\kappa_Q=1/2$, while the
matched result includes variation of $v_0$ as well.
The inclusion of recoil (blue, as opposed to purple not featuring recoil effects)
gives rise to an expected linear power correction in the pure resummed case, as can be
specifically checked in the lower inset of the left panel. After matching to fixed order,
recoil induces a marginal $\mp1$\% distortion of the spectrum below 20 GeV, which is
the leftover effect after the ${\cal O}(\as^3)$ cancellation taking place between resummation
and its expansion in \eqref{eq:add_matching}. The uncertainty bands are also very similar
across the whole phase space.

\begin{figure*}[t]
\includegraphics[width=0.49\textwidth]{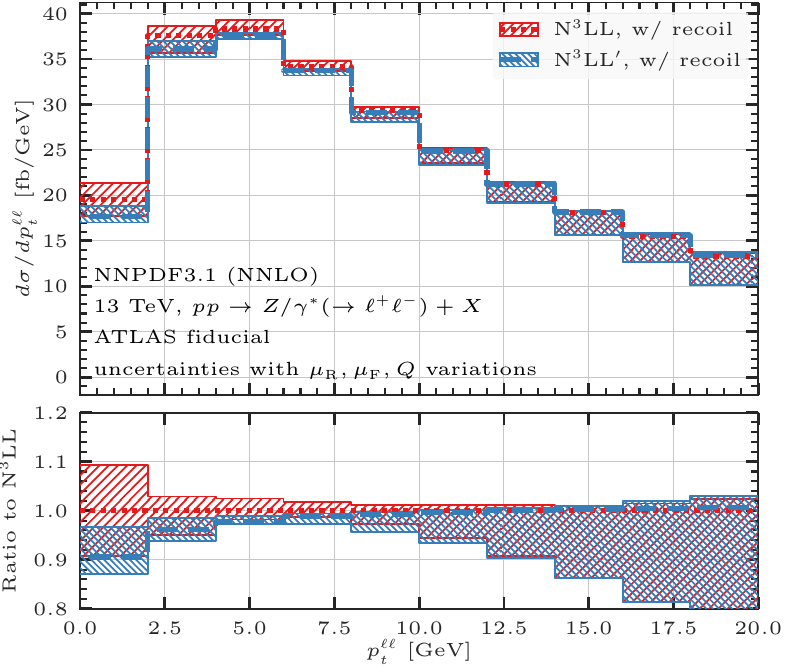}%
\includegraphics[width=0.49\textwidth]{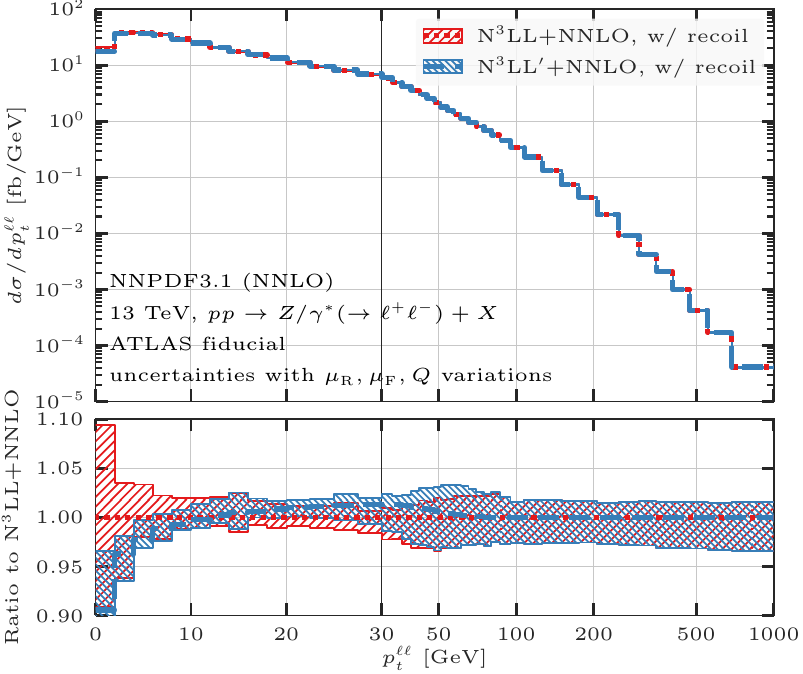}%
\caption{Left: resummed predictions at \nnnll (red) and \nnnllp (blue) for $\ptll$ in the
fiducial ATLAS setup.
Right: matched prediction at \nnnll\!+NNLO (red) and \nnnllp\!\!+NNLO (blue).
In the right plot, the $x$ axis is linear up to 30~GeV and logarithmic above.}
\label{fig:Zpt_N3LLp_vs_N3LL}
\end{figure*}

Figure \ref{fig:Zpt_N3LLp_vs_N3LL} displays a comparison, at the fiducial level
and including recoil effects, between resummed results (left panel) at \nnnll
(red) and at \nnnllp (blue) accuracy, and between matched results (right panel)
at \nnnll\!+NNLO (red) and at \nnnllp\!\!+NNLO (blue) accuracy.
All variations are relevant to central-scale values $\kappa_R=\kappa_F=1$,
$\kappa_Q=1/2$.
The inclusion of `primed' effects on the pure resummed prediction induces a
distortion in the spectrum which is less than 2\% above 5 GeV, and
that can be as large as a few percent below, which is qualitatively consistent
with (and quantitatively less pronounced than) what is shown for the
\nnllp versus \nnll comparison in the left panel of Fig.~\ref{fig:Zpt_inclusive},
featuring the same central-scale setup. The uncertainty band undergoes a
significant reduction below 10 GeV in passing from \nnnll to \nnnllp accuracy,
by up to a factor of 2 towards $\ptll\to0$.
The matched results shown on the right panel largely inherit the features just
described in the phase-space region dominated by resummation effects, whereas
for $\ptll$ above 50 GeV the prediction is dominated by the fixed-order component,
which is common to both. Overall, the \nnnllp\!\!+NNLO residual uncertainty
band is at the level of 2\,-\,3\% below 30 GeV (barring the first bin), and around
5\% above 30 GeV.

\begin{figure*}[t]
\begin{center}
\includegraphics[width=0.49\textwidth]{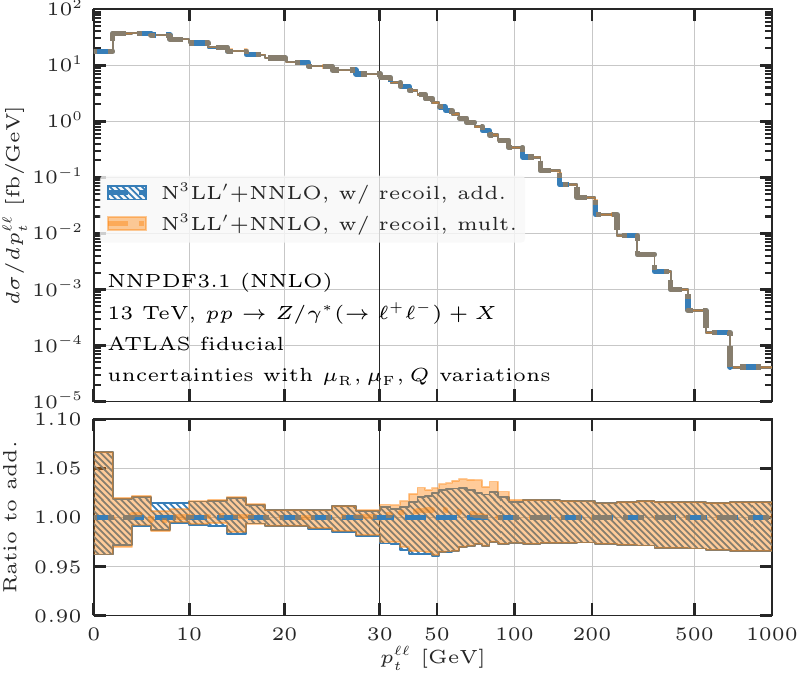}%
\end{center}
\caption{Comparison of additive (blue) and multiplicative (orange) matching prescriptions
at \nnnllp\!+NNLO, with recoil effects.
The $x$ axis is linear up to 30~GeV and logarithmic above.}
\label{fig:Zpt_match}
\end{figure*}

Fig.~\ref{fig:Zpt_match} shows a comparison of the default additive-matching
prescription defined in eq.~\eqref{eq:add_matching} (blue) with the multiplicative
matching defined in eq.~\eqref{eq:mult_matching} (orange) at the \nnnllp\!\!+NNLO
level, where both predictions include transverse-recoil effects. For reference,
the central-scale setup is $\kappa_R=\kappa_F=1$, $\kappa_Q=1/2$,
and the additive prediction is the same as in the right panel of
Fig.~\ref{fig:Zpt_N3LLp_vs_N3LL}.
The theoretical systematics related to the choice of matching family results fairly
negligible at this order, with the two predictions being essentially indistinguishable
both at central scales, and with respect to uncertainty bands.
As the envelope of the two different schemes essentially coincides with the single
uncertainty bands, we refrain from adopting it as an estimate of matching systematics,
and rather insist on the variation of parameter $v_0$ in a sensible range, such as
[2/3, 3/2] around the central $v_0$ value, as better suited to this aim.
This variation is responsible for the slight widening of the band between 30 GeV and
100 GeV, which we believe to reflect a genuine matching uncertainty in this region.

\begin{figure*}[t]
\includegraphics[width=0.49\textwidth]{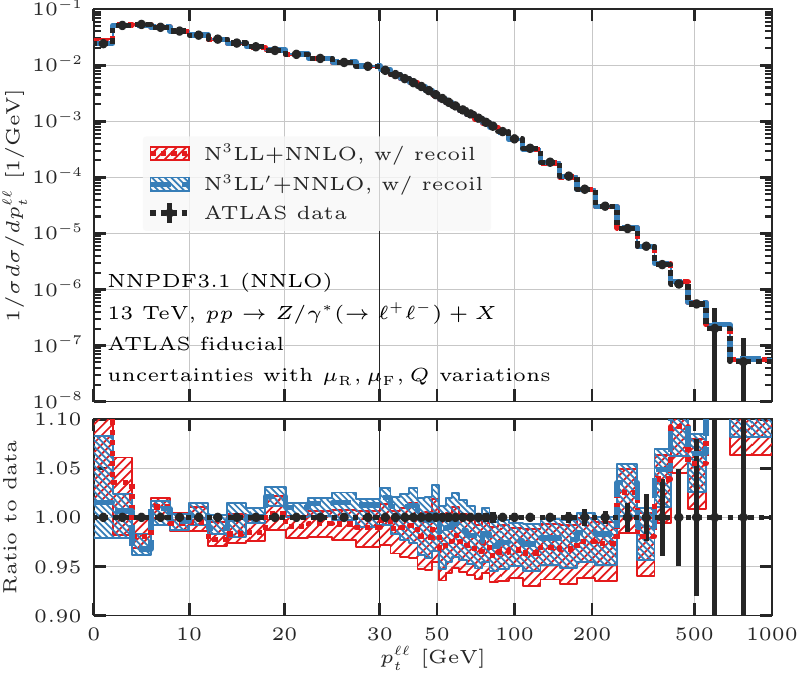}%
\includegraphics[width=0.49\textwidth]{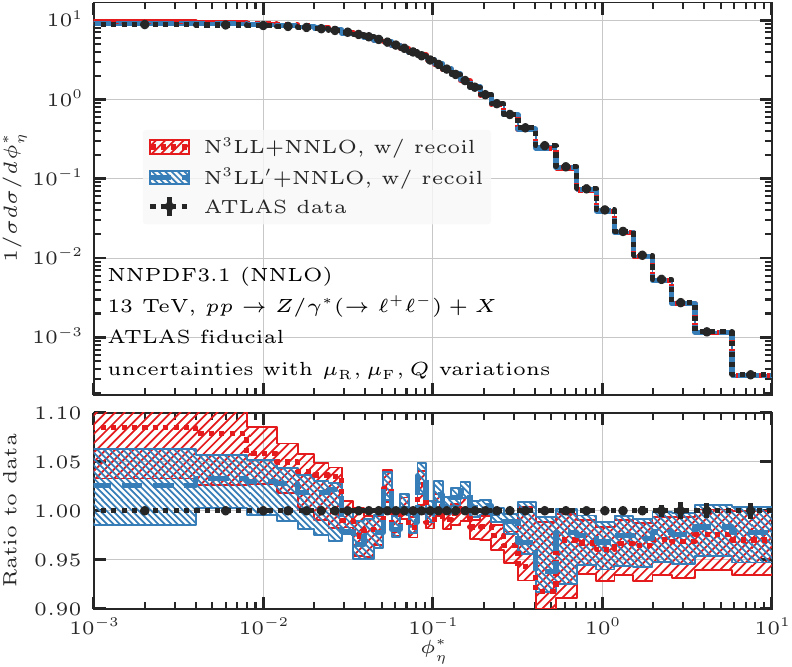}%
\caption{Comparison of matched predictions at \nnnll\!+NNLO (red) and \nnnllp+NNLO (blue)
with ATLAS data \cite{Aad:2019wmn} for $\ptll$ (left panel) and $\phs$ (right panel).
The fixed-order component is turned off below $\phs=3.4 \cdot 10^{-2}$ in the right panel,
see main text for details.
In the left plot, the $x$ axis is linear up to 30~GeV and logarithmic above.}
\label{fig:Z_theo_vs_ATLAS}
\end{figure*}

In Fig.~\ref{fig:Z_theo_vs_ATLAS} we finally compare matched predictions in the fiducial setup
to ATLAS data~\cite{Aad:2019wmn}, both for $\ptll$ (left panel) and for $\phs$ (right panel).
The left panel includes the same theoretical predictions shown in the right panel of
Fig.~\ref{fig:Zpt_N3LLp_vs_N3LL} (keeping the same colour code), which are here normalised
to their cross section in order to match the convention of the shown data.
The matched \nnnllp\!\!+NNLO predictions for $\ptll$ show a remarkable agreement with
experimental data, with a theoretical-uncertainty band down to the 2\,-\,5\% level,
essentially overlapping with data in all bins form 0 to 200 GeV (barring one
low-$\ptll$ bin, where the cancellation between the fixed-order and the expanded components
is particularly delicate, and few middle-$\ptll$ bins where the agreement is only marginal).
The inclusion of `primed' effects tends to align the shape of the theoretical prediction
to data, so that the former never departs more than 1\,-\,2\% from the latter below 200 GeV,
as opposed to the more visible relative distortion of the \nnnll\!+NNLO below 5 GeV
and above 50 GeV.
The $\phs$ results on the right panel follow by and large the same pattern just seen for
$\ptll$, with `primed' effects being relevant to improve the data-theory agreement over
the entire range, expecially at very small $\phs$, and theoretical uncertainties at or
below the $\pm3\%$ level.

We incidentally note that, due to the extremely soft and collinear regime probed
by $\phs$ data, the fixed-order component features some fluctuations at small $\phs$;
consequently, we have opted to turn it off in the first bins (up to $\phs=3.4 \cdot 10^{-2}$),
which implies that the matching formula in that region corresponds to the sole resummation
output, multiplied by $Z(v)$.
On the one hand this shows that resummation alone is capable of predicting data
remarkably well both in shape and in normalisation at very small $\phs$; on the other hand
it highlights the necessity of dedicated high-statistics fixed-order runs in order to
reliably extract information on fiducial cross sections at \nnnlo by means of slicing
techniques, especially in presence of symmetric lepton $\pt^{\ell^\pm}$ cuts.

\subsection{Higgs results}

For Higgs phenomenology we consider hadro-production at the 13 TeV LHC in
an inclusive setup, with an un-decayed Higgs boson and no cuts, as well as
in a fiducial setup, where we focus on the $H\to \gamma\gamma$ decay channel.
We employ an effective-field-theoretical (HEFT) description of the gluon-fusion
process where the the top quark running in the loops is integrated out, giving
rise to an effective $ggH$ coupling. As seen above, the hard-function coefficients
$H^{(2)}_g$ and $H^{(3)}_g$ encode the $m_{\rm top}$ dependence arising from
the Wilson coefficient of the effective vertex.
The fiducial volume is defined by the following set of cuts \cite{Aaboud:2018xdt}
\begin{eqnarray}
&&
\min(p_t^{\gamma_1}, p_t^{\gamma_2}) > 31.25~{\rm GeV} \, ,
\qquad
\max(p_t^{\gamma_1}, p_t^{\gamma_2}) > 43.75~{\rm GeV} \, ,
\nonumber\\[5pt]
&&
0 < |\eta^{\gamma_{1,2}}| < 1.37
\quad {\rm or} \quad
1.52 < |\eta^{\gamma_{1,2}}| < 2.37 \, ,
\qquad
|Y_{\gamma\gamma}| < 2.37 \, ,
\end{eqnarray}
where $p_t^{\gamma_i}$ are the transverse momenta of the two photons, $\eta^{\gamma_i}$
are their pseudo-rapidities in the hadronic centre-of-mass frame, and $Y_{\gamma\gamma}$
is the photon-pair rapidity. In the definition of the fiducial phase-space cuts
we do not include the photon-isolation requirement of \cite{Aaboud:2018xdt}, since this
would introduce additional non-global logarithmic corrections in the problem, spoiling
the formal accuracy of the resummation. However, we point out that the photon-isolation
is quite mild in this particular setup, hence it could faithfully be included at fixed
order. The photon decay is predicted in the narrow-width approximation applying a
branching ratio of $2.35 \times 10^{-3}$.

For fiducial predictions we employ parton densities from the
\texttt{PDF4LHC15\_nnlo\_mc} set \cite{Ball:2014uwa,Butterworth:2015oua,Dulat:2015mca,
Harland-Lang:2014zoa,Carrazza:2015hva,Watt:2012tq}.
Central renormalisation, factorisation, and resummation scales are set as $\mu_R = \kappa_R M_H$,
$\mu_F = \kappa_F M_H$, $Q = \kappa_Q M_H$, respectively. Theoretical-uncertainty bands
are obtained as explained in section \ref{sec:DYresults} for the Drell-Yan case.
In the inclusive setup, used solely to show the impact of running-coupling effects on
`primed' results, we employ the NNLO {\tt NNPDF3.1} PDF set~\cite{Ball:2017nwa} with
$\as(M_Z) = 0.118$.

\begin{figure*}[t]
\includegraphics[width=0.49\textwidth]{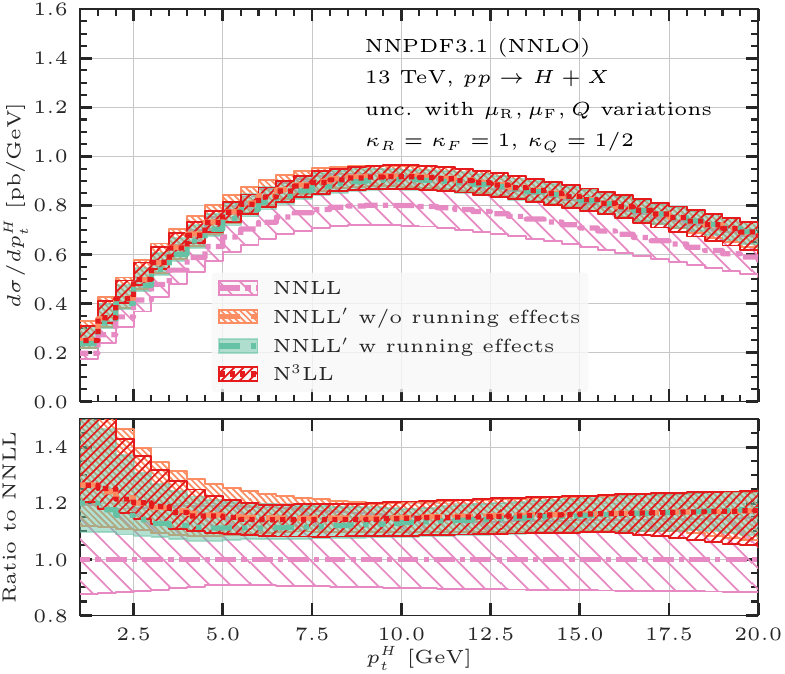}%
\includegraphics[width=0.49\textwidth]{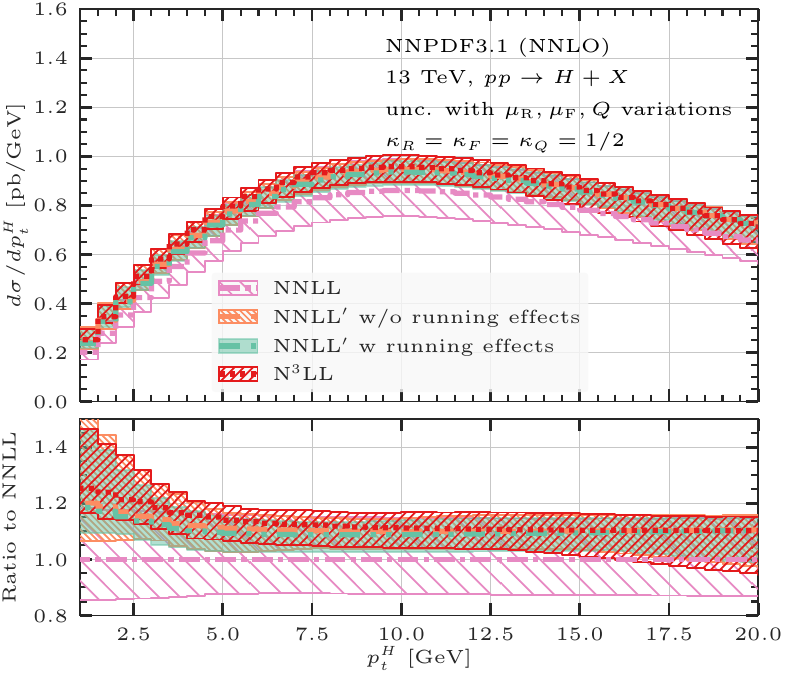}%
\caption{Resummed $\pth$ spectrum for inclusive Higgs production at \nnll, \nnllp, \nnnll.
Left panel: central scales $\kappa_R = \kappa_F = 1$, $\kappa_Q = 1/2$.
Right panel: central scales $\kappa_R = \kappa_F = \kappa_Q = 1/2$.}
\label{fig:HptInclusive}
\end{figure*}

In Fig.~\ref{fig:HptInclusive} we consider inclusive Higgs production, and show pure resummed
predictions for the Higgs transverse momentum $\pth$ at \nnll, \nnllp, and \nnnll with
central-scale choices $\kappa_R = \kappa_F = 1$, $\kappa_Q = 1/2$ (left panel), and $\kappa_R
= \kappa_F = \kappa_Q = 1/2$ (right panel).
This figure, which is the exact analogue of Fig.~\ref{fig:Zpt_inclusive} discussed above, aims
at assessing the effect of including or not running-coupling effects in `primed' results relevant
for Higgs production.
The benefit of including `primed' predictions proves significant in this case as well, but with
a different pattern with respect to Drell-Yan production.
The shape distortion in passing from \nnll to \nnllp has a slightly more limited range, mainly
extending up to 5 GeV in $\pth$; however, the normalisation of the theoretical curves is
significantly affected, with `primed' predictions correctly capturing the large $K$-factor, at
the level of 15\% at this perturbative order, which is known to arise in Higgs production.
We note the the two different \nnllp predictions are fairly similar, and remarkably closer (in
shape and normalisation) to the \nnnll one than the bare \nnll is, both in terms of central value,
and of uncertainty-band estimate. The central \nnllp prediction without running coupling tends
to be slightly closer to the central \nnnll one, while \nnllp with running coupling is slightly
more similar to \nnnll in terms of uncertainty band. In all cases does the central \nnnll prediction
lie well within the \nnllp running-coupling band, which we use as our default for the fiducial
study.

\begin{figure*}[t]
\includegraphics[width=0.49\textwidth]{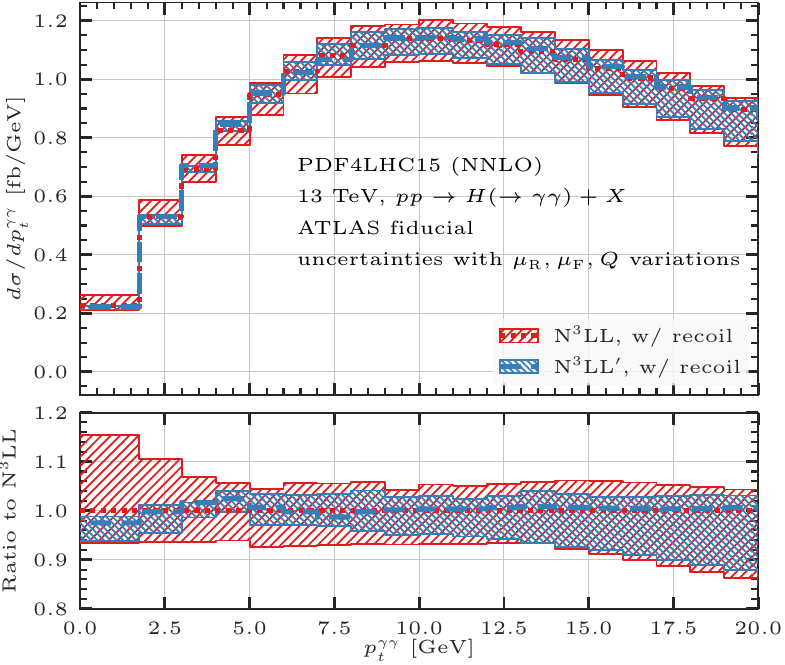}%
\includegraphics[width=0.49\textwidth]{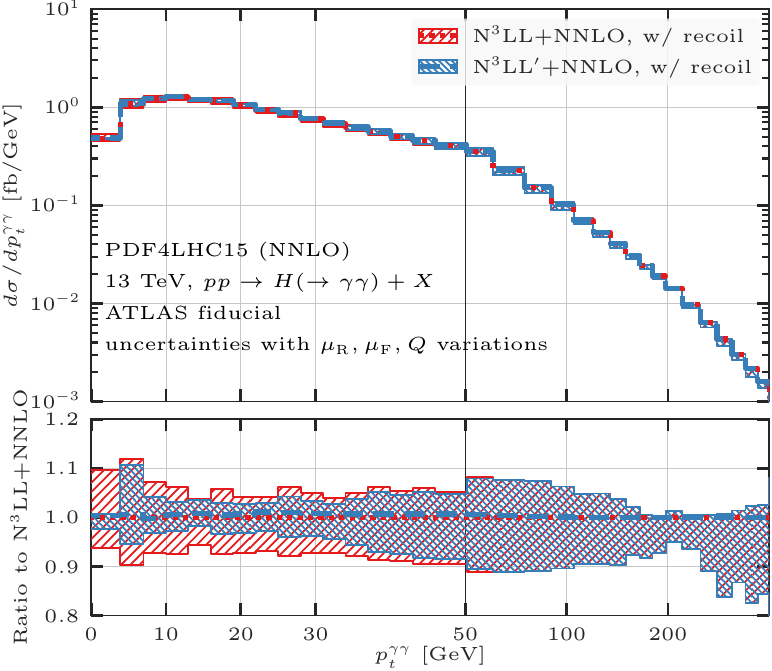}%
\caption{Left: resummed predictions at \nnnll (red) and \nnnllp (blue) for $\ptyy$ in the fiducial
ATLAS setup. Right: matched prediction at \nnnll\!+NNLO (red) and \nnnllp\!\!+NNLO (blue).
In the right plot, the $x$ axis is linear up to $p_t^{\gamma \gamma} = 50$~GeV and logarithmic above.}
\label{fig:Hpt}
\end{figure*}

Fig.~\ref{fig:Hpt} displays a comparison, relevant to the fiducial di-photon $\ptyy$ spectrum,
of \nnnllp curves (blue) agains \nnnll predictions (red), both without (left panel)
and with (right panel) additive matching to NNLO. All predictions include recoil effects,
so that this figure represents the Higgs-production analogue of Fig.~\ref{fig:Zpt_N3LLp_vs_N3LL},
but referred to central scales $\kappa_R = \kappa_F = \kappa_Q = 1/2$.
The shape distortion with respect to \nnnll predictions is more modest in the Higgs case with
respect to Drell-Yan production, partly owing to the chosen central-scale setup; moreover, the induced
$K$-factor is fairly close to unity at this order, which is sign of a good perturbative convergence.
Overall, \nnnllp predictions feature a significant reduction in theoretical
uncertainty in comparison to \nnnll ones, especially in the low-$\ptyy$ region
dominated by resummation. Residual uncertainty is as low as 5\,-\,7\% below 10 GeV,
and in the matched case it never exceeds 10\% below 40 GeV.

\begin{figure*}[t]
\begin{center}
\includegraphics[width=0.69\textwidth]{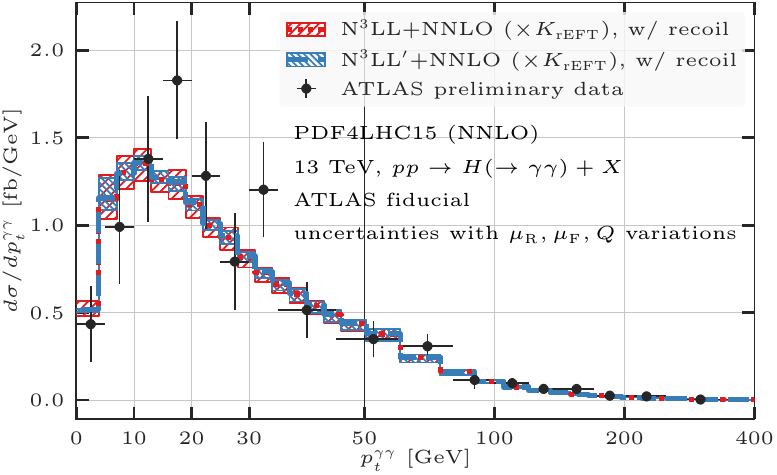}%
\end{center}
\caption{ATLAS data \cite{ATLAS-CONF-2019-029} against matched predictions at \nnnll\!+NNLO (red) and
\nnnllp\!\!+NNLO (blue) for the fiducial $\ptyy$ spectrum. Theoretical predictions are rescaled by
$K_{\rm rEFT} = 1.06584$. The $x$ axis is linear up to $p_t^{\gamma \gamma} = 50$~GeV and logarithmic above.}
\label{fig:HptATLAS}
\end{figure*}

Finally, in Figure \ref{fig:HptATLAS} we show a comparison of theoretical predictions for the fiducial
$\ptyy$ spectrum at \nnnll\!+NNLO (red) and \nnnllp\!\!+NNLO (blue) level, with recoil effects,
against ATLAS preliminary data \cite{ATLAS-CONF-2019-029}.
Theoretical predictions, based on central scales $\kappa_R = \kappa_F = \kappa_Q = 1/2$, have been
rescaled by a factor $K_{\rm rEFT} = 1.06584$ to account for the exact top-quark mass dependence
at LO.

\section{Conclusion}
\label{sec:conclusions}
In this article we have presented state-of-the-art differential predictions relevant
for colour-singlet hadro-production at the LHC within the \radish framework, up to
\nnnllp\!\!+NNLO order.
Such a level of accuracy in the resummed component is reached by supplementing the previously
available \nnnll result with the complete set of constant terms of relative order ${\cal O}(\as^3)$
with respect to the Born level. We have documented in detail how the inclusion of such
terms is achieved in \radish, as well as the validation we have performed to confirm the
correctness of their numerical implementation.
In this article we have focused on neutral Drell-Yan and Higgs production, although we stress
that the formalism used here can be straightforwardly applied to the charged Drell-Yan case
as well.

We have assessed the behaviour of `primed' predictions in inclusive Drell-Yan and Higgs
production in a comparison of two different \nnllp prescriptions (including or not
higher-order running-coupling effects, respectively) with \nnnll. This has given us confidence
on the mutual consistency of the two `primed' results, and on the reliability of their quoted
uncertainty bands, in view of comparing results based on \nnnllp predictions with experimental data.
In particular, in all considered cases are the \nnllp uncertainty bands capable of
encompassing the \nnnll central prediction, and to correctly estimate the higher-order
running-coupling ambiguity underlying the definition of `primed' accuracy.

The results presented here are fully exclusive with respect to the Born phase-space variables,
lending themselves to be flexibly adapted to the fiducial volumes considered in realistic
experimental analyses. In order to more accurately simulate the kinematics of the colour-singlet
decay products (we considered a lepton pair in the case of Drell-Yan production, and a photon pair in
the case of Higgs production), we have consistently encoded in our prediction a prescription
for the treatment of the singlet's transverse recoil against soft and collinear QCD initial-state
radiation. This includes in our results the full set of linear next-to-leading-power corrections
for azimuthally symmetric observables, such as the transverse momentum of the singlet.

The inclusion of transverse-recoil effects, which is performed at the level of differential (as
opposed to cumulative) cross sections, and the availability of ${\cal O}(\as^3)$ constant terms
in the resummed component, has led to the definition of two differential matching prescriptions,
belonging to the additive and multiplicative families, respectively.
We have compared the two schemes in Drell-Yan production, and found very good
agreement between them, showing that matching systematics are well under control. Variation of
matching parameters has anyway been conservatively included in the estimate of
the theoretical uncertainties.

Although the ingredients presented above would immediately allow us to quote numbers for the \nnnlo
fiducial Drell-Yan and Higgs cross sections, we refrain from doing so in the present article,
as in our opinion such a high-precision prediction requires dedicated high-statistics fixed-order
runs in order to avoid potential numerical biases.
This is especially the case in the context of a slicing technique in presence of symmetric cuts
on the transverse momentum of the singlet's decay products. We leave this development for future
work.

As a general upshot of the present work, which we have documented both in Drell-Yan and in Higgs
production, the inclusion of `primed' effects is highly beneficial for the stability of
the theoretical prediction, leading to a significant reduction in the residual theoretical
uncertainty. In the case of our highest-accuracy result, \nnnllp\!\!+NNLO, such a reduction
can be as large as a factor of 2 in the region dominated by resummation.

For the di-lepton $\ptll$ spectrum in Drell-Yan production, the \nnnllp\!\!+NNLO prediction is
shown to improve the comparison to ATLAS data with respect to \nnnll\!+NNLO.
The theory-data agreement is now at a remarkable level of 1\,-\,2\% below 200 GeV, and the
residual theory uncertainty is at or below the 2\,-\,5\% level in that phase-space region.
Same considerations hold for the $\phs$ observable, with the \nnnllp\!\!+NNLO band
nicely overlapping with data over the entire range, with leftover uncertainty below
$\pm3\%$.
For Higgs production we find a similar qualitative pattern, with \nnnllp predictions
featuring an uncertainty at the level of 5\% at very low di-photon $\ptyy$, and matched
\nnnllp\!\!+NNLO results well below $\pm10\%$ accuracy over the entire $\ptyy$ range.
\\

The \radish code used for the predictions shown in this paper will be made public in due
time, and the results are available from the authors upon request.

\section*{Acknowledgements}
\label{sec:ackno}
We thank Pier Francesco Monni for long-standing collaboration on the matters treated
in this article, for many inspiring discussions on related topics, and for a very
careful reading of the manuscript.
We thank Alex Huss for useful discussions and comments on the manuscript, and the
\nnlojet collaboration for providing the fixed-order results employed in this paper.
We are grateful to Claude Duhr and Bernhard Mistlberger for kindly providing us
with \nnnlo cross sections for the $pp\to\gamma^*$ process, which we used for internal
cross checks of our implementation. ER and LR acknowledge discussions with Carlo Oleari
on the axial structure of the Drell-Yan form factor.  The work of LR is supported by the
Swiss National Science Foundation (SNF) under contract 200020\_188464.

\clearpage

\appendix
\section{Parton luminosities up to \nnnllp}
\label{app:luminosities}
We report the explicit expression for the parton luminosities employed in the main text,
up to \nnnllp accuracy.
By defining the coupling factors
\begin{eqnarray}
\bar\alpha_{s(0)}
& = &
\as(\mu_R)
\, ,
\qquad
\bar\alpha_{s(1)}
\, = \,
\frac{\as(\mu_R)}{\xi}
\, ,
\qquad
\bar\alpha_{s(2)}
\, = \,
\frac{\as(\mu_R)}{\xi}
\bigg[
1 - \as(\mu_R) \, \frac{\beta_1}{\beta_0} \,
\frac{\ln\xi}{\xi}
\bigg]
\, ,
\nonumber\\
\bar\alpha_{s(3)}
& = &
\frac{\as(\mu_R)}{\xi}
\bigg[
1 - \as(\mu_R) \, \frac{\beta_1}{\beta_0} \,
\frac{\ln\xi}{\xi}
+ \,
\as^2(\mu_R) \, \frac{\beta_1^2}{\beta_0^2} \,
\frac{\ln^2\xi-\ln\xi-(1-\xi)(1-\beta_0\beta_2/\beta_1^2)}{\xi^2}
\bigg]
\, ,
\end{eqnarray}
with $\xi = 1-2\as(\mu_R)\beta_0 \ln(\mu_R/k_{t1})$,
which correspond to $\as(k_{t1})$ written in terms of $\as(\mu_R)$
at 0, 1, 2, 3 loops, and
\begin{eqnarray}
\beta_0
& = &
\frac{11 C_A - 2 n_f}{12\pi}
\, ,
\qquad
\beta_1
\, = \,
\frac{17 C_A^2 - 5 C_A n_f - 3 C_F n_f}{24\pi^2}
\, ,
\nonumber\\
\beta_2
& = &
\frac{2857 C_A^3+ (54 C_F^2 -615C_F C_A -1415 C_A^2)n_f
+(66 C_F +79 C_A) n_f^2}{3456\pi^3}
\, ,
\end{eqnarray}
the standard luminosities can be compactly written as
\begin{eqnarray}
\label{eq:luminosity-NLL0}
{\cal L}_{\rm NLL}(k_{t1})
& = &
\sum_{c, c'}
\frac{\rd|{\cal M}_{B}|_{cc'}^2}{\rd\Phi_B} \,
f_c(k_{t1},x_1) \,
f_{c'}(k_{t1},x_2)
\, ,
\end{eqnarray}
\begin{eqnarray}
\label{eq:luminosity-NNLL0}
{\cal L}_{\rm NNLL}(k_{t1})
& = &
\sum_{c, c'}
\frac{\rd|{\cal M}_{B}|_{cc'}^2}{\rd\Phi_B} \,
\sum_{i,j}
\int_{x_1}^{1}\frac{\rd z_1}{z_1}
\int_{x_2}^{1}\frac{\rd z_2}{z_2} \, 
f_i\Big(k_{t1},\frac{x_1}{z_1}\Big) \,
f_j\Big(k_{t1},\frac{x_2}{z_2}\Big)
\nonumber\\
&&
\times \,
\Bigg\{
\delta_{ci} \, \delta_{c'j} \, \delta(1-z_1) \, \delta(1-z_2)
\sum_{k=0}^1 \Big(\frac{\bar\alpha_{s(0)}}{2\pi}\Big)^k H^{(k)}(\mu_R)
\nonumber\\
&&
\quad
+ \,
\frac{\bar\alpha_{s(1)}}{2\pi}
\Big(
C_{c i}^{(1)}(z_1)\delta(1-z_2)\delta_{c'j}
+ \, \{z_1,c,i \, \leftrightarrow \, z_2,c',j\}
\Big)
\Bigg\}
\, ,
\end{eqnarray}
\begin{eqnarray}
\label{eq:luminosity-N3LL0}
{\cal L}_{\rm N^3LL}(k_{t1})
& = &
\sum_{c,c'}
\frac{\rd|{\cal M}_{B}|_{cc'}^2}{\rd\Phi_B} \, 
\sum_{i,j}
\int_{x_1}^{1}\frac{\rd z_1}{z_1}
\int_{x_2}^{1}\frac{\rd z_2}{z_2} \,
f_i\Big(k_{t1},\frac{x_1}{z_1}\Big) \,
f_j\Big(k_{t1},\frac{x_2}{z_2}\Big)
\nonumber\\
&&
\times \,
\Bigg\{
\delta_{ci} \, \delta_{c'j} \, \delta(1-z_1) \, \delta(1-z_2)
\sum_{k=0}^2 \Big(\frac{\bar\alpha_{s(0)}}{2\pi}\Big)^k H^{(k)}(\mu_R)
\nonumber\\
&&
\quad
+ \,
\frac{\bar\alpha_{s(2)}}{2\pi}
\Big(
C_{c i}^{(1)}(z_1) \, \delta(1-z_2) \, \delta_{c'j}
+ \, \{z_1,c,i \, \leftrightarrow \, z_2,c',j\}
\Big)
\nonumber\\
&&
\quad
+ \,
\Big(\frac{\bar\alpha_{s(1)}}{2\pi}\Big)^2
\Big(
C_{c i}^{(2)}(z_1) \, \delta(1-z_2) \, \delta_{c'j}
+ \{z_1,c,i \, \leftrightarrow \, z_2,c',j\}
\nonumber\\
&&
\hspace{20mm}
+ \,
C_{c i}^{(1)}(z_1) \, C_{c' j}^{(1)}(z_2) +
G_{c i}^{(1)}(z_1) \, G_{c' j}^{(1)}(z_2)
\Big)
\nonumber\\
&&
\quad
+ \,
\frac{\bar\alpha_{s(0)}\bar\alpha_{s(1)}}{(2\pi)^2}
H^{(1)}(\mu_R)
\Big(
C_{c i}^{(1)}(z_1) \, \delta(1-z_2) \, \delta_{c'j}
+ \{z_1,c,i \, \leftrightarrow \, z_2,c',j\}
\Big)
\Bigg\}
\, ,
\end{eqnarray}
\begin{eqnarray}
\label{eq:luminosity-N4LL0}
{\cal L}_{\rm N^4LL}(k_{t1})
& = &
\sum_{c,c'}
\frac{\rd|{\cal M}_{B}|_{cc'}^2}{\rd\Phi_B} \, 
\sum_{i,j}
\int_{x_1}^{1}\frac{\rd z_1}{z_1}
\int_{x_2}^{1}\frac{\rd z_2}{z_2} \,
f_i\Big(k_{t1},\frac{x_1}{z_1}\Big) \,
f_j\Big(k_{t1},\frac{x_2}{z_2}\Big)
\nonumber\\
&&
\times \,
\Bigg\{
\delta_{ci} \, \delta_{c'j} \, \delta(1-z_1) \, \delta(1-z_2)
\sum_{k=0}^3 \Big(\frac{\bar\alpha_{s(0)}}{2\pi}\Big)^k H^{(k)}(\mu_R)
\nonumber\\
&&
\quad
+ \,
\frac{\bar\alpha_{s(3)}}{2\pi}
\Big(
C_{c i}^{(1)}(z_1) \, \delta(1-z_2) \, \delta_{c'j}
+ \, \{z_1,c,i \, \leftrightarrow \, z_2,c',j\}
\Big)
\nonumber\\
&&
\quad
+ \,
\Big(\frac{\bar\alpha_{s(2)}}{2\pi}\Big)^2
\Big(
C_{c i}^{(2)}(z_1) \, \delta(1-z_2) \, \delta_{c'j}
+ \{z_1,c,i \, \leftrightarrow \, z_2,c',j\}
\nonumber\\
&&
\hspace{20mm}
+ \,
C_{c i}^{(1)}(z_1) \, C_{c' j}^{(1)}(z_2) +
G_{c i}^{(1)}(z_1) \, G_{c' j}^{(1)}(z_2)
\Big)
\nonumber\\
&&
\quad
+ \,
\frac{\bar\alpha_{s(0)}\bar\alpha_{s(2)}}{(2\pi)^2}
H^{(1)}(\mu_R)
\Big(
C_{c i}^{(1)}(z_1) \, \delta(1-z_2) \, \delta_{c'j}
+ \{z_1,c,i \, \leftrightarrow \, z_2,c',j\}
\Big)
\nonumber\\
&&
\quad
+ \,
\Big(\frac{\bar\alpha_{s(1)}}{2\pi}\Big)^3
\Big(
C_{c i}^{(3)}(z_1) \, \delta(1-z_2) \, \delta_{c'j}
+
C_{c i}^{(2)}(z_1) \, C_{c'j}^{(1)}(z_2)
\nonumber\\
&&
\hspace{20mm}
+ \,
G_{c i}^{(2)}(z_1) \, G_{c'j}^{(1)}(z_2)
+ \{z_1,c,i \, \leftrightarrow \, z_2,c',j\}
\Big)
\nonumber\\
&&
\quad
+ \,
\frac{\bar\alpha_{s(0)}\bar\alpha_{s(1)}^2}{(2\pi)^3}
H^{(1)}(\mu_R)
\Big(
C_{c i}^{(2)}(z_1) \, \delta(1-z_2) \, \delta_{c'j}
+ \{z_1,c,i \, \leftrightarrow \, z_2,c',j\}
\nonumber\\
&&
\hspace{20mm}
+ \,
C_{c i}^{(1)}(z_1) \, C_{c' j}^{(1)}(z_2) +
G_{c i}^{(1)}(z_1) \, G_{c' j}^{(1)}(z_2)
\Big)
\nonumber\\
&&
\quad
+ \,
\frac{\bar\alpha^2_{s(0)}\bar\alpha_{s(1)}}{(2\pi)^3}
H^{(2)}(\mu_R)
\Big(
C_{c i}^{(1)}(z_1) \, \delta(1-z_2) \, \delta_{c'j}
+ \{z_1,c,i \, \leftrightarrow \, z_2,c',j\}
\Big)
\Bigg\}
\, ,
\end{eqnarray}
where $H^{(0)}(\mu_R)=1$, $x_{1,2} = \re^{\pm Y} M/\sqrt s$, and $Y$ is the
Born-level rapidity of the colour singlet in the centre-of-mass frame of the
collision, with energy $\sqrt s$.

As for the luminosities relevant for `primed' predictions, they assume a different
functional form for the running or non-running options, described in the main
text. In the running case, we just set ${\cal L}_{{\rm N}^k{\rm LL}'} =
{\cal L}_{{\rm N}^{k+1}{\rm LL}}$; in the non-running case,
${\cal L}_{{\rm N}^k{\rm LL}'} = \hat{\cal L}_{{\rm N}^{k+1}{\rm LL}}$,
where the $\hat{\cal L}_{{\rm N}^{k+1}{\rm LL}}$ corresponds to a luminosity
${\cal L}_{{\rm N}^{k+1}{\rm LL}}$ with the replacement $\bar\alpha_{s(j)}
\to \bar\alpha_{s(j-1)}$ for $j>0$.

\section{V-A structure of the form factor in the neutral Drell-Yan process}
\label{app:axial}
In this subsection we discuss the subtleties arising in the extraction
of the hard-virtual corrections to the form factor for the neutral-current
Drell-Yan process
\begin{equation}
\label{eq:dy_main}
q(p_1)\ \bar{q}(p_2)
\, \to \,
Z/\gamma^*
\, \to\,
\ell^-(p_3)\ \ell^+(p_4)
\, ,
\end{equation}
that we will denote with the shortcut $Z/\gamma^*$, although in the
following the dependence upon the final-state leptonic momenta is
explicitly taken into account.

Before discussing how the hard-virtual coefficients in
eq.~\eqref{eq:coeff-fun} are extracted from the loop corrections to
the form factor, it is useful to recall the structure of the Drell-Yan
tree-level amplitude expressed in terms of spinor currents.
The fermion-antifermion-photon, and fermion-antifermion-$Z$ vertices are
defined as
\begin{equation}
- i |e| Q_f \gamma^\mu
\, ,
\hspace{1cm}
- i |e| (V_f + A_f \gamma_5) \gamma^\mu
\, ,
\end{equation}
respectively, where $Q_f$ is the charge of the fermion in units of the positron
charge $|e|$, and
\begin{eqnarray}
V_f
\, = \,
\frac{T^3_f - 2 Q_f \sin^2\theta_W}{2\sin\theta_W \cos\theta_W}
\, ,
\qquad
A_f \, = \,
\frac{T^3_f}{2\sin\theta_W\cos\theta_W}
\, ,
\end{eqnarray}
$\theta_W$ being the weak mixing angle and $T^3_f = \pm 1/2$
the weak isospin quantum number of the fermion type $f$.
The fermionic currents relevant for the process in eq.~\eqref{eq:dy_main} are
\begin{equation}
J_{V;A}^{\mu(0)}(f)
\, = \,
\bar{F}\ [ 1 ;\gamma_5 ] \gamma^\mu F
\, ,
\end{equation}
where, for the initial-state quark current ($f=q$), the Dirac spinors read
$\bar{F}=\bar{v}(p_2)$ and ${F}=u(p_1)$, whereas, for the leptonic current ($f=\ell$),
$\bar{F}=\bar{u}(p_3)$ and ${F}=v(p_4)$.

The tree-level amplitude can be written as
\begin{eqnarray}
\label{eq:alo}
\Amp{0}
& = &
\frac{1}{q^2} (Q_q J^{(0)}_V(q)) \cdot (Q_\ell J^{(0)}_V(\ell))
\nonumber\\
&&
+ \, \frac{1}{(q^2-\mz^2) + i \mz\gammaz}
\Big( V_q J^{(0)}_{V}(q) +  A_q J^{(0)}_{A}(q) \Big) \cdot
\Big( V_\ell J^{(0)}_{V}(\ell) +  A_\ell J^{(0)}_{A}(\ell) \Big)
\, ,
\end{eqnarray}
where $q=p_1+p_2$, and the symbol `$\cdot$' represents the Lorentz
dot product (among currents, in this case). The superscript on the
currents indicates the loop order at which they are computed.
By denoting the products of currents of fermion line $f$ with
\begin{equation}
[J\hspace{-0.15cm}J^{(ij)}_{XY}(f)]^{\mubar\mu}
\, = \,
[J^{\mubar (i)}_{X}(f)]^{\dag} \,
[J^{\mu (j)}_{Y}(f)]
\, ,
\qquad
XY \, \in \, \lg VV, \, AA, \, AV, \, VA \rg
\, ,
\end{equation}
the tree-level squared amplitude reads (up to global factors not
relevant for the present discussion)
\begin{equation}
\label{eq:ampsqlo}
|\Amp{0}|^2
\, \propto \,
|A_\gamma|^2 + |A_Z|^2 + 2 \Re (A_\gamma^* \, A_Z)
\, ,
\end{equation}
where\footnote{Upon summing over fermion polarisations, one has
  $\JJq{00}{AA}^{\mubar\mu} = \JJq{00}{VV}^{\mubar\mu}$ and $\JJq{00}{AV}^{\mubar\mu} = \JJq{00}{VA}^{\mubar\mu}$,
  and the same holds for leptonic currents. Crucially,
  $\JJq{00}{VV}^{\mubar\mu}\JJl{00}{AV}^{\mubar\mu} = \JJl{00}{VV}^{\mubar\mu}\JJq{00}{AV}^{\mubar\mu} = 0$.}
\begin{eqnarray}
  &&|A_\gamma|^2 \, = \, \frac{1}{q^4} \ Q_q^2 Q_\ell^2 \ \JJq{00}{VV}^{\mubar\mu} \JJl{00}{VV}_{\mubar\mu}\,, \nonumber\\
  &&|A_Z|^2 \, = \, \frac{1}{(q^2-\mz^2)^2 + (\mz\gammaz)^2} \nonumber\\
  && \hspace{2cm} \lq V_q^2 \JJq{00}{VV} + A_q^2 \JJq{00}{AA} + A_q V_q \Big( \JJq{00}{VA} + \JJq{00}{AV}\Big) \rq^{\mubar\mu} \nonumber\\
  && \hspace{2cm} \lq V_\ell^2 \JJl{00}{VV} + A_\ell^2 \JJl{00}{AA}  + A_\ell V_\ell \Big(
  \JJl{00}{VA}  + \JJl{00}{AV} \Big) \rq_{\mubar\mu} \nonumber\\
  &&\phantom{|A_Z|^2 } \, = \, \frac{1}{(q^2-\mz^2)^2 + (\mz\gammaz)^2} \Big{\{}
  \( V_q^2 + A_q^2 \) \( V_\ell^2 + A_\ell^2 \) \JJq{00}{VV}^{\mubar\mu} \JJl{00}{VV}_{\mubar\mu}
  \nonumber\\ 
  &&\hspace{5.3cm} + \, 4 A_q V_q A_\ell V_\ell \, \JJq{00}{VA}^{\mubar\mu} \JJl{00}{VA}_{\mubar\mu}\Big{\}} \,, \nonumber\\
  &&2 \Re (A_\gamma^* \, A_Z) \, = \, \frac{2 (q^2 -\mz^2)}{q^2 [(q^2-\mz^2)^2 + (\mz\gammaz)^2]} \
  \Big[ Q_q V_q \JJq{00}{VV} + Q_q  A_q \JJq{00}{VA} \Big]^{\mubar\mu} \nonumber\\
  && \hspace{6.3cm} \Big[ Q_\ell V_\ell \JJl{00}{VV} + Q_\ell A_\ell \JJl{00}{VA} \Big]_{\mubar\mu} \nonumber\\
  && \phantom{2 \Re (A_\gamma^* A_Z)} = \frac{2 (q^2 -\mz^2)}{q^2 [(q^2-\mz^2)^2 + (\mz\gammaz)^2]} \Big{\{}
  Q_q V_q Q_\ell V_\ell \JJq{00}{VV}^{\mubar\mu} \JJl{00}{VV}_{\mubar\mu} \nonumber\\
  &&\hspace{6.3cm} + \, Q_q A_q Q_\ell A_\ell \JJq{00}{VA}^{\mubar\mu} \JJl{00}{VA}_{\mubar\mu} \Big{\}} \, .
\end{eqnarray}

The $H^{(n)}(\mu_R)$ coefficients in eq.~\eqref{eq:coeff-fun} are of
pure hard-virtual origin, and they are obtained from the finite parts
of the \MSbar{}-renormalised loop corrections (`r,f' in the
equations below) to the tree-level amplitude.
For $Z/\gamma^*$ production, if one solely considers loop diagrams where
the external quark line is directly connected to the $Z/\gamma^*$ vertex,
the full tree-level squared amplitude~\eqref{eq:ampsqlo} can be factored
in front of the hard function, namely up to three loops one has
\begin{equation}
\label{eq:Hexp}
\Big{|} \sum_{n=0}^3 \left(\frac{\as}{2\pi}\right)^n \, \Amp{n}\Big{|}^2_{\text{r,f}} =
|\Amp{0}|^2 \, \sum_{n=0}^3 \left(\frac{\as}{2\pi}\right)^n \, H^{(n)} + {\cal O}(\as^4)\,,
\end{equation}
where $\Amp{n}$ has the same structure as the one given
in~\eqref{eq:alo}, but it includes the ${\cal O}(\as^n)$
vertex corrections to the fermionic currents $J_{V;A}^\mu(q)$,
\begin{equation}
J_{V;A}^\mu(q)
\, = \,
\sum_{n=0}^3
\( \frac{\as}{2\pi}\)^n J^{(n)}_{V;A}(q)
\, .
\end{equation}

To convince oneself that this is indeed the case, one can consider the
first-order coefficient $H^{(1)}$ of the hard-virtual function, that
is given by $\{2 \Re({\Amp{0}}^{*} \Amp{1} )\}_{\text{r,f}}$.  Within
this contribution, the hadronic tensor features Dirac traces involving
the product of two quark vector currents $\JJq{01}{VV}$, of two quark
axial-vector currents $\JJq{01}{AA}$, and mixed terms, such as
$\JJq{01}{VA}$.
The radiative corrections to these three current correlators are the
same, and they can be extracted from the quark form factor, despite
the latter quantity is defined as the coupling of a virtual photon to
a massless quark-antiquark pair, and hence, by definition, it contains
the radiative corrections to $J_V^\mu(q)$, i.e. to the vector form factor.
The physical reason why the radiative corrections to the massless vector
and axial-vector form factor are the same ultimately stems from chirality
conservation in QCD.
Algorithmically, this a consequence of the properties of the $\gamma_5$
matrix: the Dirac trace in $\JJq{01}{AA}$ gives the same result as the
one in $\JJq{01}{VV}$, therefore the $A_q^2$ term receives exactly the
same radiative corrections of, e.g., the term
$Q_q^2$~\cite{Hamberg:1990np}. The computation of the radiative
corrections to the terms proportional to $V_q A_q$ and $Q_q A_q$ is
more delicate, as they involve a fermionic trace with just one
$\gamma_5$ matrix, hence they require a consistent treatment of
the axial-vector current in $d$ dimensions, which requires care from two
loops~\cite{Melnikov:2006kv,Larin:1993tq}.  Eventually, as happens at
tree level, the only non-vanishing contraction of $\JJq{01}{VA}$ is with
$\JJl{00}{VA}$, yielding terms proportional to $V_q A_q V_\ell A_\ell$
(and $Q_q A_q Q_\ell A_\ell$) and with exactly the same radiative
correction as the term $\JJq{01}{VV}$ $\JJl{00}{VV}$.
The same property holds also at two and three loops.

In the light of the above discussion, for $Z/\gamma^*$ production, the
tree-level squared amplitude~\eqref{eq:ampsqlo} can be factored out,
as in eq.~\eqref{eq:Hexp}, and the $H^{(n)}$ coefficient can be
obtained from the quark form factor at $n$ loops.

Starting from NNLO, the vertex corrections to $Z/\gamma^*$ production
contain graphs where the external off-shell gauge boson does not
couple directly to the external quark line with flavour $f=q$, but
rather to an internal closed fermion loop.
\begin{figure*}[t]
  \begin{center}
    \includegraphics[width=0.9\textwidth]{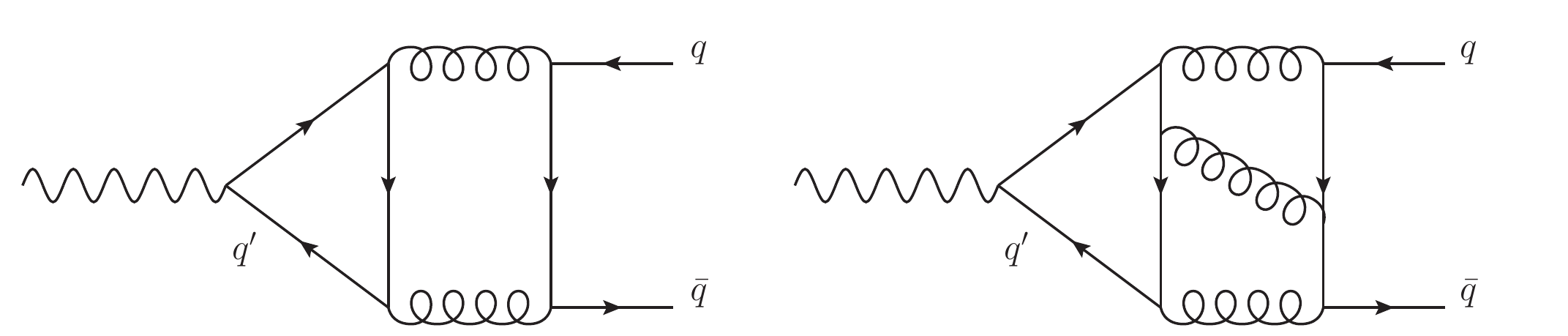}
  \end{center}
  \caption{Example of Feynman diagrams contributing to the `singlet'
    contribution at two and three loops.
  }
  \label{fig:singlet}
\end{figure*}
In Fig.~\ref{fig:singlet} we show representative diagrams at two and
three loops.
For such terms, customarily referred to as `singlet' contributions,
the factorisation in \eqref{eq:Hexp} is violated.
Up to \nnnlo, $Z/\gamma^*$ production features two singlet contributions,
namely
\begin{itemize}
\item[(a)] non-vanishing finite corrections entering the axial-vector current
but not the vector one, starting from two loops;
\item[(b)] corrections to the vector and axial-vector currents not factorising
the tree-level form factor, at three loops.
\end{itemize}

As for contribution (a), at two loops the singlet correction to the vector
current vanishes identically for each quark running in the fermion loop,
by means of Furry's theorem.
Since the axial-vector coupling is proportional to $T^3_f$,
within a given generation,
the singlet correction to the latter also vanishes,
provided the two quarks are degenerate in mass.
Such a cancellation does not take place exactly in the third family.
The leftover contribution, finite owing to the fact that the Standard
Model is anomaly free, and vanishing in the $m_{\rm top}\to m_{\rm bottom}$
limit, has been computed in ref.~\cite{Dicus:1985wx}.
For a given external quark line $q$, the radiative correction is
proportional to $A_q \sum_{q'} A_{q'} J(m_{q'},\mz)$ (eq.~(7),
Ref.~\cite{Dicus:1985wx}). As this is a correction to $H^{(2)}$ that
is proportional to $\JJq{00}{AA}$ but whose coupling is not proportional
to $A_q^2$, it does not factorise $|\Amp{0}|^2$. 
To our knowledge, these contributions are typically not included in resummed calculations.
In our implementation, we have not included these axial corrections to the
vertex.

We also stress that at ${\cal O}(\as^2)$ there are other terms of this type
arising from the `real-virtual' interference of the
$\ell^+\ell^-$+1 jet matrix elements.
Similarly to the two-loop singlet axial corrections to the vertex, these terms are UV- and IR-finite,
and vanish for each mass-degenerate quark family running in the fermion
loop. Such `real-virtual' corrections are also absent from our prediction.

The singlet contribution (b) to the quark form factor at three-loops
has been computed in Refs.~\cite{Baikov:2009bg,Lee:2010cga,Gehrmann:2010ue}.
In our notation, it contributes to the third-order expansion of
$J^\mu_V(q)$ as
\begin{equation}
\(\frac{\as}{2\pi}\)^3 J^{(3)}_V(q)
\,\, \ni \,\,
\(\frac{\as}{2\pi}\)^3 J^{(3s)}_V(q)
\,\, \equiv \,\,
\(\frac{\as}{2\pi}\)^3 C_3^s \, J_V(q)
\, ,
\end{equation}
where
\begin{eqnarray}
C_3^s
& = &
C_F \frac{N_c^2-4}{8N_c} \,
\( 4 - \frac{80}3 \zeta_5 + \frac{14}3 \zeta_3 + 10 \zeta_2 - \frac25 \zeta_2^2 \)
\, .
\end{eqnarray}

The $J^{(3s)}_V(q)$ current couples to the external quark line
in the loop amplitude through $\sum_{q'}Q_{q'}$ (or $\sum_{q'}V_{q'}$)
for $\gamma^*$ (or $Z$) exchange, where index $q'$ labels all possible
quark flavours running in the closed fermion loop.
The singlet contribution enters the hard-virtual coefficient through
the interference term $2 \Re({\Amp{0}}^{*} \Amp{3s})$
(we drop the subscript $\text{r,f}$ as $J^{(3s)}_V(q)$ is UV- and
IR-finite), where
\begin{eqnarray}
\label{eq:a3s}
\Amp{3s}
& = &
\frac{\sum_{q'} Q_{q'} Q_\ell}{q^2} J^{(3s)}_V(q) \cdot  J_V(\ell) + \nonumber\\
&&
+ \, \frac{\sum_{q'}V_{q'}}{(q^2-\mz^2) + i \mz\gammaz} J^{(3s)}_{V}(q)
\cdot \Big( V_\ell J_{V}(\ell) +  A_\ell J_{A}(\ell) \Big)\, .
\end{eqnarray}
We note that, in case of $\gamma^*$ production alone, one could collect
a term $\sum_{q'}Q_{q'}/Q_q$ in the first line of eq.~\eqref{eq:a3s},
thereby expressing the full result in a factorised form, as done in
refs.~\cite{Baikov:2009bg,Lee:2010cga,Gehrmann:2010ue}, which is not
possible when considering both $\gamma^*$ and $Z$ channels. 

By taking the interference with the tree-level amplitude~\eqref{eq:alo},
one gets
\begin{equation}
\label{eq:H3s}
2 \Re({\Amp{0}}^{*} \Amp{3s} )
\, = \,
2\lg |A^{(03s)}_\gamma|^2 + |A^{(03s)}_Z|^2 + I^{(03s)}_{\gamma^*/Z} \rg\,,
\end{equation}
where
\begin{eqnarray}
|A^{(03s)}_\gamma|^2 & = & \frac{C_3^s\sum_{q'} Q_{q'}}{q^4} \ Q_q Q_\ell^2 \
\JJq{00}{VV}^{\mubar\mu} \JJl{00}{VV}_{\mubar\mu}
\, ,
\\
|A^{(03s)}_Z|^2 
& = &
\frac{C_3^s \sum_{q'} V_{q'}}{(q^2-\mz^2)^2 + \mz^2\gammaz^2}
\,
\Big\{
V_q \(V_\ell^2 + A_\ell^2\)
\JJq{00}{VV}^{\mubar\mu} \, \JJl{00}{VV}_{\mubar\mu}
\nonumber\\
&&
\hspace{35mm}
+ \,
2A_qA_\ell V_\ell \,
\JJq{00}{VA}^{\mubar\mu} \, \JJl{00}{VA}_{\mubar\mu}
\Big\}
\, ,
\nonumber\\
I^{(03s)}_{\gamma^*/Z}
& = &
\frac{C_3^s (q^2-\mz^2)}{q^2[(q^2-\mz^2)^2 + \mz^2\gammaz^2]} \
\Big\{
Q_\ell V_\ell
\Big(
Q_q \sum_{q'} V_{q'} +
V_q \sum_{q'} Q_{q'}
\Big)
\JJq{00}{VV}^{\mubar\mu} \JJl{00}{VV}_{\mubar\mu}
\nonumber\\
&&
\hspace{35mm}
+ \,
Q_\ell A_q A_\ell \sum_{q'} Q_{q'} \, \JJq{00}{VA}^{\mubar\mu} \JJl{00}{VA}_{\mubar\mu}
\Big\}
\, .
\end{eqnarray}
In our implementation we included, at ${\cal O}(\as^3)$ the singlet
contribution as in eq.~\eqref{eq:H3s}, with the sum over $q'$ running
over the 5 massless flavours.
We have not included in our calculation the singlet contribution to the
massless axial-vector quark form factor at three loops, that has been
computed very recently in \cite{Gehrmann:2021ahy}. We leave this update
for a future development, as this contribution is expected to be
numerically negligible for the results presented in this article.

\bibliographystyle{JHEP}
\bibliography{N3LLp}

\end{document}